\documentclass{aa} 
\pdfoutput=1
\usepackage{natbib}
\bibpunct{(}{)}{;}{a}{}{,}
\usepackage{graphicx}
\usepackage{color}
\usepackage{amsmath}  %
\usepackage[varg]{txfonts}  
\usepackage{url} 
\usepackage{enumitem}
\setlist{noitemsep,nosep} 
\usepackage{fixltx2e}   
\usepackage[figuresright]{rotating}
\renewcommand{\deg}{\ensuremath{^\circ}}
\newcommand{\new}[1]{#1} 

\begin{document}

\title{Extragalactic archeology with the GHOSTS Survey I.}
\subtitle{Age-resolved disk structure of nearby low-mass galaxies}
\author{David Streich\inst{\ref{AIP}}
\and Roelof S. de Jong\inst{\ref{AIP}}
\and Jeremy Bailin\inst{\ref{UA}}
\and Eric F. Bell\inst{\ref{UMich}}
\and Benne W. Holwerda\inst{\ref{leiden}}
\and Ivan Minchev\inst{\ref{AIP}}
\and Antonela Monachesi\inst{\ref{MPA}}
\and David J. Radburn-Smith\inst{\ref{UW}}
}
\institute{Leibniz-Institut f\"ur Astrophysik Potsdam (AIP), An der Sternwarte 16, 14482 Potsdam, Germany, \email{dstreich@aip.de}\label{AIP}
\and Department of Physics \& Astronomy, University of Alabama, Box 870324, Tuscaloosa, AL 35487, USA\label{UA}
\and Department of Astronomy, University of Michigan, 304 West Hall, 1085 S. University Ave., Ann Arbor, MI 48109, USA\label{UMich}
\and University of Leiden, Sterrenwacht Leiden, Niels Bohrweg 2, NL-2333 CA Leiden, The Netherlands\label{leiden}
\and Max Planck Institut für Astrophysik, Karl-Schwarzschild-Str. 1, 85748 Garching, Germany\label{MPA}
\and Department of Astronomy, University of Washington, Box 351580, Seattle, WA 98195, USA\label{UW}
}
\date{Received .../ Accepted ...}

\abstract{}
{We study the individual evolution histories of three nearby low-mass edge-on galaxies (IC\,5052, NGC\,4244, and NGC\,5023).}
{Using resolved stellar populations, we constructed star count density maps for populations of different ages and analyzed the change of structural parameters with stellar age within each galaxy.}
{We do not detect a separate thick disk in any of the three galaxies, even though our observations cover a wider range in equivalent surface brightness than any integrated light study. While scale
heights increase with age, each population can be well described by a single disk. Two of the galaxies contain a very weak additional component, which we identify as the faint halo. The mass of these faint halos is lower than 1\% of the mass of the disk.\\
The three galaxies show low vertical heating rates, which are much lower than the heating rate of the Milky Way. This indicates that heating agents, such as giant molecular clouds and spiral structure, are weak in low-mass galaxies.\\
All populations in the three galaxies exhibit no or only little flaring. While this finding is consistent with previous integrated light studies, it poses strong constraints on galaxy simulations, where strong flaring is often found as a result of interactions or radial migration.}
{}


\maketitle

\section{Introduction}

Studies of galaxy evolution have always had to solve the problem of the very long timescales involved: Galaxies only change very slowly over millions and billions of years. The study of galaxy evolution must therefore rely on indirect methods and measurements.

Fortunately, the finite speed of light 
opens a way to look into the past. By observing galaxies far away, we automatically look at galaxies in the past. This allows us to study the evolution of galaxy properties from the early Universe to the present time. This leads to a good knowledge of the evolution of the overall galaxy populations, but it does not help with the study of the evolution of individual galaxies. The question what the predecessor of the Milky Way, or of any current galaxy in general, has looked like cannot be answered through such studies.

A different approach to studying galaxy evolution is the detailed analysis of current galaxies in order to distinguish their individual histories. The detailed knowledge of the kinematics, ages, and chemistry of all stellar populations of a galaxy today would allow inference on its formation and evolution. This approach is often called galactic archeology or near-field cosmology. But even without the full knowledge of the kinematical and chemical distribution of the stars, galaxies contain much information about their past. Integrated colors (and their gradients) contain information about the ages and metallicities of the underlying stellar distribution, integrated spectra give valuable information about the ages, kinematics, and metallicities and even the temporal evolution of them \citep[e.g.][]{ocvirk08}, and deep color-magnitude
diagrams allow determining the star formation history (SFR) and metal enrichment function.

Most processes in the evolution of galaxies also leave their signature on the structure of galaxies; an active merger history,
for instance, will lead to a strong bulge, while the quiescent accretion of gas forms a thin galactic disk. In this paper we aim to better understand galaxy evolution by analyzing the vertical and radial structure of stellar populations in nearby galaxy disks.

\subsection{Vertical structure of disks}
From simple modeling of the vertical structure of galactic disks,
it is expected that it follows a sech$^2$ density profile when an isothermal disk is assumed \citep{vanderkruit81a}, but observationally, a centrally more peaked profile such as a sech or exponential profile\footnote{Often the more general sech$^{2/n}(n z/z_0)$ is used, which also includes an exponential for $n\rightarrow \infty$. These functions differ only in the midplane, they are all asymptotically exponential for large z.} was found to be more appropriate for the stellar distribution \citep{vanderkruit88,degrijs97b}. This deviation from the isothermal model can be explained by a mixture of stellar populations with different velocity dispersions, as they are observed in the Milky Way, where the velocity dispersion increases with stellar age \citep{wielen77,carlberg85}.

Observations show that a single-disk profile is often not sufficient to describe the disk, but that a second component with a larger scale
height is necessary. These so-called thick disks were first discovered in S0 galaxies \citep{burstein79,tsikoudi79}, then in many other galaxies 
and in our own Milky Way \citep{gilmore83}. Later they were found to be ubiquitous \citep{pohlen04,yoachim06,comeron11b}.

Thick disks were usually seen as a distinct component. Many properties of the thick disk could be determined
in the Milky Way: It is kinematically hot \citep[e.g.,][]{nissen95,girard06}, old, metal poor, and alpha enhanced \citep[e.g.,][]{fuhrmann08}. Recently, the picture of a clear distinct thick disk has been challenged by \citet{bovy12a,bovy12b}. They proposed the disk to be a superposition of many ``mono-abundance population disks'', which can be each described by a single-disk model.

In many studies it was found that thick disks do not only have larger scale heights, but also larger scale lengths \citep[][and others]{robin96,yoachim06,juric08}, but this has also been questioned for the MW  by \citet{bensby11} and \citet{bovy12b} as well as
by recent simulations \citep{stinson13,bird13,minchev14}.

The possible formation processes of the thick disk are still unclear. Thick disks may
\begin{itemize}
\setlength{\itemsep}{0pt}
\item[-] form as thick disks, for example, in very massive star formation aggregates \citep{kroupa02} or turbulent clumpy disks \citep{bournaud09} such as would appear after gas-rich mergers \citep{brook04};
\item[-] be created from a pre-existing thin disk that is thickened through internal heating or scattering by giant molecular clouds or spiral arms\footnote{but this effect was shown to be too weak to form the observed thick disks \citep{villumsen85}.};
\item[-] be created from a pre-existing disk by redistributing stars through outward radial migration \citep{schoenrich09,loebman11}\footnote{We
note that various more recent simulations \citep{minchev12b,martig14b,veraciro14} have cast doubt on the heating effect of radial migration.};
\item[-] result from external heating of an earlier disk by minor mergers \citep{quinn93} or dark matter halo bombardment \citep{kazantzidis08};
\item[-] be made of accreted material from satellite galaxies \citep{statler88,abadi03}.
\end{itemize}
Obviously, a combination of these processes might also play a role, with different ratios in different types of galaxies.

\subsection{Radial structure of disks}
Disks are thin, rotationally supported structures. They show a radial light profile that is (close to) exponential  \citep{patterson40,devaucouleurs59,freeman70}. However, some disks are truncated in the outer parts \citep{vanderkruit79}, which means that outside the break radius (typically at a radius of 1.5-6.0 scale lengths) a different exponential is needed to describe the surface brightness profile. In most cases (60\%) the outer exponential is steeper than than the inner one, but for 30\% it becomes shallower. Only a minority of 10\% does not show a break in the radial light profile at all \citep{pohlen06}.

Disks form through the dissipational collapse of a cooling gas cloud that forms stars after collapsing. How this leads to an exponential light profile is a still-unsolved problem. The two prevailing ideas are that (1) the exponential profile reflects the initial angular momentum distribution of the gas cloud \citep{freeman70,larson76} and (2) the viscosity of the gas leads to a redistribution of angular momentum that results in an exponential profile \citep{lin87}. More recently, \citet{elmegreen13} suggested that stellar scattering off of transient mass clumps in the disk naturally results in an exponential profile.

A widely accepted idea is that disks form from inside out: at first, gas with low angular momentum gathers in the center and forms stars, and later gas with higher angular momentum gathers around this and forms more stars. This behavior is seen in many different models \citep[e.g.,][]{larson76,white91,burkert92,mo98,naab06,brook06} and leads to negative age and metallicity gradients \citep[e.g.,][]{matteucci89,chiappini97,boissier99,prantzos00}. Observational evidence for the inside-out formation of disks can be directly found in age gradients \citep{munozmateos07,macarthur09,williams09b}, in metallicity gradients \citep{macarthur09}, color gradients \citep{dejong96c}, and in the change in galaxy sizes with redshift \citep{patel13,vanderwel14} . 

The interpretation of radial gradients has to take into account that stars can change their radial position in the galaxy with time. It has long been known that scattering effects can blur and heat galactic disks. But it has been found only relatively
recently that stars can also change their radial position and still keep the circular nature of their orbits \citep{sellwood02}. This radial migration process dramatically changes the chemical evolution and the metallicity gradients \citep{schoenrich09} and leads to a radial extension of disks \citep{roskar08a,sanchez09}. It might also cause the formation of a thick-disk component \citep{loebman11}.

While the physical reason for the exponential disk profiles is still unclear, the reasons for a break in the radial profiles are even less well understood. The breaks occur at a similar surface brightness in all galaxies \citep{kregel04a}, at the same radius for all heights above the plane, and for all populations \citep{pohlen07,dejong07b}. They can be observed even at redshifts of $z\approx1$ \citep{perez04,trujillo05}. Truncated disks have a color minimum at the break radius, while antitruncated, that
is, disks whose scale length is greater beyond the break, and unbroken disks only show a flattening of their color profile at large radii, both in the local Universe and at higher redshifts ($z\approx1$) \citep{bakos08,azzollini08a,azzollini08b}.

The nature of breaks is also still discussed. \citet{bakos08} claimed that a break is only due to a change in stellar populations and that the mass profile is unbroken, but this is in contrast to results from the GHOSTS survey \citep{dejong07b,radburn12}, which show breaks in star counts of different populations. The break in simulations is also connected to a steepening of the stellar mass profile \citep{roskar08a,sanchez09}. These simulations predict a minimum of the age distribution at the break radius and a smooth metallicity profile.

The numerous break formation models can be roughly divided into two groups. The first connects the break to a break in star formation, either due to a limited gas distribution \citep{vanderkruit87} or to a star formation threshold \citep{fall80,kennicutt89,dopita94,schaye04,elmegreen06}. The second group contains models that dynamically redistribute stars after their formation, either as a result of secular angular momentum redistribution \citep{debattista06,foyle08,minchev12a} or of tidal interactions \citep{gnedin03,kazantzidis08}.

\paragraph{}
We here examine the structure of different stellar populations with distinct ages to study the temporal evolution of galaxy disks. In Sect.~\ref{sec:methods} we explain the data and methods, we present our results in Sect.~\ref{sec:results}, discuss them in Sect.~\ref{sec:discussion}, and conclude with a summary in Sect.~\ref{sec:summary}.
\section{Data and methods}
\label{sec:methods}
\subsection{GHOSTS Survey}
\label{sec:ghosts}
\begin{table*}[!ht]
\centering
\caption{Details of the galaxy sample.}
 \begin{tabular}{lccccccccc}
\hline \hline
  name      & RA(J2000.0)\tablefootmark{1} & DEC(J2000.0)\tablefootmark{1}& z$_0$\tablefootmark{1}&PA\tablefootmark{2}&Incl.\tablefootmark{2} \tablefootmark{a}&morph.\tablefootmark{2}& V$_{max}$\tablefootmark{2}&(m$-$M)$_0$\tablefootmark{3}&dist\tablefootmark{3}    \\ 
           & (\deg)     & (\deg)    & (\arcsec)& (\deg) & (\deg) &       &  (km/s)  &(mag)      & (Mpc)   \\ 
\hline
  \object{IC\,5052}  & 313.006792 & -69.19331 & 13.35 & -38.0  & 90.0   & 7.1  & 79.8      & 28.76     & 5.6     \\
  \object{NGC\,5023} & 198.052500 & 44.041222 & 9.29  & 27.9   & 90.0   & 6.0  & 80.3      & 29.06     & 6.5     \\
  \object{NGC\,4244} & 184.373583 & 37.807111 & 22.14 & 42.2   & 88.0   & 6.1  & 89.1      & 28.21     & 4.4     \\
  \object{NGC\,4631} & 190.533375 & 32.541500 & 13.74 & 82.6   & 85.0   & 6.5  & 138.9     & 29.34     & 7.4     \\
  \object{NGC\,891}  &  35.639224 & 42.349146 & 11.86 & 22.8   & 88.0   & 3.0  & 212.2     & 29.80     & 9.1     \\
  \object{NGC\,7814} &   0.812042 & 16.145417 &  -    & 134.4  & 70.6\tablefootmark{d}   & 2.0  & 230.9     & 30.80     & 14.4    \\
  \object{NGC\,4565} & 189.086584 & 25.987674 & 13.67 & -44.8  & 90.0   & 3.2  & 244.9     & 30.38     & 11.9    \\
\hline
 \end{tabular}
\tablefoot{\tablefoottext{1}{Right ascension, declination, and scale heights from K-Band fits by \citet{seth05a}};\tablefoottext{2}{position angle, inclination, morphology index, and rotational velocity from HyperLEDA \citep{paturel03}}; \tablefoottext{3}{distance modulus and distance from \citet{radburn11}}.\\
\tablefoottext{a}{While we list the inclination value from HyperLEDA here, our selection of edge-on galaxies was made based on a visual inspection of the images. Therefore we have included NGC\,7814 because of its thin and straight dust lane, but excluded NGC\,253 because of its visible spiral structure.}}
\label{tab:galaxydata}
\end{table*}

GHOSTS is an extensive multicycle HST survey that images the resolved stellar populations in the halos and outer disks of 17 nearby disk galaxies. The galaxies in the GHOSTS sample span a wide range of morphologies, from Sab to Sd, and masses (with $V_{rot}$ ranging from $80$\,km/s to $230$\,km/s). \new{A detailed description of the survey is given in \citet{radburn11} and the data are publicly available at the Mikulski Archive for Space Telescopes (MAST)\footnote{\url{http://archive.stsci.edu/pub/hlsp/ghosts/}}.}

The survey was performed using the cameras ACS and WFC3 and filters F814W and F606W. These filters were chosen because they have a high throughput and because red giants, which are supposed to be the majority of (bright) stars in the assumed old halos, have the highest flux in this wavelength range. 

The observations were designed so that red giant branch stars could be well resolved; the observations typically reach at least a signal-to-noise ratio (S/N) of 10 at 1\,mag below the tip of the red giant branch. For the nearer galaxies ($D<5$\,Mpc) this requirement was obtained within the limits for HST SNAP programs, while for farther galaxies GO programs were run. The dedicated GHOSTS observations were complemented with all archived observations fulfilling the same requirements. The current GHOSTS database contains data from cycles 12 to 21.

\subsubsection{GHOSTS data reduction}
All details of the observations and the data reduction can be found in \citet[][for the ACS data]{radburn11} and \citet[][for the WFC3 data]{monachesi15b}. Here, we only summarize the main points of the data reduction.

After basic data reduction (bias subtraction, flat-fielding, cosmic ray and bad pixel identification, drizzling), the program DOLPHOT, which is a modified version of HSTphot \citep{dolphin00}, is used to identify and photometrically measure the stars. We performed PSF photometry using the \textsc{tiny tim} point-spread functions (PSF). Together with the positions and magnitudes of the stars, DOLPHOT reports many photometric parameters for diagnostic purposes, such as sharpness, roundness, crowding, and the S/N. These parameters can be used to distinguish between stars and other objects. 

We defined a set of culls on the diagnostic parameters that ensured that a maximal number of stars and a minimal number of contaminants was detected. To estimate the number of contaminants that are expected for a given set of culls, observations from the Hubble archive were chosen that are aimed at high-redshift objects at high galactic latitudes. These observations are expected to be free of any detectable stars, and thus the culls should minimize the detections in these fields. At the same time, the culls were requested to maximize the recovery fraction of artificial stars that are placed into the same images. We call the culls we chose with those two requirements sparse field culls because they are optimized for fields with low star count numbers.

A second set of culls was optimized for a high recovery fraction of artificial stars in very crowded regions. This is called the crowded field cull. Compared to the sparse field cull, it has about 60\% more contaminants, but at the same time, the recovery fraction of stars in crowded regions is more than twice as high as for the sparse field culls. Since the typical number of contaminants is very small (about a few dozen per ACS field) compared to the number of stars in crowded regions (a few thousand per ACS field), the larger number of real detections outweighs the increased number of contaminants. For more details on determining optimal culls we refer to the appendix of \citet{radburn11}. 
In the following we always use the crowded field culls, if not stated otherwise.

To further increase the reliability of our star catalogs, we used SExtractor \citep{bertin96} to create a segmentation mask that masks out all extended sources such as background galaxies or nearby bright stars. 

\subsubsection{Sample of edge-on galaxies}
The GHOSTS sample contains seven edge-on galaxies that have data on their disks\footnote{An eighth galaxy, NGC\,5907, turned out to be farther away (D$=16.8$\,Mpc) than previously measured. Therefore its CMD is not deep enough for our analysis, and it was excluded from further observations due to the long exposure times that would have been needed.}, which we included in our analysis (see Table~\ref{tab:galaxydata}). In this first paper we examine the three low-mass galaxies (75\,km/s~$<V_{rot}<100$\,km/s). The three massive galaxies ($V_{rot}>200$\,km/s) will be covered in a subsequent paper. The intermediate-mass galaxy NGC\,4631 is analyzed in an additional paper because the strong effects of the interaction with its neighbor NGC\,4627 require a different
treatment.

\subsection{Age selection in the CMD}
\label{sec:ageselection}
Color-magnitude diagrams (CMD) allow us to separate the stars into populations of different ages. The CMDs of the GHOSTS galaxies show clearly distinguishable structures that belong to different stellar populations (see the CMDs in Figs.~\ref{fig:CMDs_ic5052} to~\ref{fig:CMDs_ngc5023} in the appendix). These populations are \new{the upper main sequence, the helium-burning branches, the asymptotic giant branch, and the red giant branch.}

\begin{figure*}[!ht]
\centering
\begin{minipage}{0.49\textwidth}
 \includegraphics[width=0.99\textwidth]{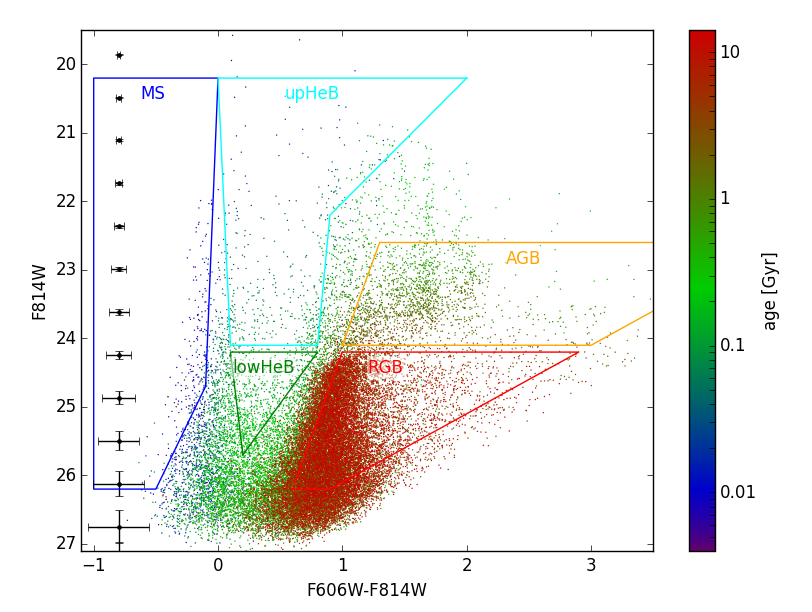}
\end{minipage}
\begin{minipage}{0.49\textwidth}
 \includegraphics[width=0.99\textwidth]{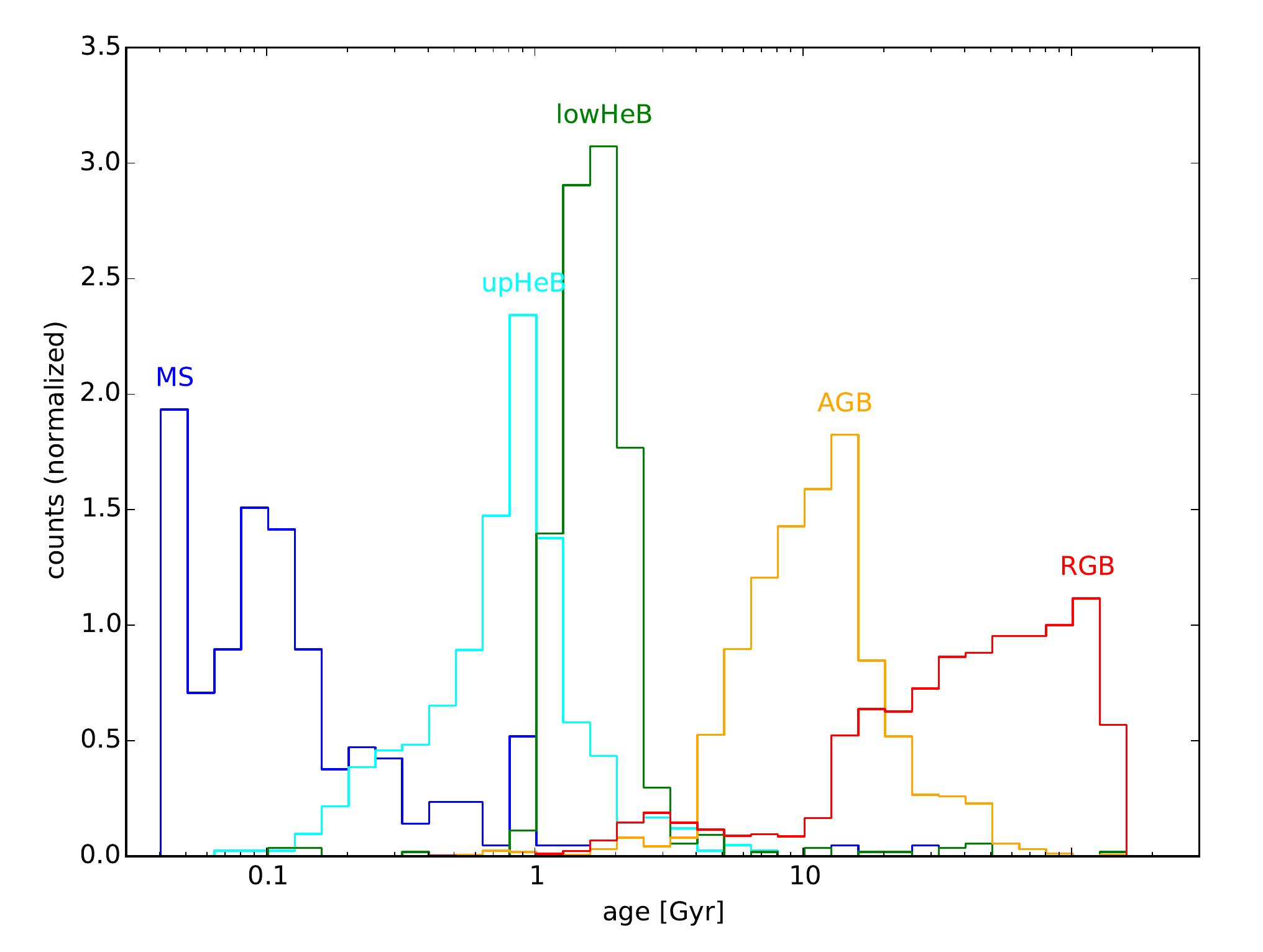}
\end{minipage}
\caption{\emph{Left panel}: An artificial CMD generated with a constant star formation rate and a flat metallicity distribution function ($-2.2<[Z]<0.2$). The colored boxes are the population boxes as described in the text. Stars are colored according to their age (see color bar). \emph{Right panel}: The age distribution of stars in the five population boxes in the left figure.}
\label{fig:cmd-ages}
\end{figure*}

To define age-separated stellar populations, we defined five CMD bins (see Fig. \ref{fig:cmd-ages}, left). These CMD bins were designed to cover different ages with as little overlap in age as possible (see Fig. \ref{fig:cmd-ages}, right). To define the bins and determine their age distributions, we used a synthetic CMD created with MATCH \citep{dolphin97,dolphin02} and using the Padova isochrone set \citep{girardi10,marigo08}. A constant star formation rate (from log(t)=10.15 to log(t)=6.6) and a flat metallicity distribution function (between [Z]=-2.2 and [Z]=0.2) were used.

The populations defined by these CMD bins are as follows.
\begin{description}
\item[Main sequence] (MS, blue): contains \new{massive core hydrogen-burning stars that are} mainly younger than 40\,Myr, but stars up to 300\,Myr can contribute to this population.
\item[Upper helium-burning branches] (upHeB, cyan): contains \new{massive core helium-burning stars with an age of} mainly between 40\,Myr and 150\,Myr, with a weaker contribution from stars between 25\,Myr and 300\,Myr. 
\item[Lower helium-burning branches] (lowHeB, green): contains \new{intermediate-mass core helium-burning stars with ages} mainly between 100\,Myr and 400\,Myr, with only little contamination by different ages.
\item[Asymptotic giant branch] (AGB, yellow): contains \new{intermediate-mass shell helium-burning stars with ages} mainly between 0.5\,Gyr and 2\,Gyr, but it might also contain a few stars as young as 350\,Myr or as old as 6.5\,Gyr 
\item[Red giant branch] (RGB, red): contains \new{low-mass shell hydrogen-burning stars that are} mainly older than 3.0\,Gyr but also younger AGB stars, and even lower HeB stars may contribute here.
\end{description}

Throughout this paper, we use the median ages of these populations (see Table~\ref{tab:population_properties}) for all plots showing the ages of the populations, and their error bars encompass the range between the 10 percentile and the 90 percentile of the age distributions.

\begin{table}[!ht]
\centering
\caption{Properties of the age distribution of the defined stellar populations, derived from models assuming a constant SFR and a flat metallicity distribution.}
 \begin{tabular}{cccccc}
\hline \hline
                   & MS & upHeB & lowHeB        & AGB   & RGB   \\ 
mean(age) [Gyr]  & 0.028        & 0.089 & 0.21  & 1.36  & 5.07 \\
std(age) [Gyr]   & 0.041        & 0.066 & 0.19  & 0.94  & 3.96 \\ 
median(age) [Gyr]& 0.010        & 0.082 & 0.18  & 1.16  & 4.41 \\
10 percentile    & 0.005        & 0.037 & 0.12  & 0.52  & 0.81 \\
90 percentile    & 0.037        & 0.146 & 0.27  & 2.52  & 11.5 \\
\hline
 \end{tabular}
\label{tab:population_properties}
\end{table}

\subsection{Creating stellar density maps}
\label{sec:densitymaps}

We used the star catalogs of the GHOSTS survey (see Sect.~\ref{sec:ghosts}) to create star count density maps for each of the populations defined in Sect.~\ref{sec:ageselection}. The raw star count maps are simple two-dimensional histograms of the stars in each CMD selection box. The bin size of the maps is (7.2\arcsec)$^2$. 
These raw star count maps were then corrected for incompleteness effects, using completeness maps created from the artificial star tests. They were also corrected for masked regions in the original images. We used the average of all fields in overlapping regions of different fields. 
Finally, bins with a completeness lower than 0.5 and bins whose
areas were masked to more than 40\% in the original image were excluded?.

\paragraph{Completeness correction}
Not all stars that are in our field can be reliably measured. Stars near the detection limit might be missed because of statistical variations in the photon flux, and stars in crowded or bright sky regions might also be missed. 
We quantified the completeness of our observations through the recovery fraction from the artificial star tests. For each population the fraction of recovered artificial stars was calculated in each spatial bin, that is, we created a completeness map for each population. This is different from the approach in the GHOSTS DR1 \citep{radburn11}, where the completeness was presented as a function of magnitude, color, and sky brightness, but not as a function of position. The change in the approach was necessary because we found that the completeness can vary within a field independent of sky brightness. In particular, we found that the completeness depends on the y coordinate of the HST images, with a decreasing completeness toward the chip gap. This is probably caused by CTE effects, which affect the central regions of an image more than the edges.

\subsection{Fitting methods}
\label{sec:fittingmethods}

To quantify the structural parameters of the disks, we fit different galaxy models to the star count maps. Because the star count maps often contain very low numbers of stars per pixel, the quality of the fit has to be calculated with a Poissonian likelihood estimator instead of the often used $\chi^2$. Such a Poissonian likelihood estimator was first proposed by \citet{cash79} and has strongly been argued for by \citet{dolphin02}. It can be derived in the same way as the $\chi^2$ statistics, but starting from the Poissonian probability function
\begin{align}
P_i = \frac{m_i^{n_i}}{n_i!}\text{e}^{-m_i},
\end{align}
with $m_i$ denoting the expected counts in pixel $i$ in the model and $n_i$ the observed counts in that pixel.
Then the maximum likelihood model can be obtained by minimizing the Poissonian likelihood ratio (PLR)\footnote{The last term in the sum only depends on the data, which of course do not change during the fit. Therefore it can be omitted without effecting the best-fit model, but including this term gives the advantage that each term under the sum is greater than zero (which is important for some numerical minimization routines) and that for large $n_i$ the PLR converges to the same value as a $\chi^2$.}
\begin{align}
PLR=-2\ln P = -2\sum_i\left(n_i\ln m_i-m_i+n_i(1-\ln n_i)\right). 
\end{align}

\paragraph{Determining scale heights:} 
To analyze the vertical structure of the population at different radii from the galaxy center, we fit an isothermal sheet model \citep{vanderkruit81a} to the stellar surface density profiles. 
The model reads 
\begin{align}
\label{model1D}
n(z) = n_0 \operatorname{sech}^2\left(\frac{z-z_c}{z_0}\right)+n_{bg},
\end{align}
where $n_0$ is the star count density in the midplane, $z_c$ the center of the stellar distribution relative to the galaxy midplane (defined through the center coordinates and the position angle of the galaxy), $z_0$ the scale height, and $n_{bg}$ the surface density of contaminants. 

We used an isothermal model even though many observations found a more peaked function to better fit the vertical light profile of galaxies. The isothermal model was preferable because we fit our model to distinct stellar populations of well-defined ages, for which the isothermal assumption is more justified than for the overall stellar content of a galaxy.

\paragraph{2D fits:} 

To measure scale lengths, scale heights, and break radii, we fit a simple model of an edge-on disk with a broken exponential as radial profile to the 2D maps of star count density:
\begin{align}
\label{model2D}
n(x,z) = n_{bg} + n_0 \text{sech}^2\left(\frac{z-z_c}{z_0}\right) n(x)
\end{align}
with 
\begin{align}
n(x)= \left.|x-x_c\right.| 
\begin{cases} \text{K}_1\left(\frac{|x-x_c|}{h_{r,i}}\right) & \text{for } |x-x_c|<r_{b} \\ 
\frac{\text{K}_1(r_{b}/h_{r,i})}{\text{K}_1(r_{b}/h_{r,o})} \text{K}_1\left(\frac{|x-x_c|}{h_{r,o}}\right) & \text{for } |x-x_c|>r_{b}
\end{cases},
\end{align}
where $x_c$ is the x-coordinate of the galaxy center, $h_{r,i}$ the inner and $h_{r,o}$ the outer scale length, and $r_b$ the break radius. 
K$_1(x)$ is the modified Bessel function of the second kind and $x \text{K}_1(x)$ is the projected surface density of an exponential disk seen edge-on, that is\footnote{This derivation assumes an infinite exponential disk, while we actually modeled a truncated disk. Therefore the model is not fully self-consistent.}
\begin{align}
\text{I}(x) = \int\limits_{-\infty}^{\infty}\exp\left(-\frac{\sqrt{x^2+y^2}}{h_r}\right)\text{d}y =
2\int\limits_{x}^{\infty} \frac{r\exp(-\frac{r}{h_r})}{\sqrt{r^2-x^2}}\text{d}r =
2x \text{K}_1\left(\frac{x}{h_r}\right).
\end{align}

This model assumes a constant scale height along the disk. We show in Sect.~\ref{sec:results} that this assumption is justified.
We also note that the model in Eq.~\ref{model2D} is a single-disk model, meaning that it does not contain any thick-disk, bulge, or halo components. This will of course limit its applicability; but for the low-mass galaxies in our sample, this model is sufficient to describe the distribution of stars in each population.

For the actual 2D fits we used the program IMFIT\footnote{\url{http://www.mpe.mpg.de/~erwin/code/imfit}} \citep{erwin14}, which permits using the Poisson likelihood ratio statistics for the minimization process. IMFIT is also designed to make the inclusion of additional image functions simple, and we have extended it with the broken edge-on disk model of Eq.~\ref{model2D}.

\new{The error bars shown in this paper were calculated with IMFIT using a bootstrap analysis of the given data. This means that all error bars only include statistical uncertainties.}

\subsection{Spitzer data}
\label{sec:spitzer-data}

For a comparison of the star count maps from GHOSTS with integrated light observations we used data from the InfraRed Array Camera \citep[IRAC;][]{fazio04} of the Spitzer Space Telescope. The data were reduced within the \textit{Spitzer Edge-On Disk Galaxies Survey} project \citep{holwerda06}. We here briefly describe
the data and the reduction process.

The data contain mosaics of 32 edge-on disk galaxies in all four IRAC channels. They were taken in a dedicated GO program (GO\,20268: The Formation of Dust Lanes in Nearby Edge-on Disk Galaxies; PI: R. S. de Jong) and were complemented with archival data. The observing strategy and data reduction were set up in such a way that the final data products were equivalent in quality to the data products from the Spitzer Infrared Nearby Galaxy Survey \citep[SINGS;][]{kennicutt03}.

The IRAC Basic Calibrated Data (BCD), together with the corresponding uncertainty (BUNC) and individual bad pixel mask (BDMSK), were retrieved with {\it Leopard} \citep{leopard}. The data reduction was performed with the standard Spitzer Science Center pipeline for raw IRAC data; the mosaics were created with the MOPEX software \citep{mopex}. The reduction process includes bias and flatfield corrections, conversion from engineering to scientific units, addition of WCS coordinates, combination of single exposures into a mosaic, and it removes cosmic rays. The final images are aligned with the galaxy major axes and have a pixel size of 0.75\,arcsec. In addition to the images, noise maps and mask files were created.

The IRAC channels~1 and~2, with effective wavelengths of $\lambda\approx3.6\,\mu m$ and $\lambda\approx4.5\,\mu m$, respectively, mainly trace stellar emission and have the advantage of being nearly unaffected by dust. We used channel~1 to determine the structural parameters of the overall stellar population in our galaxies by fitting the edge-on disk model (Eq.~\ref{model2D}) to the [3.6] data.

The IRAC channels~3 and~4, with effective wavelengths of $\lambda\approx5.7\,\mu m$ and $\lambda\approx8.0\,\mu m$, respectively, trace the emission of hot dust and polycyclic aromatic hydrocarbon molecules, but also include contributions from stellar emission. We used the data from channels~1 and~2 to estimate the stellar contribution to the light in channel~4 \citep[as described in][]{pahre04} and subtracted it from the channel~4 data to create images of the nonstellar emission in the galaxies. We fit the same edge-on disk model as in Eq. \ref{model2D} to these nonstellar images to derive estimates of the scale height and -length of the dust in our galaxies.

\section{Results}
\label{sec:results}
In this section we first study the vertical stellar density profiles at different galactocentric distances, followed by the radial distributions at different heights above the midplane. The section concludes with a discussion of the observations and a comparison with integrated light observations.

\subsection{Vertical profiles}
\label{sec:lowmass-verticalprofs}

\begin{sidewaysfigure*}
    \centering
    \begin{minipage}[t]{0.33\linewidth}
        \centering
        \includegraphics[width=0.99\textwidth]{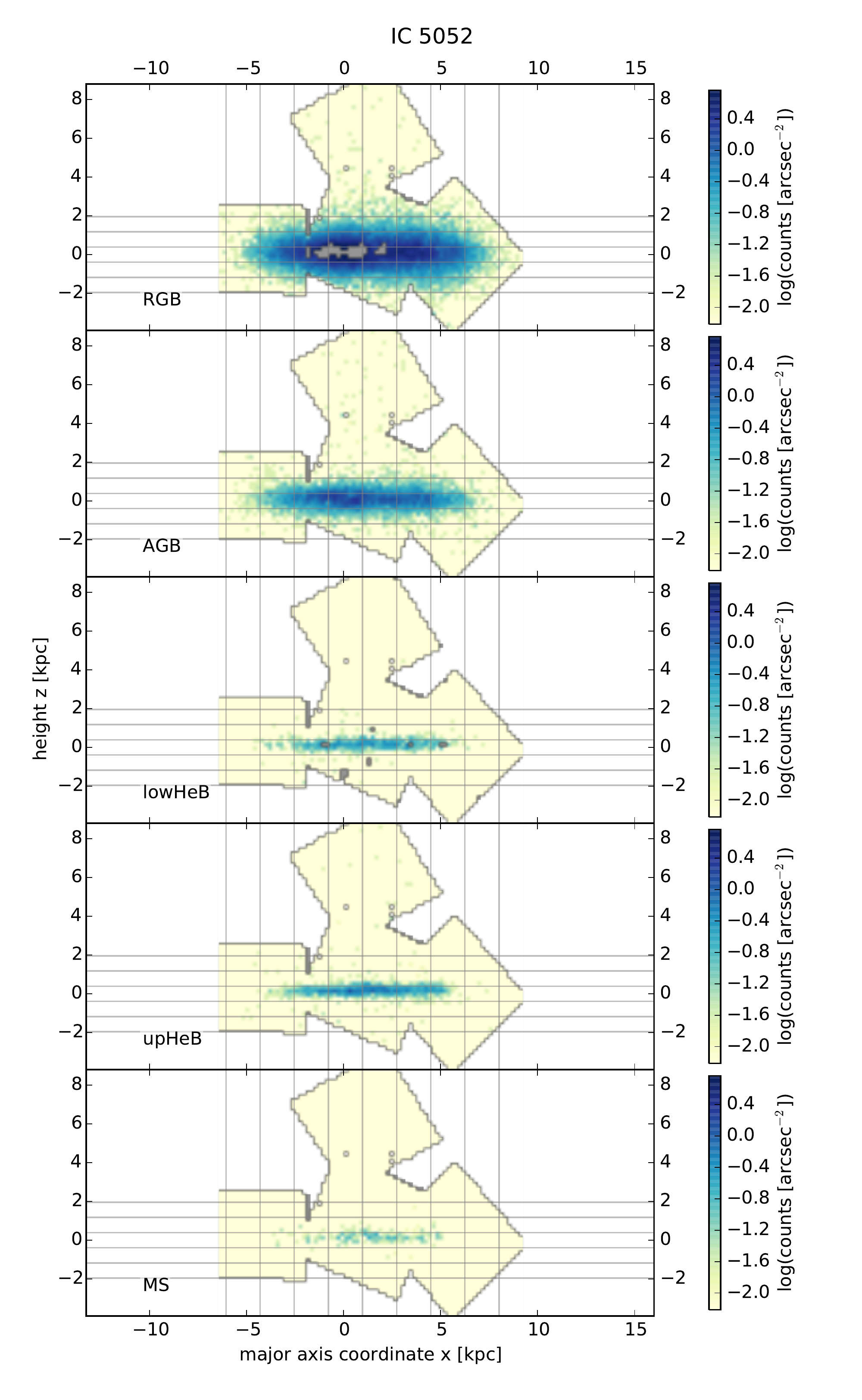}
    \end{minipage}%
    \begin{minipage}[t]{0.33\linewidth}
        \centering
        \includegraphics[width=0.99\textwidth]{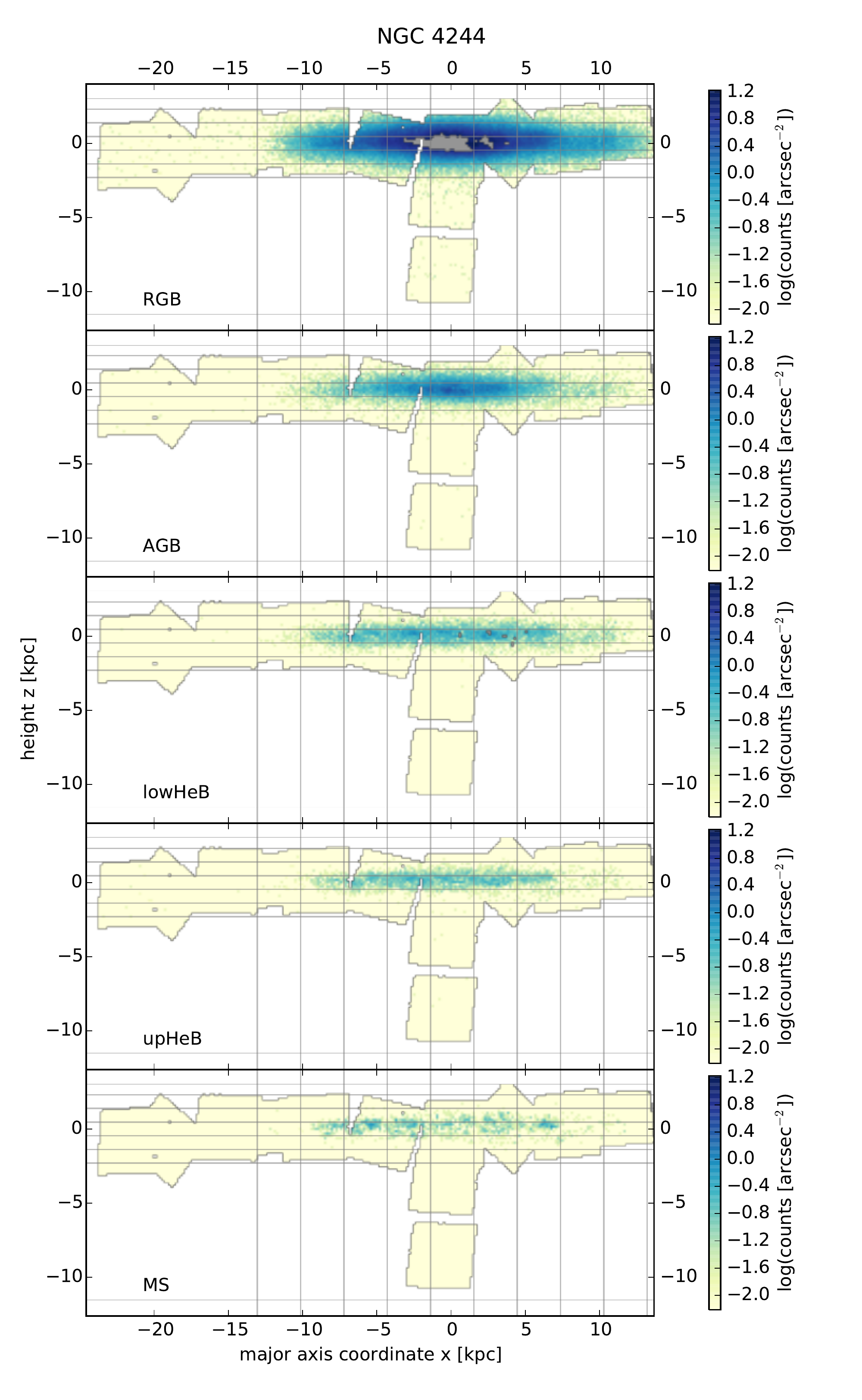}
    \end{minipage}
    \begin{minipage}[t]{0.33\linewidth}
        \centering
        \includegraphics[width=0.99\textwidth]{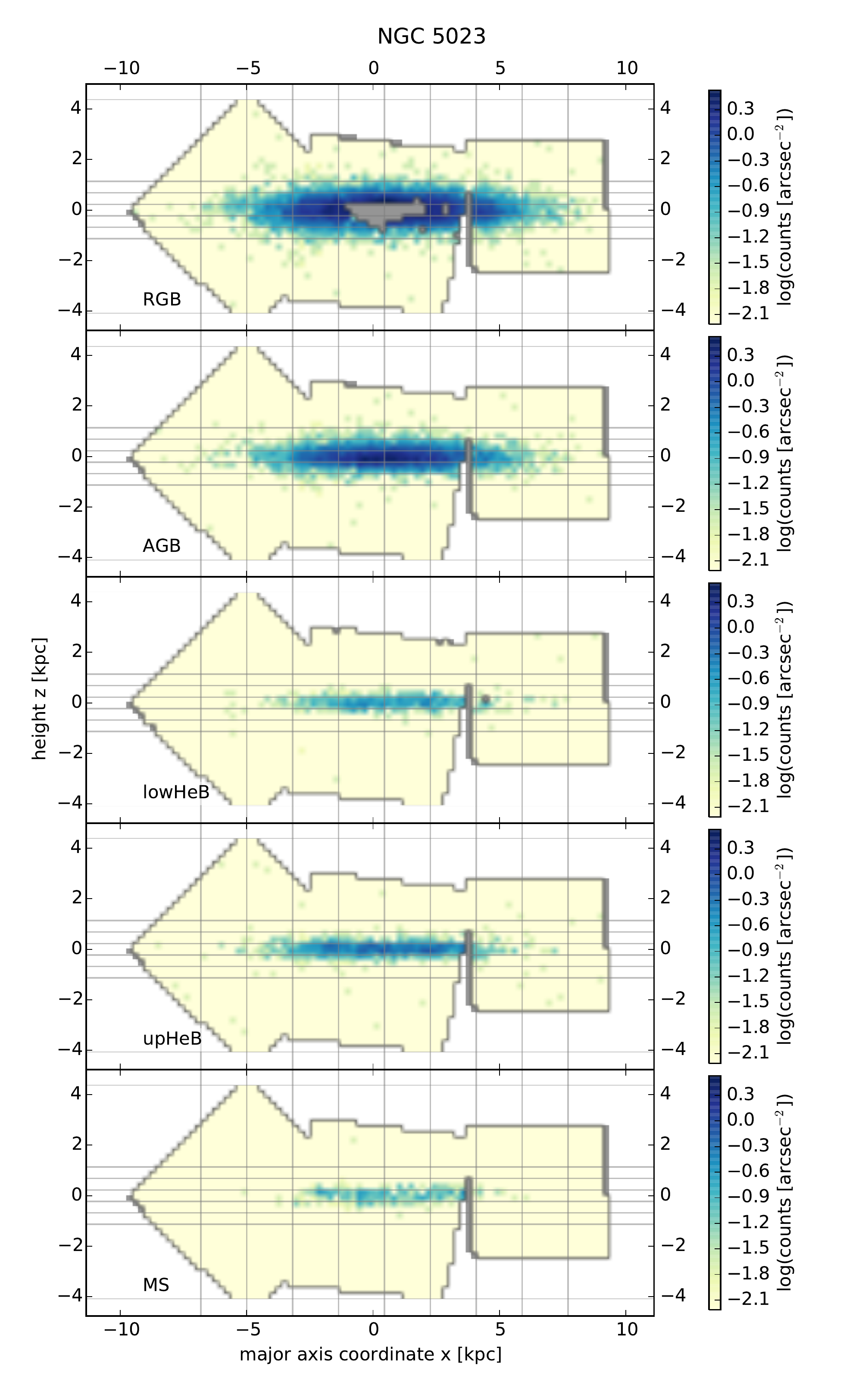}
    \end{minipage}%
    \caption{Star count surface density maps (logarithmic scale) of stellar populations in three low-mass edge-on galaxies: IC\,5052, NGC\,4244, and NGC\,5023 (left to right). Gray areas show the regions that were masked because of our crowding limit. Vertical and horizontal gray lines are the bin edges for the extraction of the vertical and horizontal profiles in Figs.~\ref{fig:vprofs1} and~\ref{fig:rprofs1}.}
    \label{fig:popmaps1}
\end{sidewaysfigure*}

We split the observed stars into the five age groups described in Sect.~\ref{sec:ageselection}. The stellar density maps for all of these populations are shown in Fig.~\ref{fig:popmaps1}. It is obvious that older populations are more extended than young ones. While this is some extent also visible in the radial direction, it is clearly evident in the vertical direction.

\begin{sidewaysfigure*}
    \centering
    \begin{minipage}[t]{0.33\linewidth}
        \centering
        \includegraphics[width=0.99\textwidth]{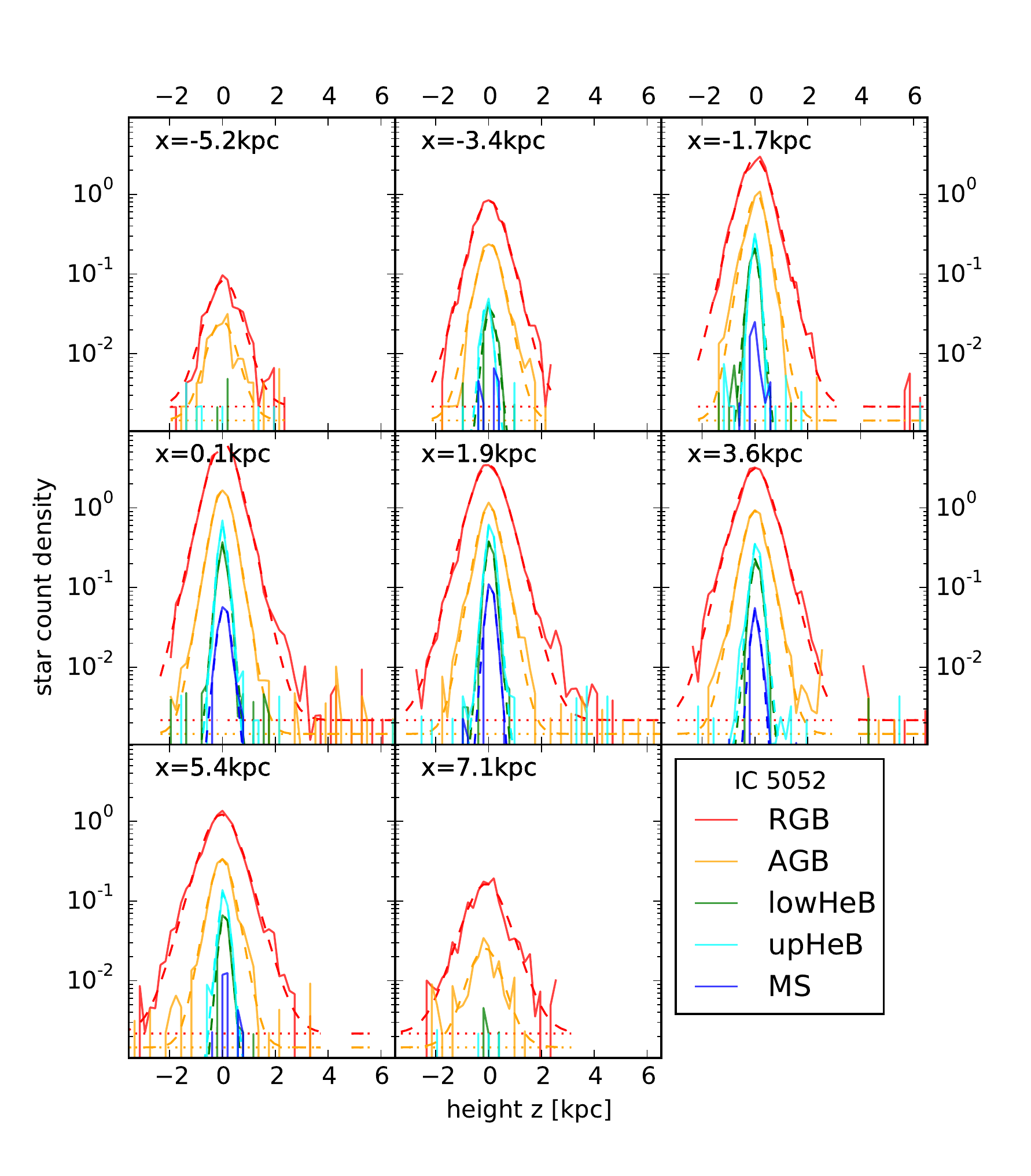}
    \end{minipage}%
    \begin{minipage}[t]{0.33\linewidth}
        \centering
        \includegraphics[width=0.99\textwidth]{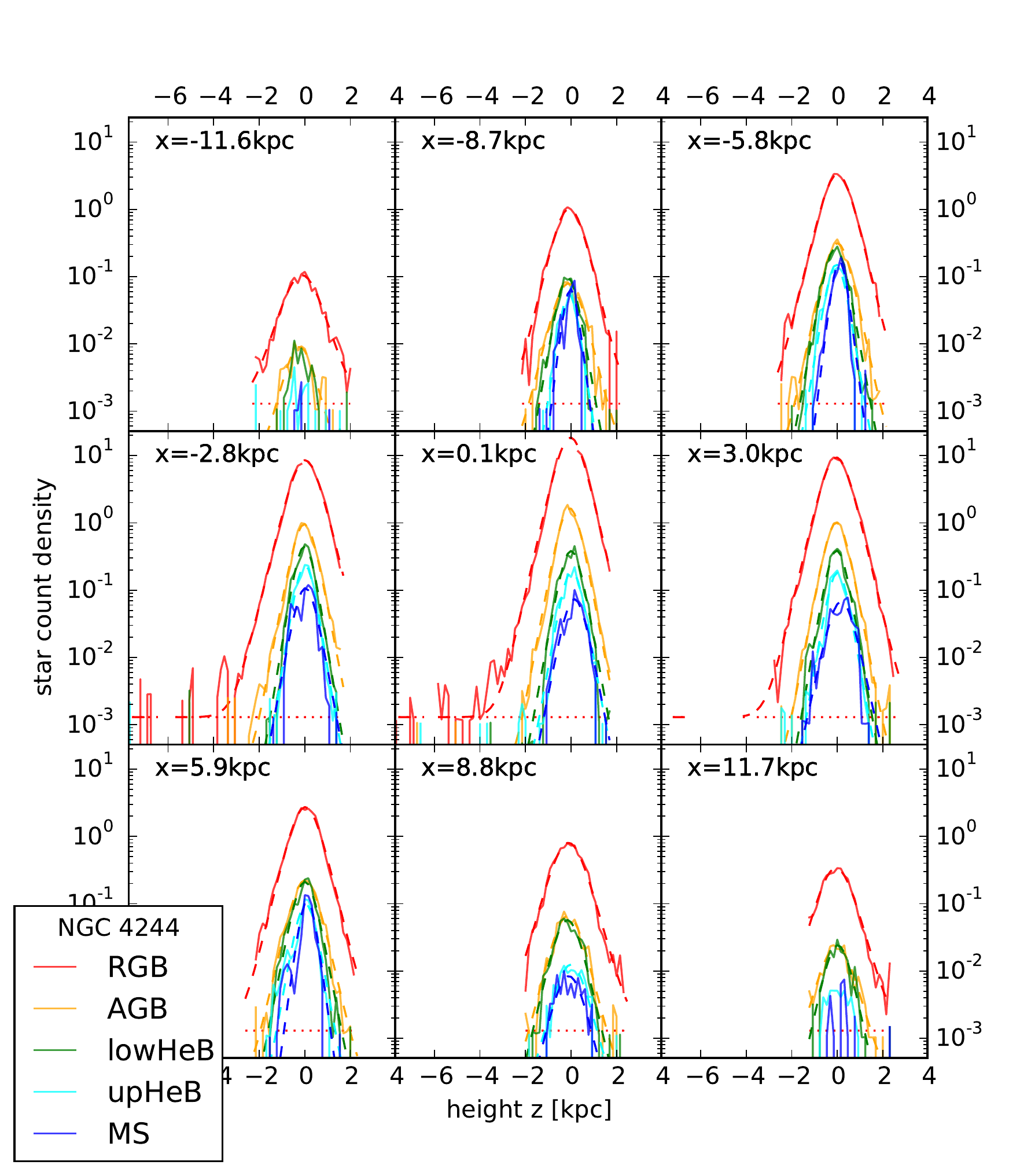}
    \end{minipage}
    \begin{minipage}[t]{0.33\linewidth}
        \centering
        \includegraphics[width=0.99\textwidth]{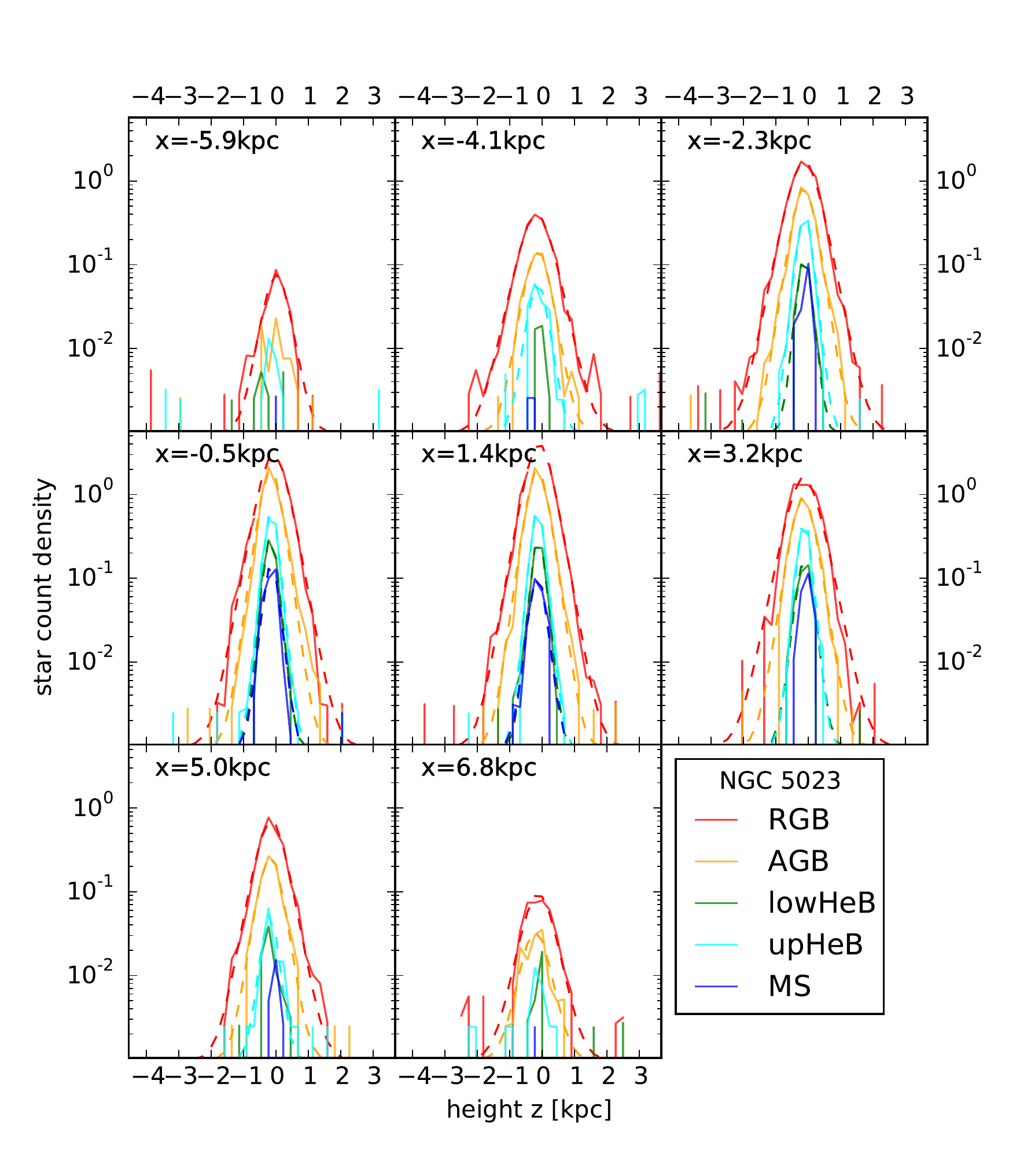}
    \end{minipage}%
    \caption{Vertical density profiles of the five populations (red - RGB; yellow - AGB, green - lower HeB, cyan - upper HeB, blue - MS) at different radii in three low-mass edge-on galaxies: IC\,5052, NGC\,4244, and NGC\,5023 (left to right). Solid lines are the data, dashed lines are the best sech$^2$ fits. The boundaries of the regions from which the profiles are extracted are shown in Fig.~\ref{fig:popmaps1} as vertical gray lines.}    \label{fig:vprofs1}
\end{sidewaysfigure*}

The thickness of the disks can be well examined in Fig.~\ref{fig:vprofs1}. Vertical star count profiles are shown there at different radial positions within the galaxies. The RGB stars (show in red) are the dominant population in all our galaxies at all radii; they have the highest star count density and the largest vertical extent. To quantify the thickness of each population, we fit a sech$^2$ model to each radial bin (see Eq.~\ref{model1D}); the fits are shown as a dotted line in Fig.~\ref{fig:vprofs1}. Remarkably, each profile can be fit well by a single sech$^2$ profile (plus background). There is no need to add a second component, that
is, a thick disk (for more details on this, see Sect.~\ref{sec:thickdisk?}). 
It is also noteworthy that most profiles show a clear peak in the center and that only the RGB profiles near the galaxy centers have a central dip. Such dips are often observed in vertical profile studies \citep[e.g.,][]{seth05b} and are usually attributed to the extincting effects of dust. We discuss this matter in
more detail in Sect.~\ref{sec:dusteffects}.

\begin{figure}[!tb]
 \centering
  \includegraphics[width=3.6in]{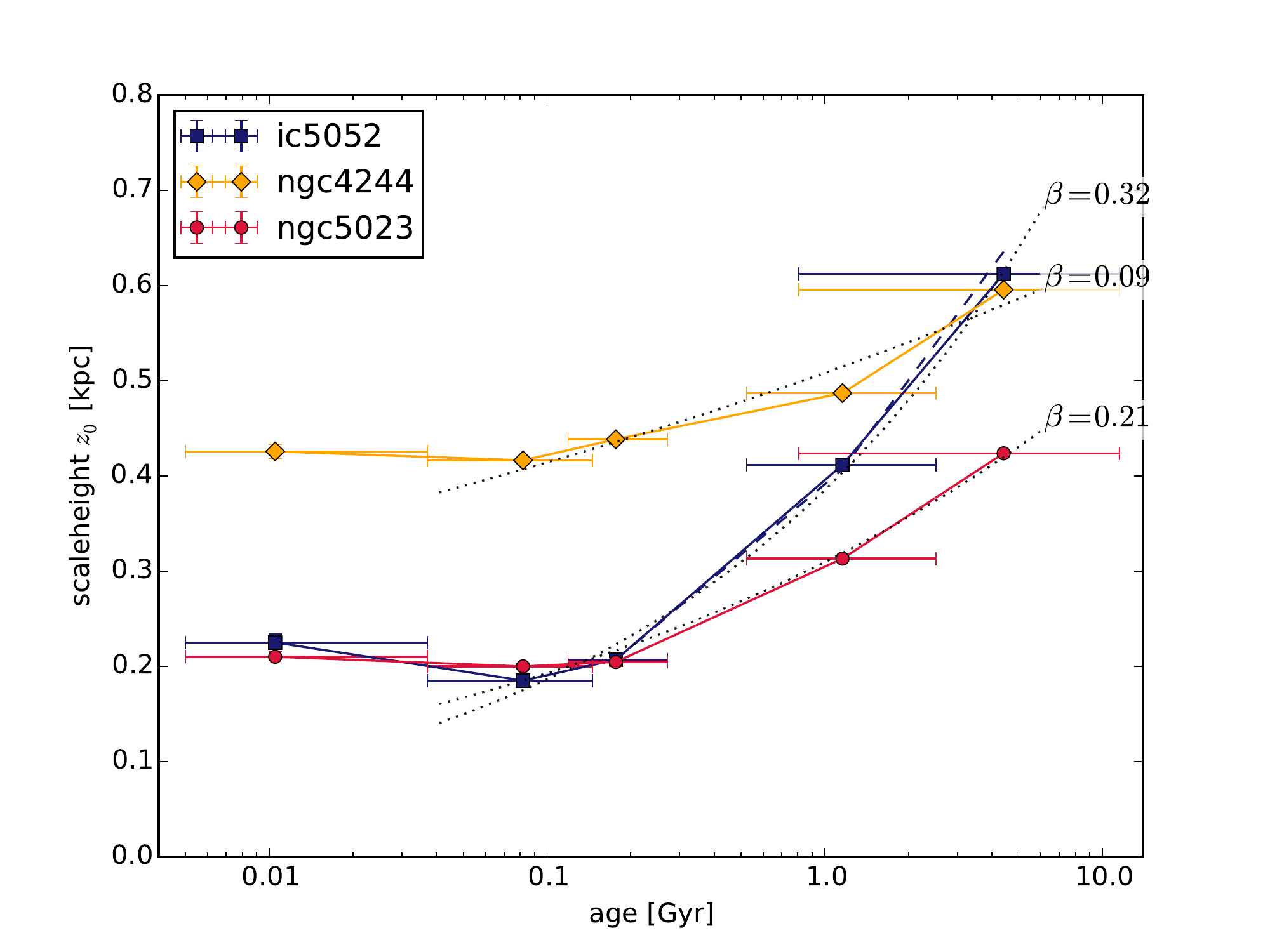}
 \caption[Dependance of scaleheight on age]{Change of the average scale height with stellar age. The blue dashed line shows the results of fits of IC\,5052 including an additional spheroidal component. The dotted lines are power-law $z_0\propto t^\beta$ fits to the data, excluding the youngest population; the power-law indices $\beta$ are listed on the right.}
 \label{fig:scaleheightevolution}
\end{figure}

The radially averaged scale heights that\new{ are determined through two-dimensional fits are shown as} as a function of age in Fig.~\ref{fig:scaleheightevolution}. In general, an increase of scaleheights with age is visible: While the three youngest populations (MS, upper HeB, and lower HeB) have approximately the same scale height, this scale height is smaller than the AGB scale height, which is even smaller than the scale height of the RGB stars.
The relative amount of this increase differs from galaxy to galaxy; in IC\,5052 the scale height increases by 100\% (from young to AGB) and 50\% (from AGB to RGB), in NGC\,5023 by 50\% and 33\%,
and in NGC\,4244 by only 20\%   from young to AGB and from AGB to RGB.

\begin{figure}[!tb]
    \centering
        \includegraphics[width=3.6in]{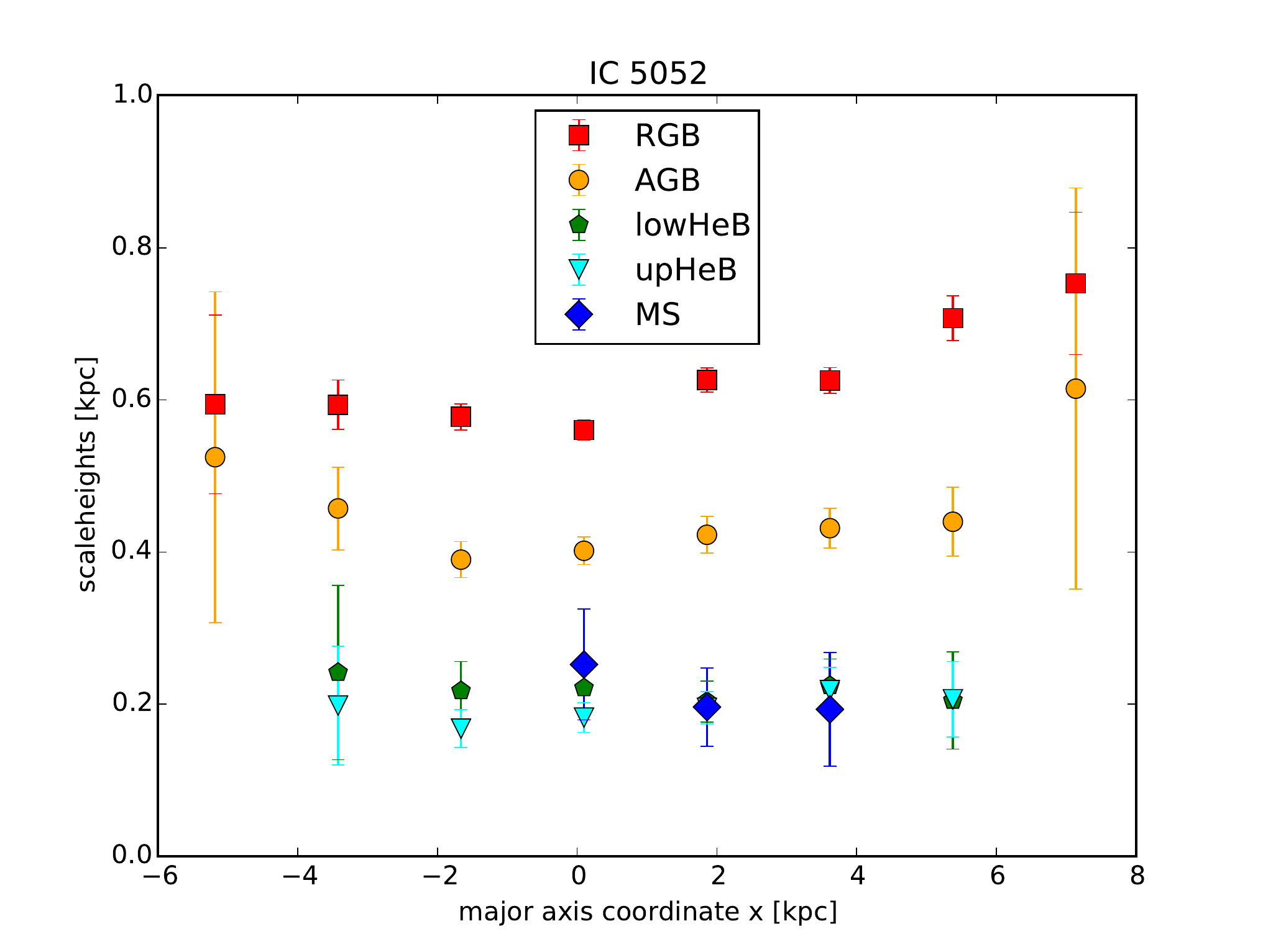}
        \includegraphics[width=3.6in]{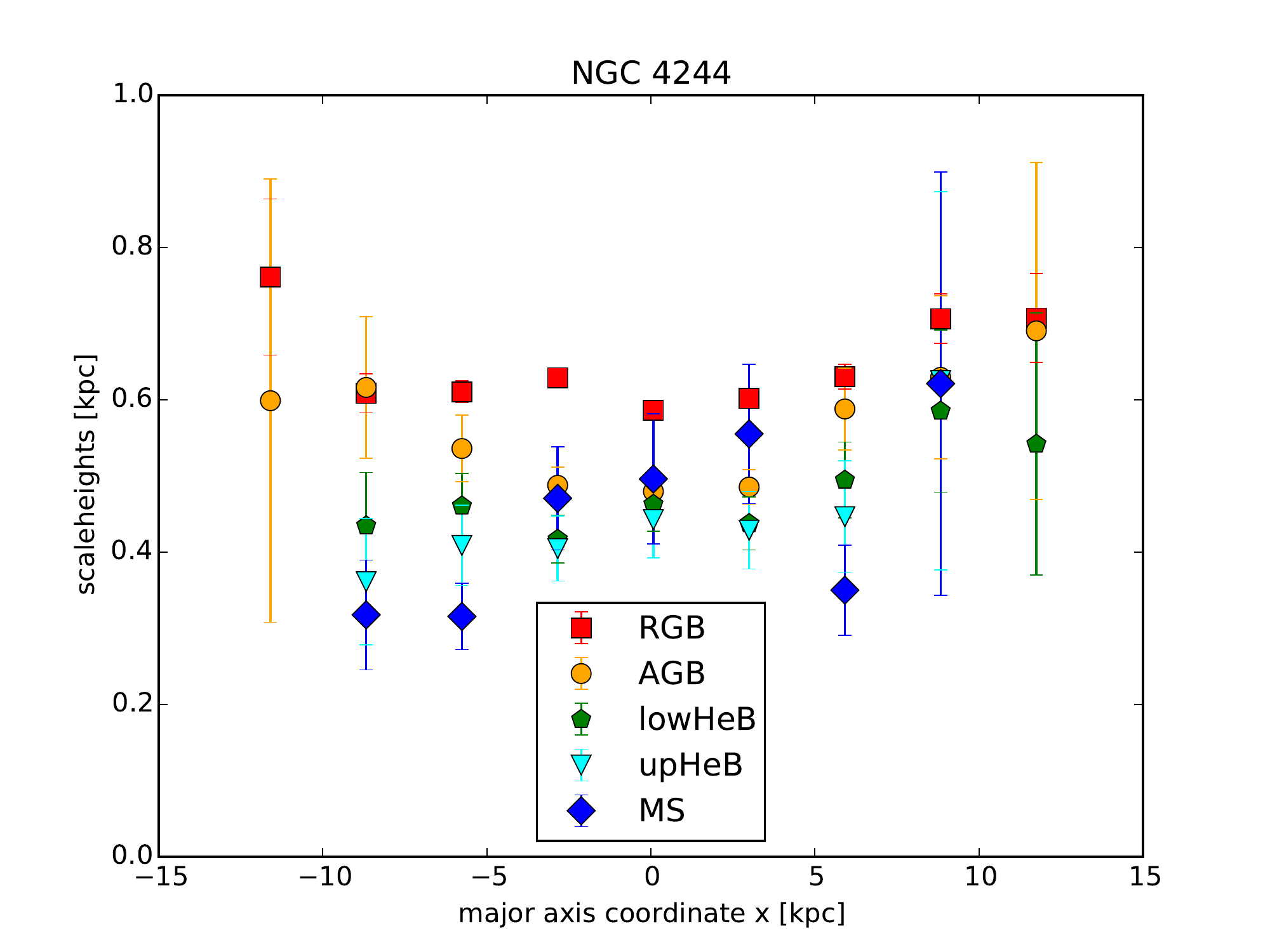}
        \includegraphics[width=3.6in]{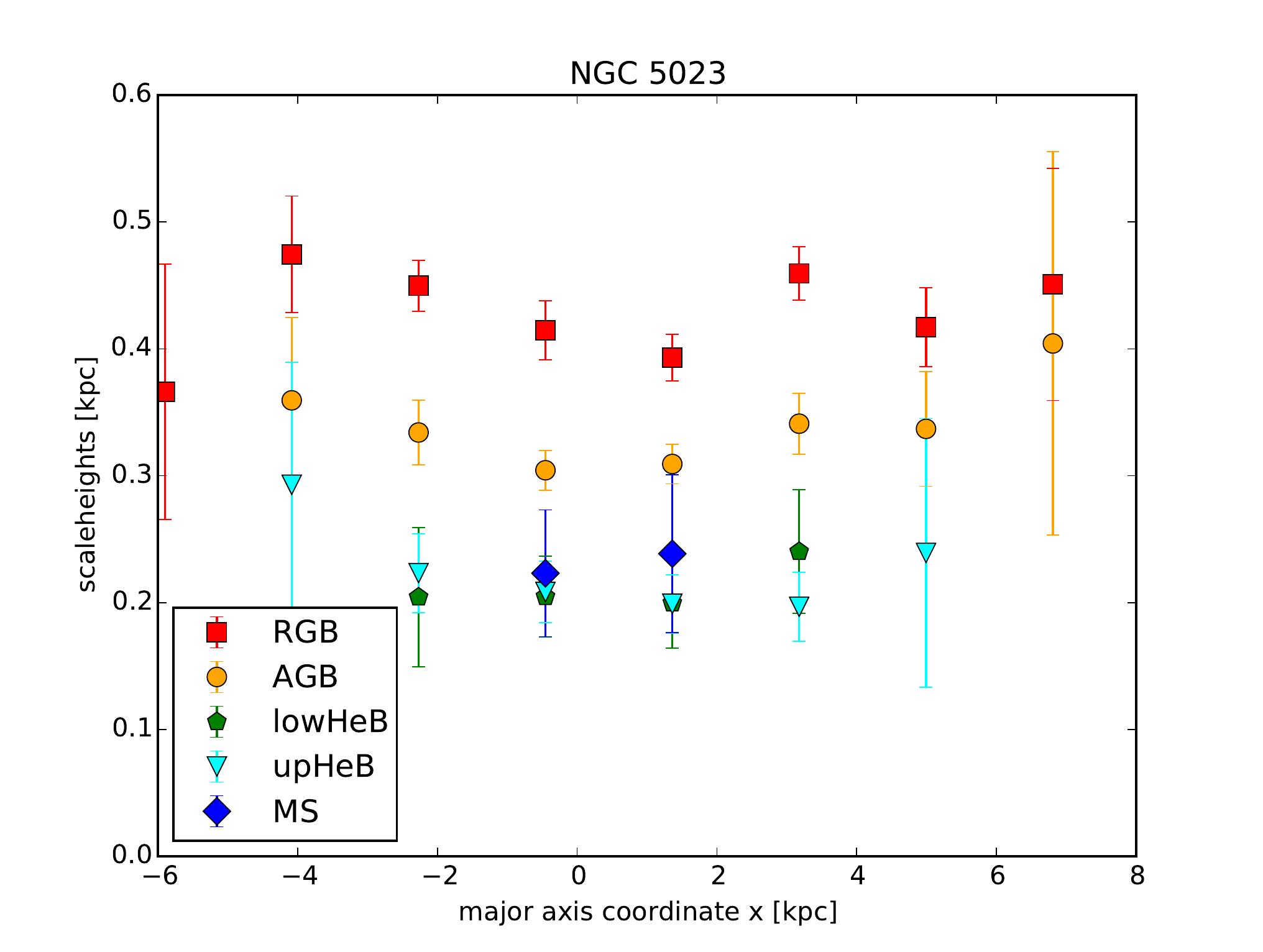}
    \caption{Disk scale heights at different radial position in the three low-mass edge-on galaxies IC\,5052, NGC\,4244, and NGC\,5023 (left to right). Colors are the same as in Fig.~\ref{fig:vprofs1}.}
    \label{fig:scaleheights1}
\end{figure}

We measured the scale heights at different radial positions within the galaxies, as can also be seen in Figs.~\ref{fig:vprofs1} and~\ref{fig:scaleheights1}. Figure~\ref{fig:scaleheights1} nicely shows that for most populations the scale heights change very little with radial position within each galaxy. While for the three young populations the scale heightx are constant along the disk, the older populations show some mild flaring. We quantified the strength of the flaring by fitting a straight line to the scale heights as a function of projected radius. The slopes of these regression lines are given in Table~\ref{tab:flaring}.

\begin{table}[!ht]
\centering
\caption{Increase of scale heights with projected radius.}
 \begin{tabular}{llccc}
\hline \hline
galaxy & population & slope & rel. flaring & $R_{max}$\\ 
          &      & [pc/kpc] & [\%]      & [kpc]\\\hline
IC\,5052  & RGB  & 22$\pm$5 & 28$\pm$ 7 & 7.1 \\
          & AGB  & 10$\pm$4 & 18$\pm$ 7 & 7.1 \\
NGC\,4244 & RGB  & ~7$\pm$3 & 14$\pm$ 5 & 11.9 \\
          & AGB  & 13$\pm$3 & 31$\pm$ 9 & 11.9 \\
NGC\,5023 & RGB  & ~9$\pm$7 & 14$\pm$13 & 6.8 \\
          & AGB  & 12$\pm$2 & 28$\pm$ 6 & 6.8 \\
\hline 
 \end{tabular} 
\tablefoot{Amount of flaring in the older populations of the three galaxies (younger populations do not flare). The column "slope" gives the absolute flaring $dz_0/d R$, the column "rel. flaring" the relative increase of scale height $(z_0(R_{max})-z_0(R$=$0))/z_0(R$=$0)$ over the full observed range in radii, and the column $R_{max}$ the maximum galactocentric radius to which a scale height was measured.} 
\label{tab:flaring}
\end{table}
The slopes in the young populations (MS, upHeB, and lowHeB) are consistent with zero flaring. The intermediate and old populations in all three galaxies have very similar slopes of $\approx 10$\,parsec per kiloparsec, except for the RGB in IC\,5052, which has almost 30\,pc/kpc.

\subsection{Radial profiles and 2D fits}

\begin{figure*}
    \centering
     \begin{minipage}[t]{0.99\textwidth}
        \centering
        \includegraphics[height=0.31\textheight]{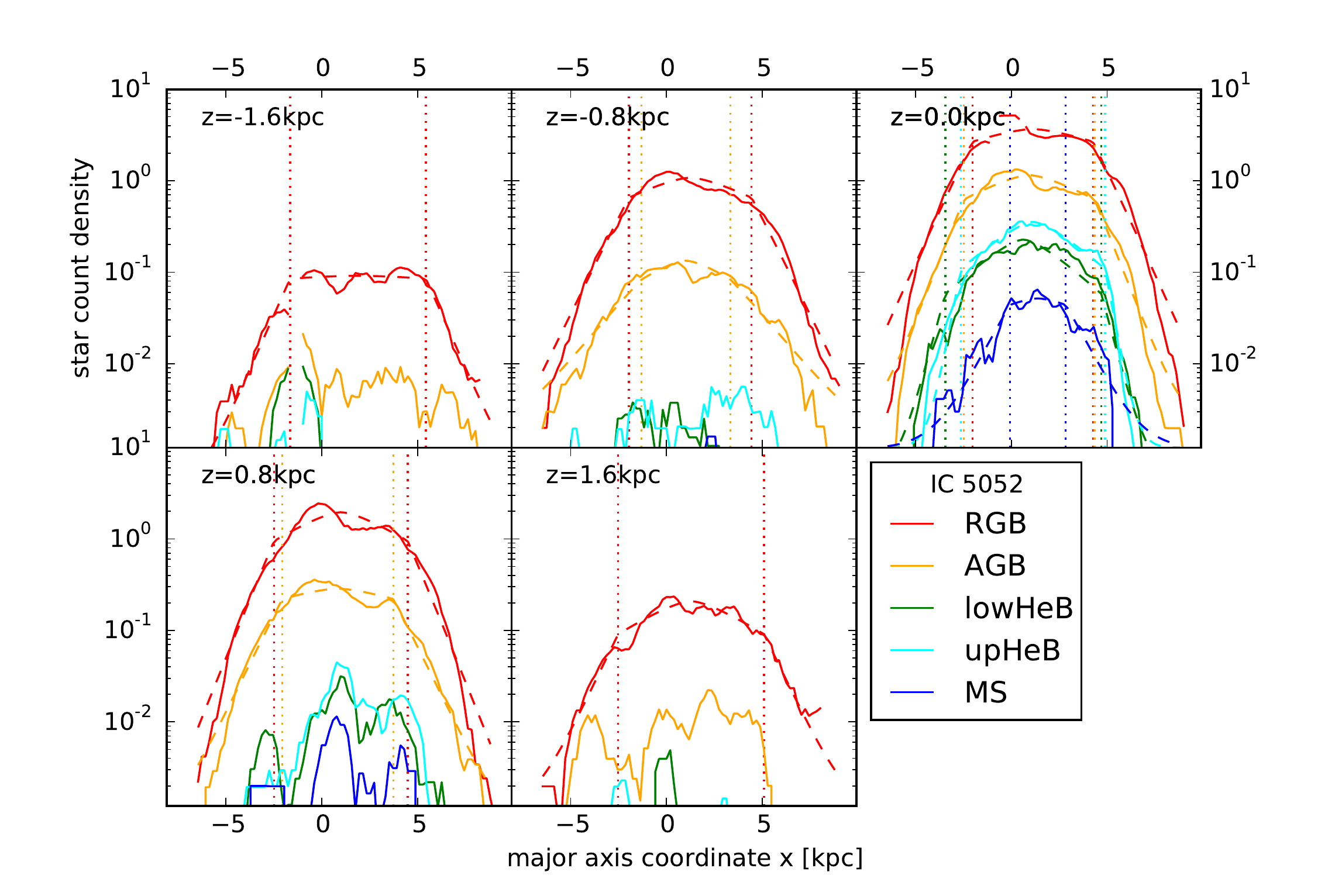} 
    \end{minipage}%
    ~\\ 
     \begin{minipage}[t]{0.99\textwidth}
        \centering
        \includegraphics[height=0.31\textheight]{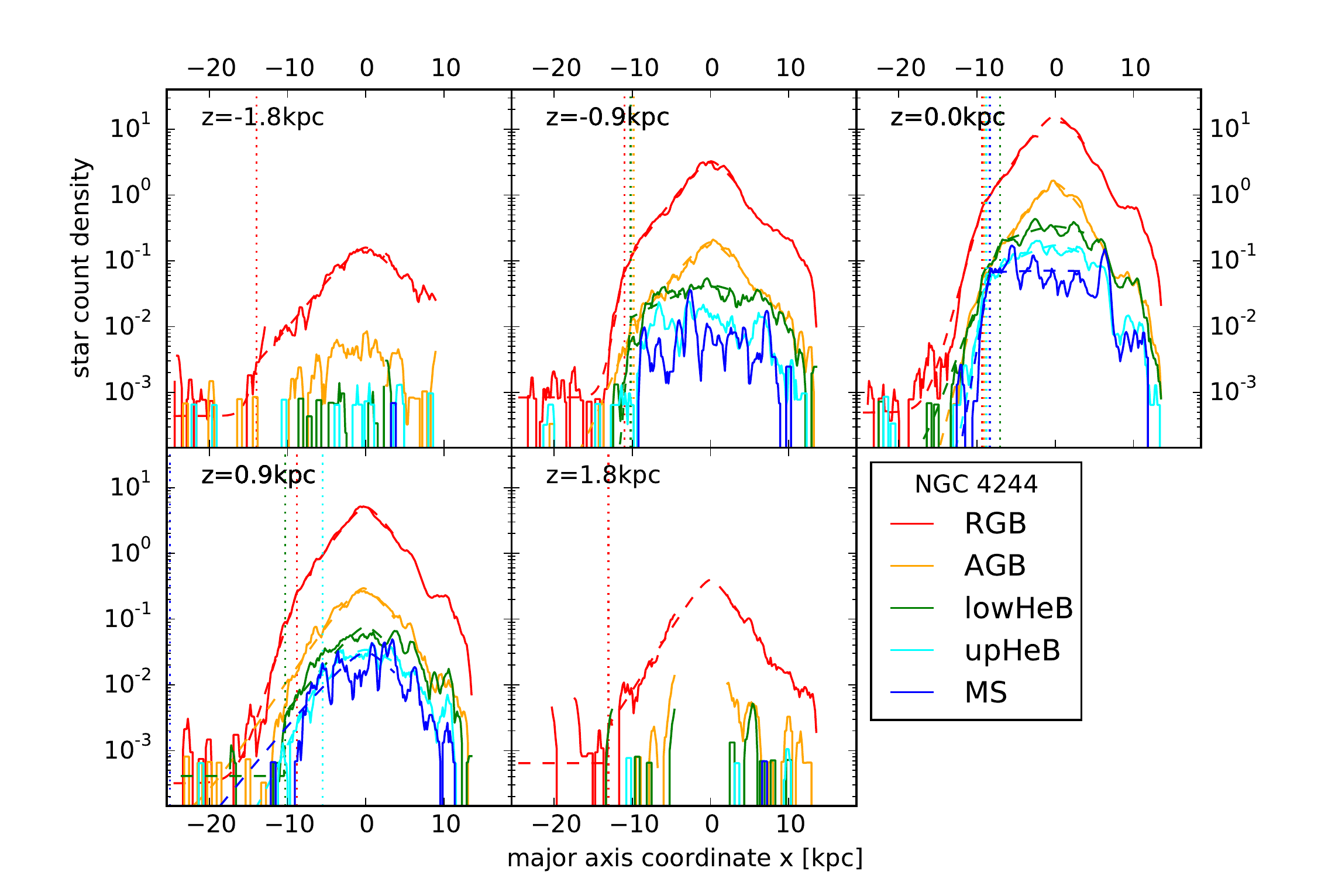} 
    \end{minipage}
    ~ \\
     \begin{minipage}[t]{0.99\textwidth}
        \centering
         \includegraphics[height=0.31\textheight]{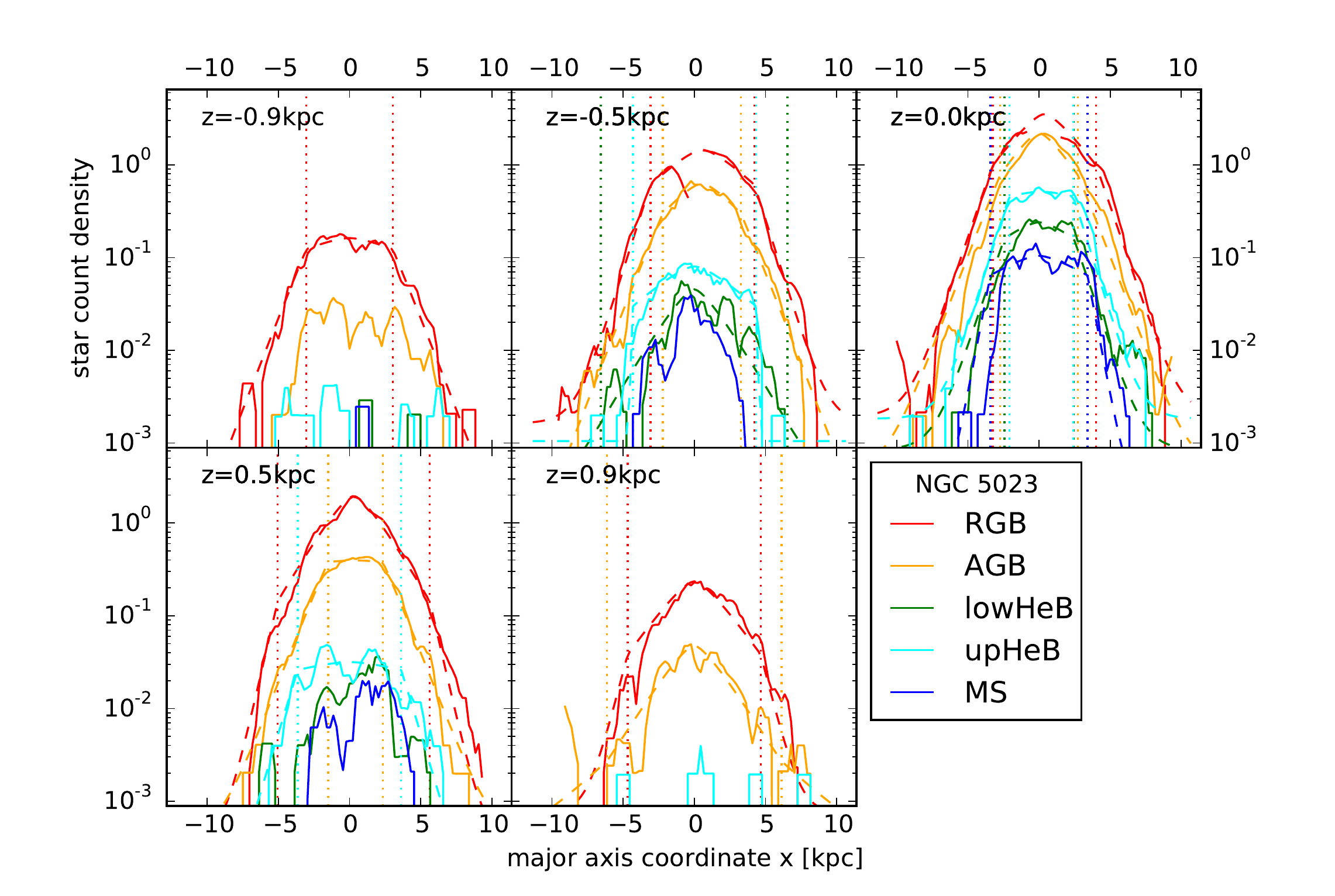} 
    \end{minipage}%
    \caption{Radial density profiles of the five populations (red - RGB; yellow - AGB, green - lower HeB; cyan - upper HeB; blue - MS) at different heights above the plane in three low-mass edge-on galaxies: IC\,5052, NGC\,4244, and NGC\,5023 (top to bottom). Solid lines are the data, dashed lines are the best fits. Vertical dotted lines show the break radii of the fits.
    The boundaries of the regions from which the profiles are extracted are shown in Fig.~\ref{fig:popmaps1} as horizontal gray lines.
    }
    \label{fig:rprofs1}
\end{figure*}

We also extracted the radial profiles from our data. The profiles at different heights from the midplane are plotted in Fig.~\ref{fig:rprofs1}. While the profiles do have some irregularities, they can \new{roughly} be described as broken exponentials, with steeper slopes outside the break.

\begin{figure}[!tb]
    \centering
        \includegraphics[width=3.6in]{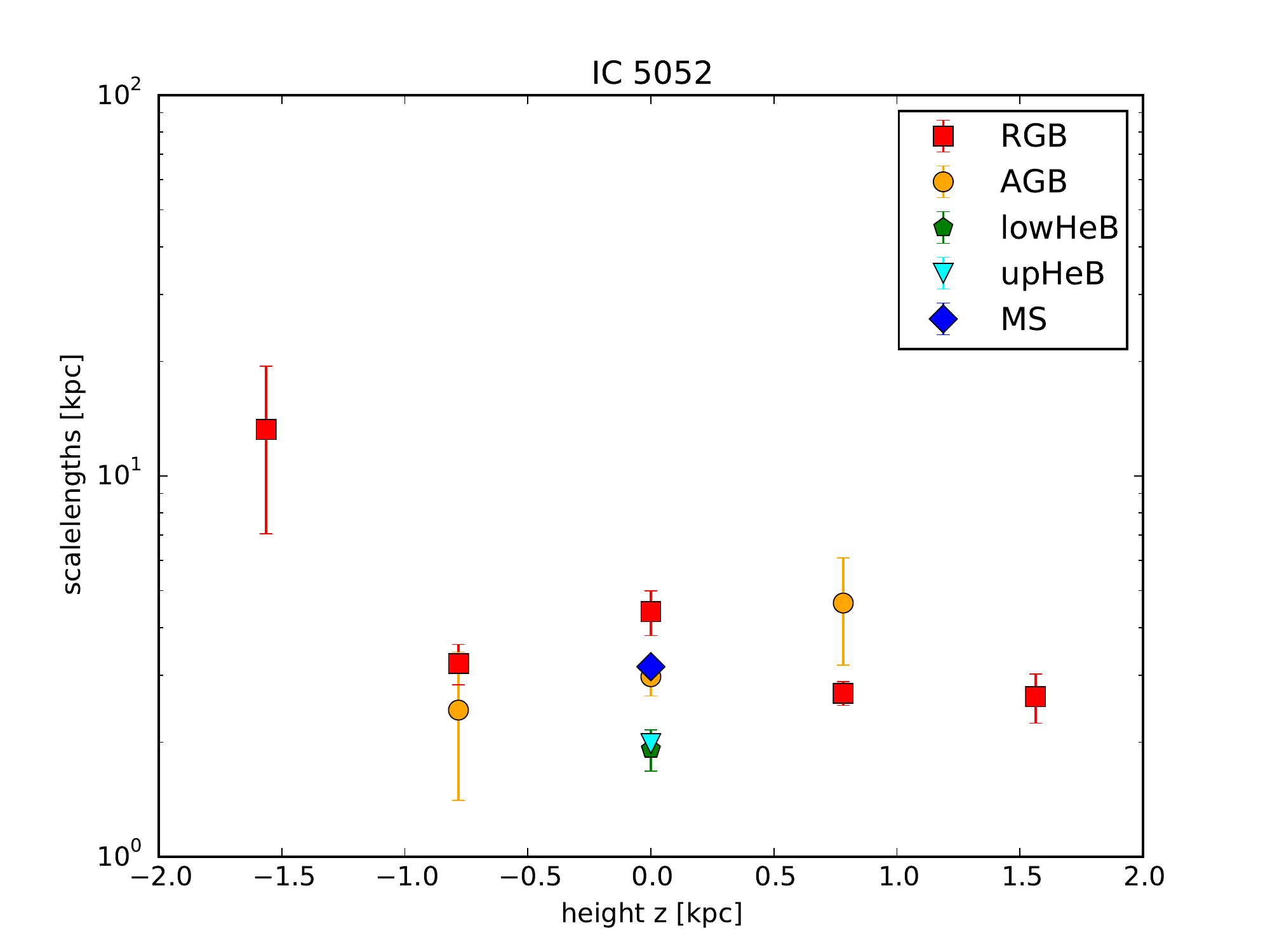}
        \includegraphics[width=3.6in]{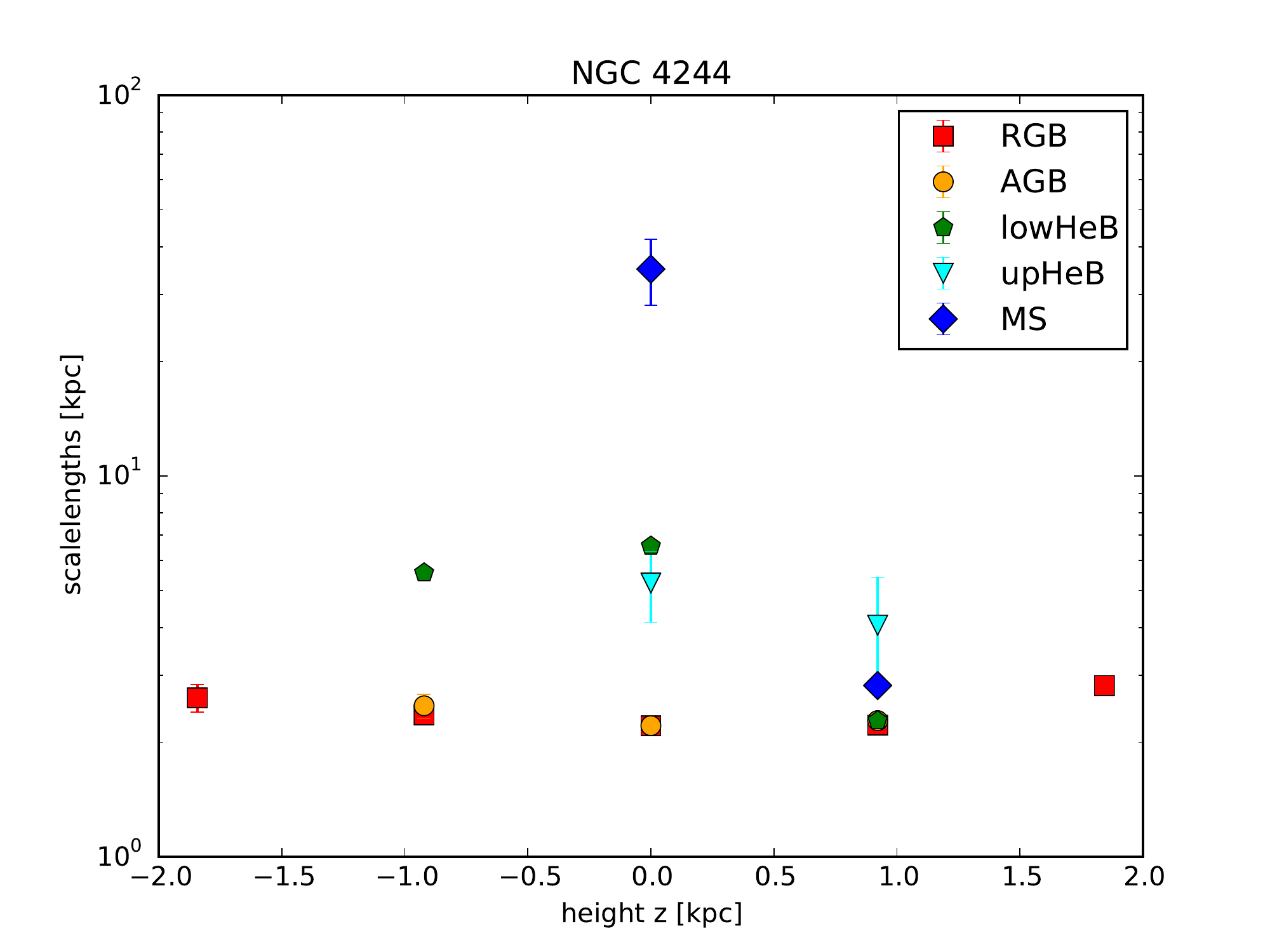}
        \includegraphics[width=3.6in]{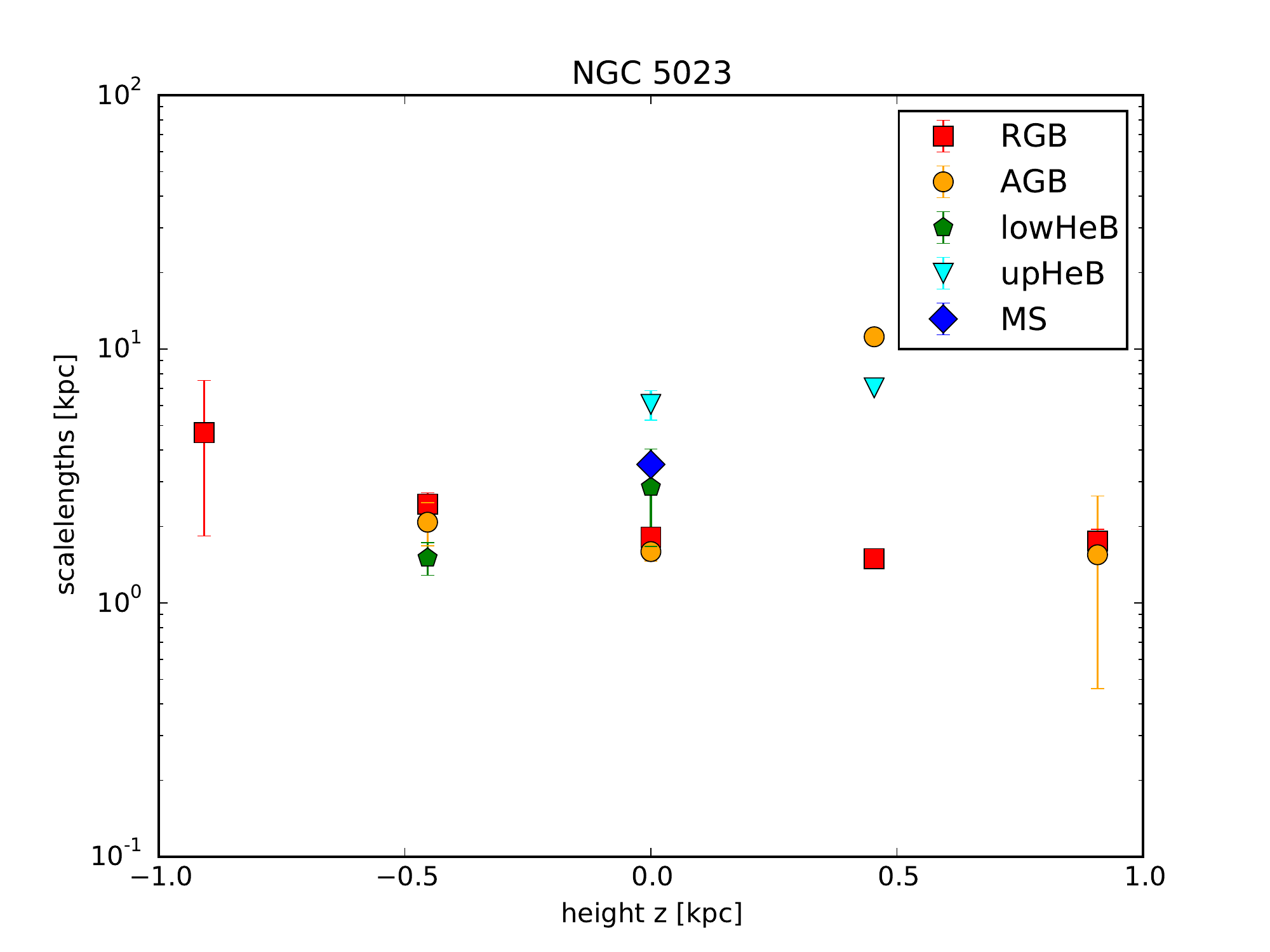}
    \caption{Disk scale lengths at different positions above
and below the plane in the three low-mass edge-on galaxies IC\,5052, NGC\,4244, and NGC\,5023 (left to right). Colors are the same as in Fig.~\ref{fig:vprofs1}.}
    \label{fig:scalelengths}
\end{figure}

\new{Figure~\ref{fig:scalelengths} shows that the scale lengths of AGB and RGB populations are approximately constant with height above the plane. For younger populations the height dependence could not be measured because there are too few younger stars at large heights.}

\begin{figure}[!tb]
 \centering
  \includegraphics[width=3.6in]{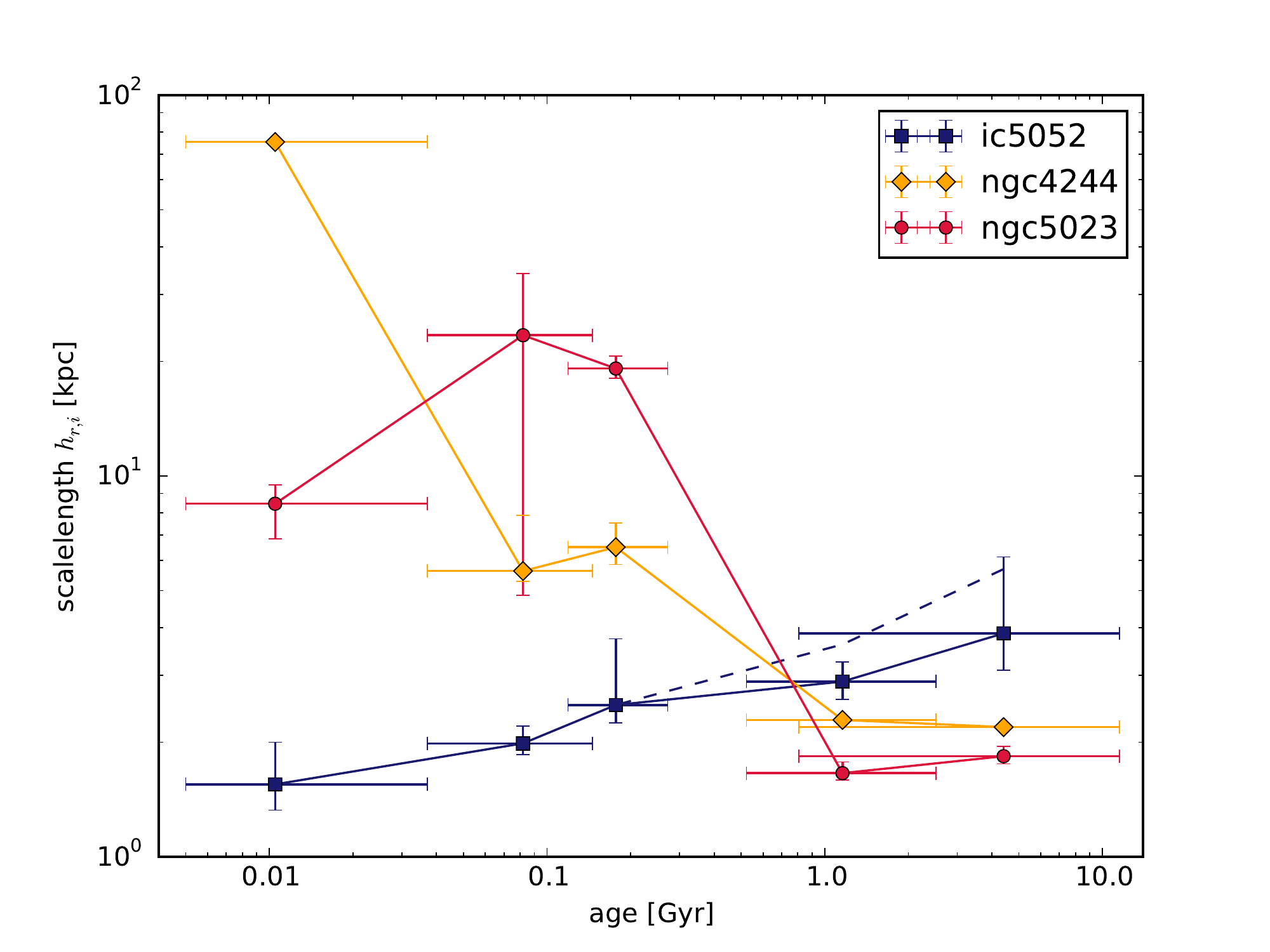}
 \caption[Dependance of scalelength on age]{Change of the average scale length (inside the break) with stellar age. The blue dashed line shows the fit results for IC\,5052 including an additional spheroidal component.}
 \label{fig:scalelengthevolution}
\end{figure}

\begin{figure}[!tb]
 \centering
 \includegraphics[width=3.6in]{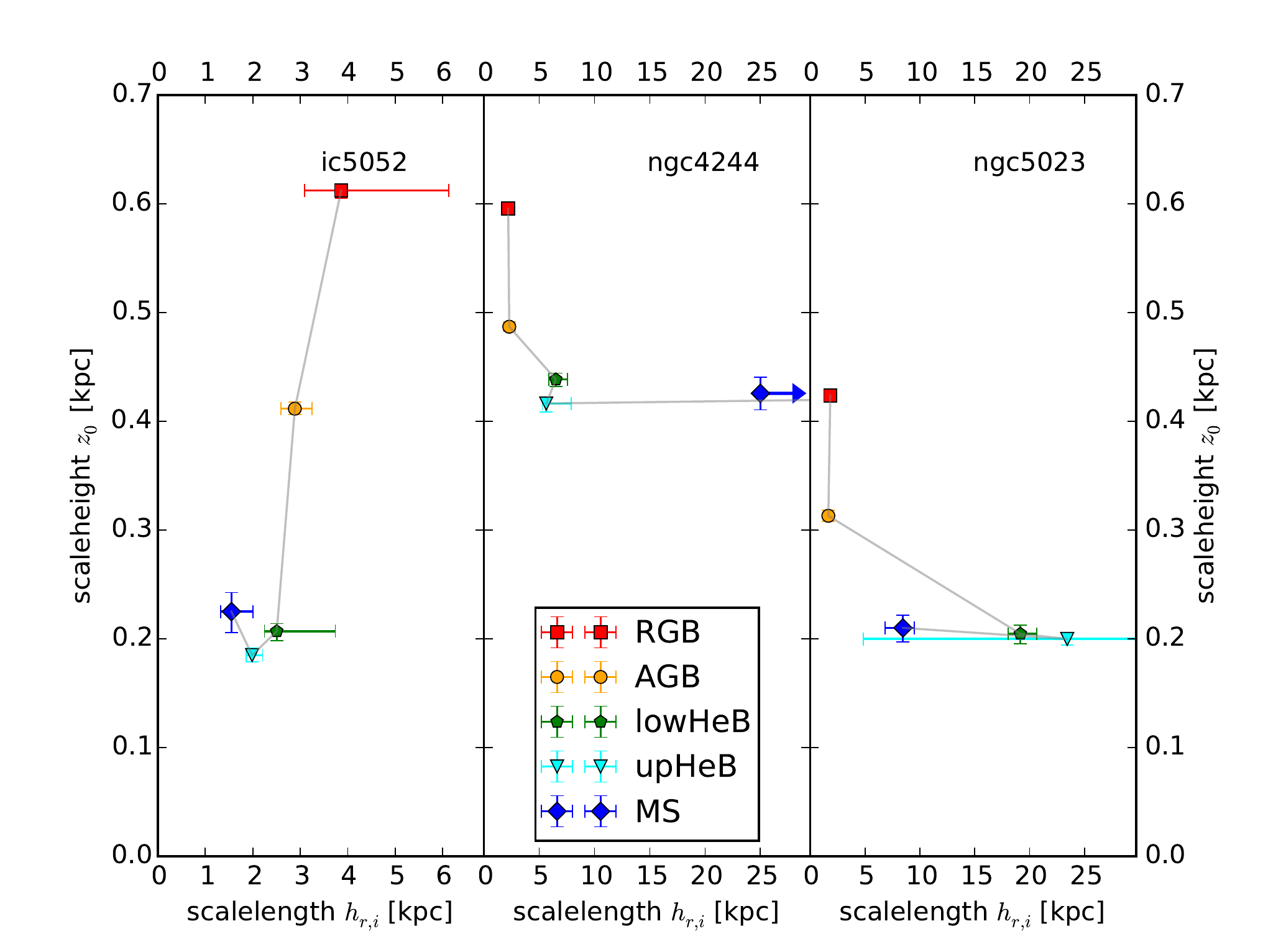}
 \caption{Scale heights vs. scale lengths (note the different scales on the x-axes).}
 \label{fig:scales}
\end{figure}
In two of the galaxies, NGC\,4244 and NGC\,5023, the \new{average} scale lengths (\new{which were determined through the 2D~fits}) decrease with age, meaning that older populations are more centrally concentrated than younger populations. IC\,5052 shows the opposite trend, and the older populations have flatter profiles than the young populations (see Figs.~\ref{fig:scalelengths} and~\ref{fig:scalelengthevolution} and Table~\ref{tab:2Dfit_results}).
From combining the results for scale length and height,  it is
clear that younger populations are in general thin and radially extended, while older populations are thicker and radially more compact, as we show in Fig.~\ref{fig:scales}. This is what is seen in the MW for mono-abundance populations \citep[e.g.,][]{bovy12b}.

\begin{figure}[!tb]
 \centering
 \includegraphics[width=3.6in]{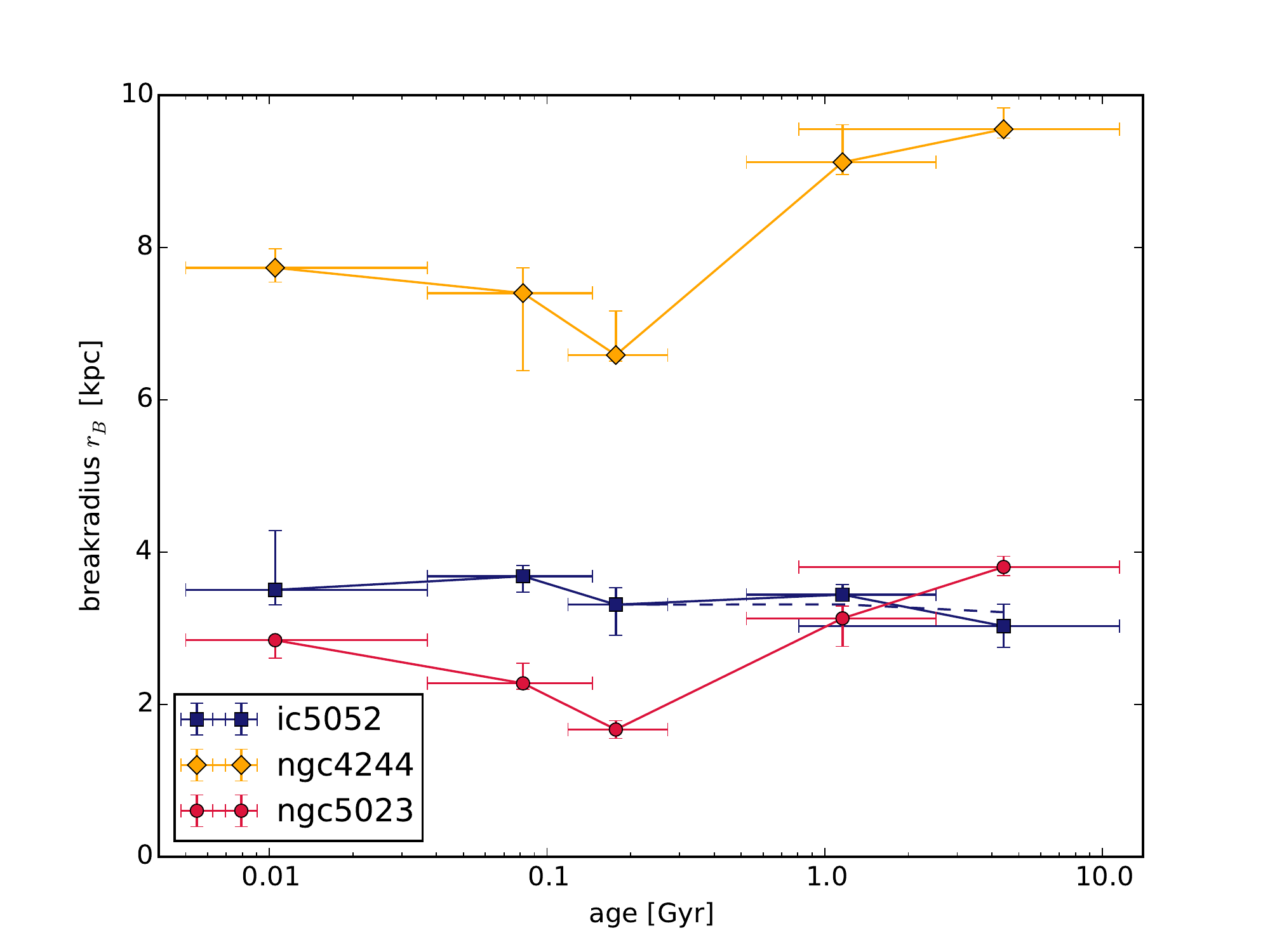}
 \caption[Dependance of break radii on age]{Change of the break radius with stellar age. The blue dashed line shows the fit results for IC\,5052 including an additional spheroidal component.}
 \label{fig:breakradiusevolution}
\end{figure}

\paragraph{Breaks in radial profiles:}
The break radii for all populations within a galaxy are approximately the same (within a kpc, see Fig.~\ref{fig:breakradiusevolution}), and there is no trend with age. 

\begin{figure}[!tb]
 \centering
 \includegraphics[width=3.6in]{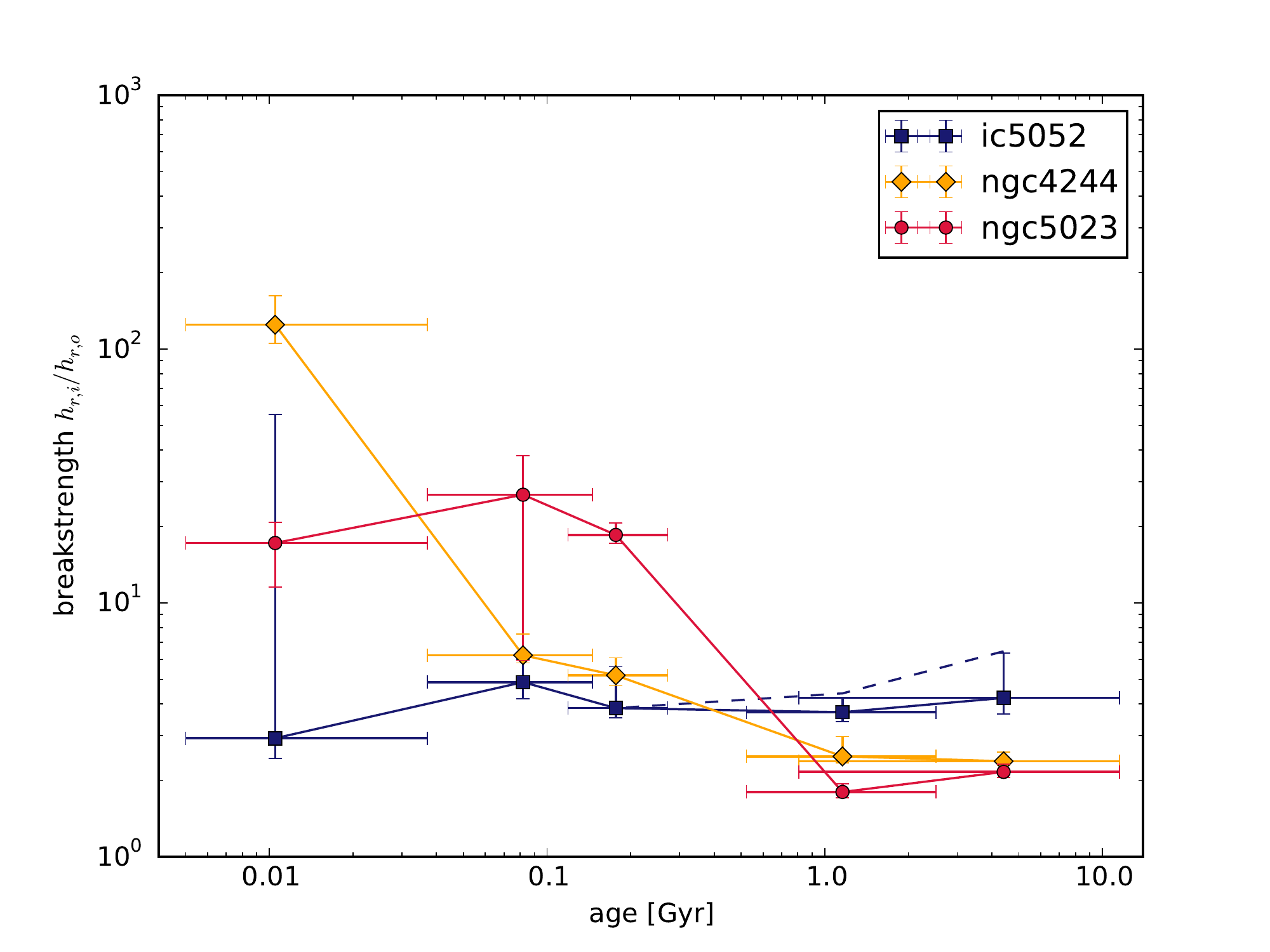}
 \caption[Dependance of break strength on age]{Change of the break strength, i.e., the ratio of inner to outer scale length, with stellar age. The blue dashed line shows the fir results for IC\,5052 including an additional spheroidal component.}
 \label{fig:breakstrengths}
\end{figure}

Outside the break, the younger populations have steeper profiles than older populations. This leads directly to a stronger break for younger populations; their profiles are rather flat inside the break and very steep outside, while for older populations the scale length changes only little (by about a factor 2-3) across the break (see Fig.~\ref{fig:breakstrengths}).

\begin{table}[!ht]
\centering
\caption[Scaleheights and scalelengths of the different populations]{Scale
heights and scale lengths (inside the break) from the 2D fits of the different populations.}
 \begin{tabular}{lcccccc}
\hline \hline
population &\multicolumn{2}{c}{IC\,5052}&\multicolumn{2}{c}{NGC\,4244}&\multicolumn{2}{c}{NGC\,5023}  \\ 
           &   $h_r$   &  $z_0$       &   $h_r$   &  $z_0$        &   $h_r$   &  $z_0$      \\
\hline
    MS     &    1.55   &   0.23       &   75.36   &  0.43         &   8.46    &  0.21       \\
 upHeB     &    1.99   &   0.19       &    5.63   &  0.42         &  23.42    &  0.20       \\
 lowHeB    &    2.50   &   0.21       &    6.50   &  0.44         &  19.15    &  0.20       \\
   AGB     &    2.89   &   0.41       &    2.29   &  0.49         &   1.66    &  0.31       \\
   RGB     &    3.86   &   0.61       &    2.19   &  0.60         &   1.84    &  0.42       \\
halo RGB   &     -     &   2.30       &     -     &  3.04         &    -      &    -        \\
\hline
[3.6$\mu$m]&    2.14   &   0.43       &    1.84   &  0.46         &   1.24    &  0.30        \\ 
PAH [8$\mu$m]&  1.11   &   0.38       &    1.77   &  0.57         &   1.08    &  0.35        \\
\hline 
 \end{tabular} 
\tablefoot{All values in kpc.} 
\label{tab:2Dfit_results}
\end{table}

\subsection{Discussion of observed profiles}

\subsubsection{Irregularities in the profiles}\label{sec:irregularities}

\begin{figure}[!tb]
 \centering
 \includegraphics[width=3.5in]{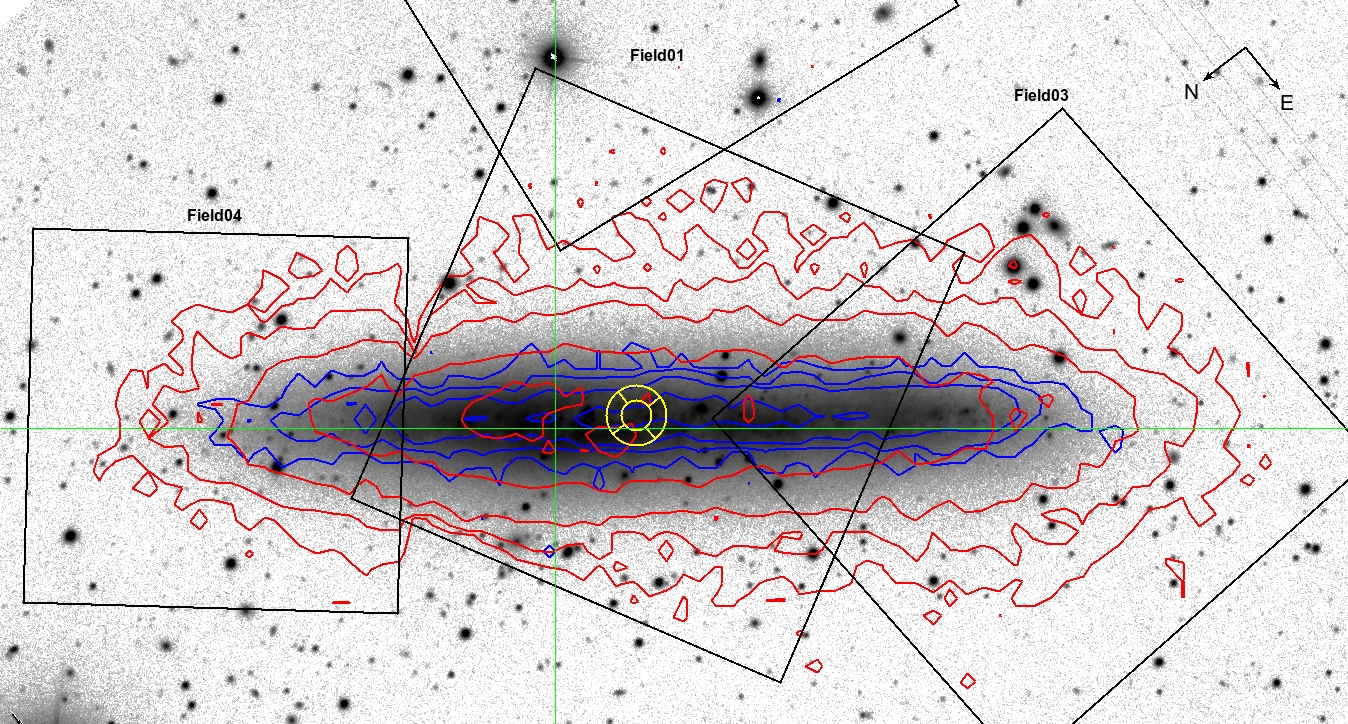} \caption[Image of the asymmetric disk of IC\,5052]{R-band image of IC\,5052 \citep{meurer06}. The density of RGB stars is shown as red contours, the density of young stars as blue contours. The yellow reticle indicates the dynamic center of the \ion{H}{I} gas \citep{peters13}, the green cross shows the central position of the galaxy from HyperLEDA. The black boxes show the (approximate) coverage of our GHOSTS fields.}
 \label{fig:ic5052centers}
\end{figure}

\paragraph{IC\,5052} This galaxy appears to be lopsided. In Fig.~\ref{fig:rprofs1} the centers of all disk fits for IC\,5052 are removed by about 1-2\,kpc from the literature value\footnote{from HyperLEDA, \url{http://leda.univ-lyon1.fr/}} of the galaxy center, which is indicated as the zero point of the x-axis in the figure. The AGB and RGB profiles have their maximum near the literature value of the center, but the center of the fit is dominated by the central position between the breaks. 
In Fig.~\ref{fig:ic5052centers}, the contours of young stars and RGB stars of IC\,5052 are shown in an R-band image \citep{meurer06}. The innermost RGB contours coincide well with the brightest region of the R-band image and with the HyperLEDA position of IC\,5052. In contrast, the center of the outer RGB and the young star contours lie farther to the southeast and coincide better with the dynamical center of the galaxy, taken from HI observation \citep{peters13}. 

In the residual images from our two-dimensional fitted models (see Fig~\ref{fig:ic5052-2Dfits}) the overdensity appears approximately circular. We therefore repeated the fits of the RGB and AGB populations in IC\,5052 with the overdensity described by an additional Sersic
component and evaluated the structural parameters of the disk again. While the scale heights and the break, the radii of the two component fits are approximately the same as in the one-component fit, the scale lengths are significantly larger. The results of the two component fits are shown in Figs.~\ref{fig:scaleheightevolution}, \ref{fig:scalelengthevolution}, 
\ref{fig:breakradiusevolution}, and \ref{fig:breakstrengths} as a dashed line. 

\paragraph{NGC\,4244} This galaxy also has an asymmetric radial profile (see Fig.~\ref{fig:rprofs1}). At the northeastern \new{(in our plots, at the left)} end of the disk, all populations show
a break at about 9\,kpc \citep[see also][]{dejong07b}. On the southwestern \new{(right)} side the break occurs at about 7\,kpc, followed by a plateau between 8\,kpc and 12\,kpc, where the profile again breaks. These structures cannot be modeled by our simple edge-on galaxy models, and thus we only used the northeastern part for our fit (up to $x<4$\,kpc).

\subsubsection{Additional thick-disk or halo component?}
\label{sec:thickdisk?}

It is now widely accepted that galaxies contain multiple (disk) components. In contrast to this, as evident in Fig.~\ref{fig:vprofs1}, the vertical profiles of our low-mass galaxies could be fitted well by single-disk models. This is remarkable because more than three orders of magnitudes in surface density is covered between the central and the outermost parts of the galaxies; this corresponds to about eight magnitudes in surface brightness. This is significantly deeper than most of the integrated light observations that have been used to detect the thick disks. 

However, the profiles presented so far were not optimized for low-density regions. As we described in Sect.~\ref{sec:ghosts}, GHOSTS has two sets of culls for optimal star selection, one optimized for crowded regions, the other optimized for sparse regions. In previous sections we used the crowded field culls to optimally sample the inner regions of the galactic disks. We now use the sparse field culls, which are optimized for low-density regions.

Additionally, the fine binning of the data we used before made it impossible to detect structures below a surface density of 0.019\,arcsec$^{-2}$ (which equals one star per bin). By using broader bins, we can further extend the analysis to fainter regions. In the following we use \new{a 200\arcsec~wide strip along the minor axis of each galaxy and adjust the vertcal bin size from $\approx7\arcsec$ at highest densities (near the midplane) to $\approx40\arcsec$ at lowest densities to create deeper vertical profiles} and search for additional structural components.

\begin{figure}
    \centering
    \includegraphics[width=3.6in]{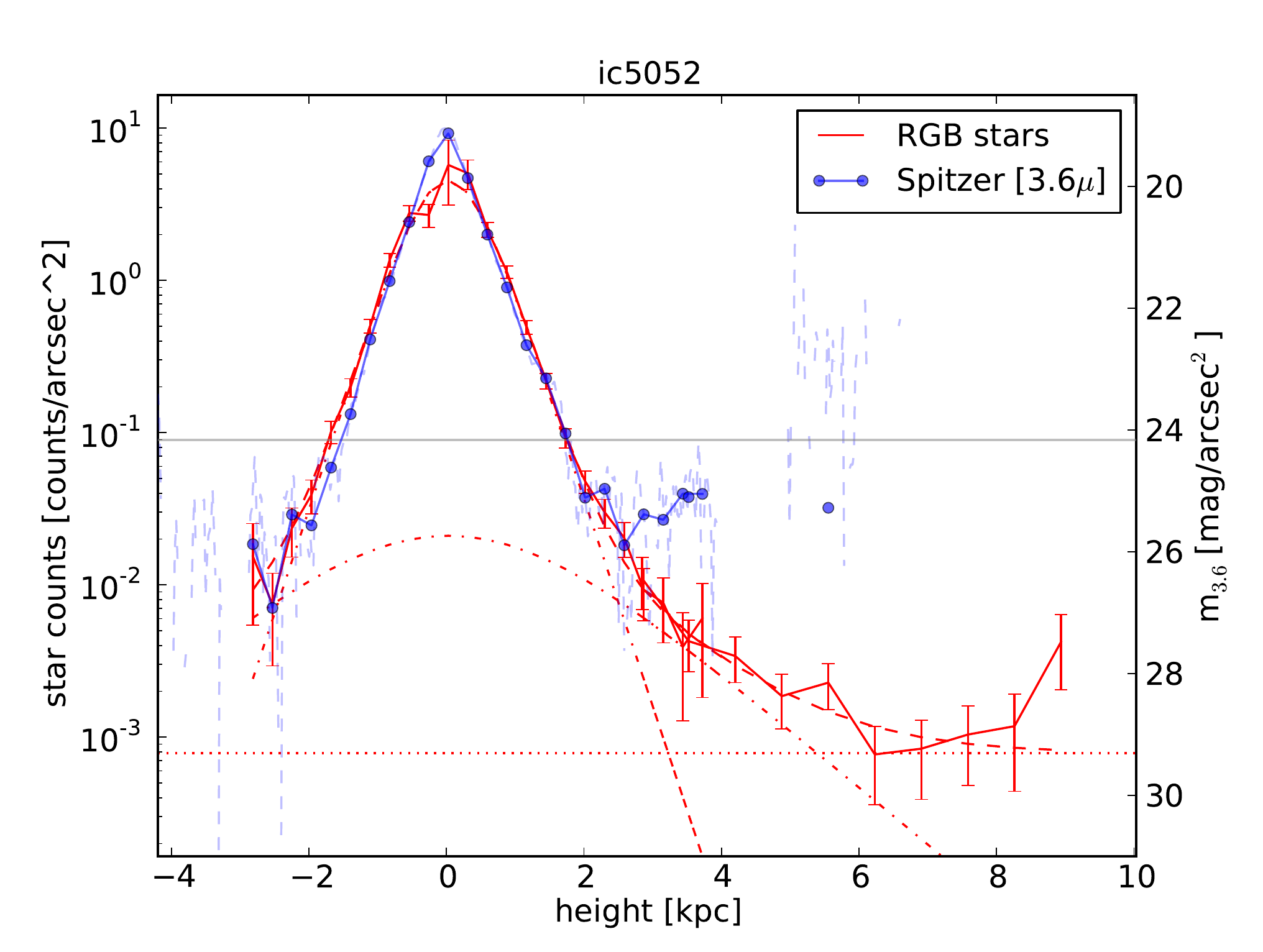}
    \includegraphics[width=3.6in]{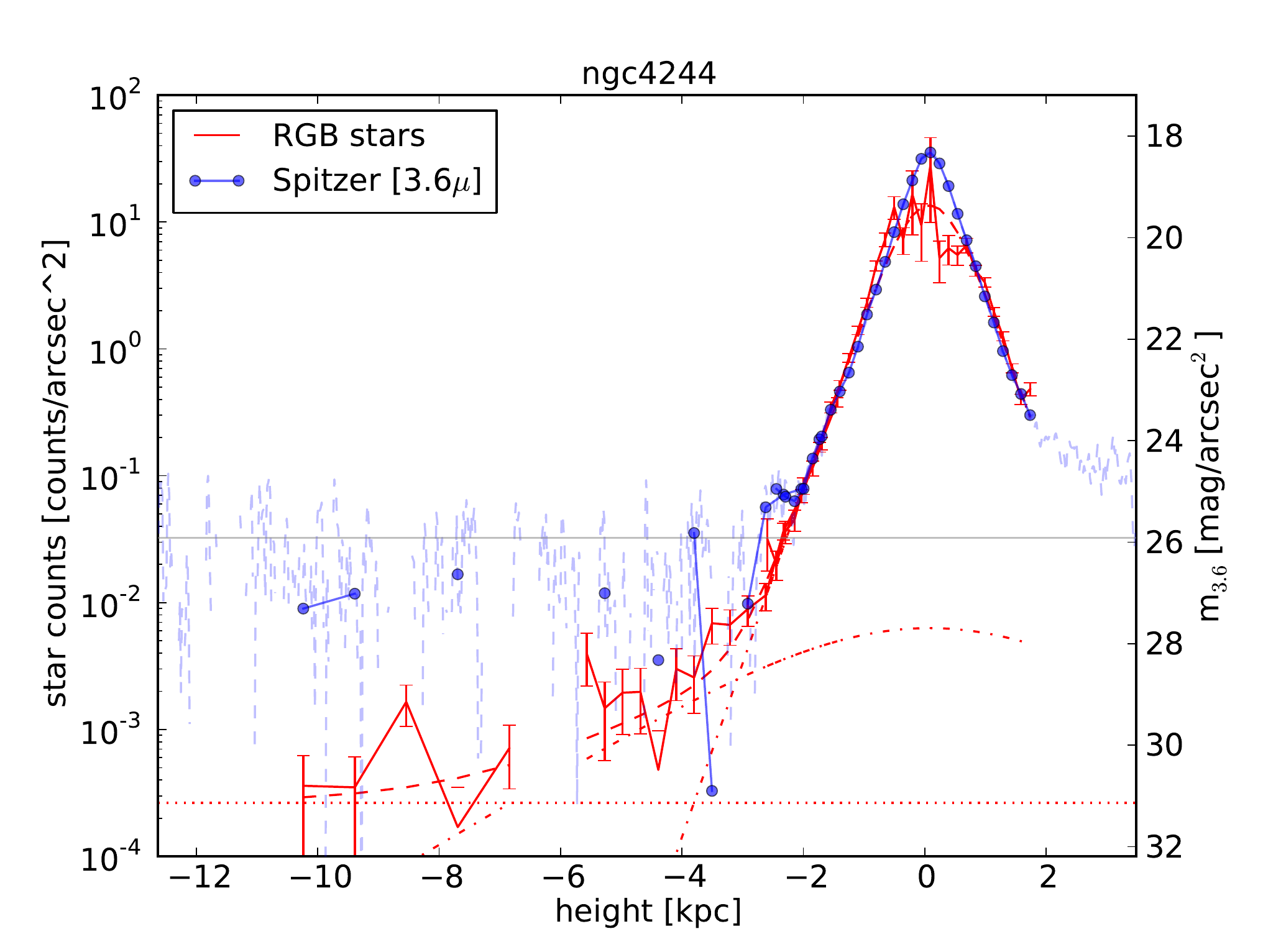}
    \includegraphics[width=3.6in]{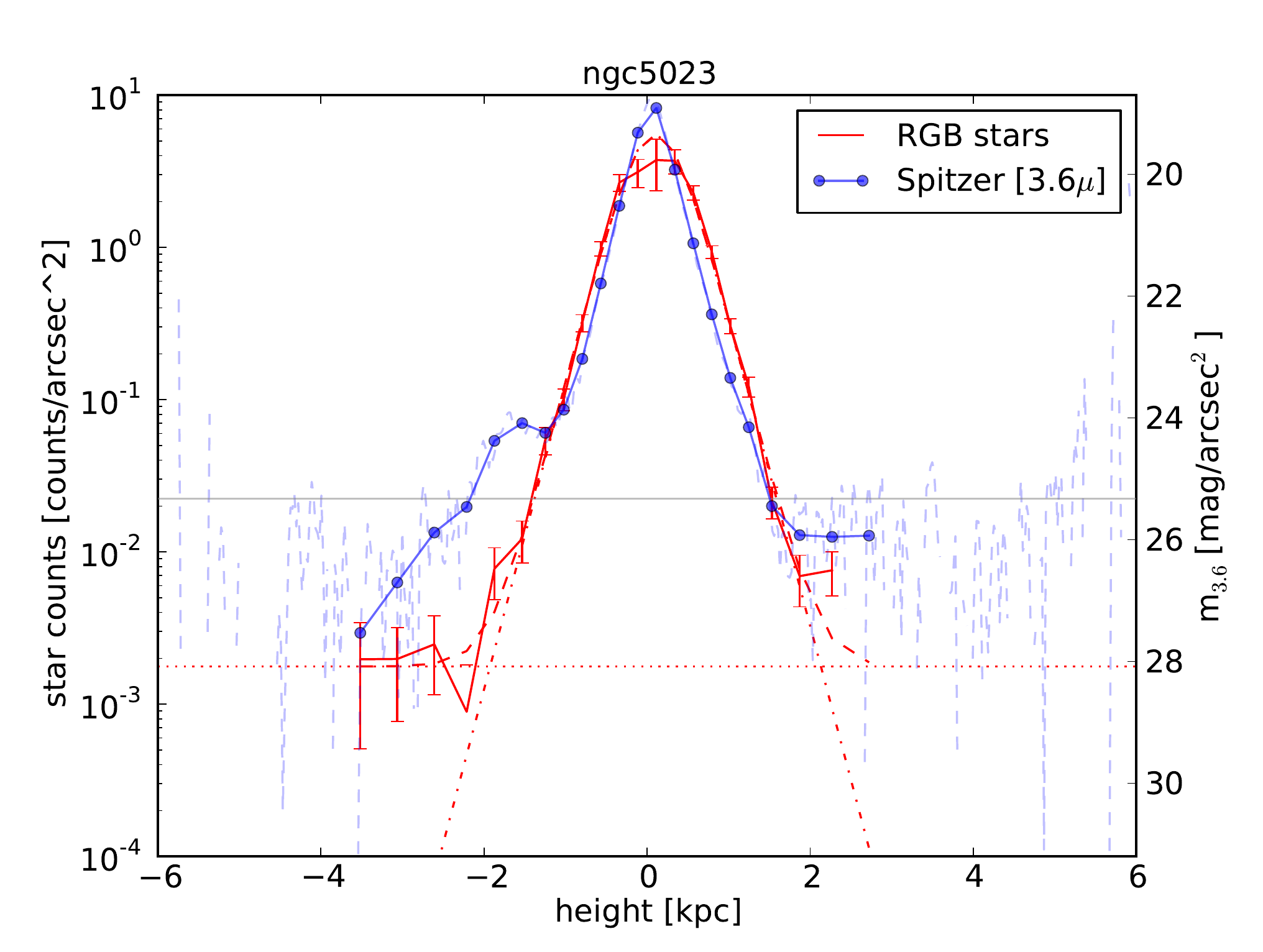}
  \caption[Deep vertical RGB star count profiles and comparison with Spitzer data]{Surface number density profiles of RGB stars in the central regions of the three galaxies (shown in red), the dashed lines are the best-fit models, the dash-dotted lines show the different components, two sech$^2$ functions, and a constant for background contamination. The light blue dashed line is the (background-subtracted) surface brightness in the Spitzer 3.6$\mu$m filter, the solid blue line shows the Spitzer data averaged to the same bin size as the star count data, and the horizontal gray line plots the estimated noise level of the background of the Spitzer data.}
  \label{fig:deepvprofiles}
\end{figure}

The result of this is shown in Fig.~\ref{fig:deepvprofiles}. Compared to Fig.~\ref{fig:vprofs1}, we now reach about one order of magnitude deeper. A faint second component can now be detected in IC\,5052 and NGC\,4244. Since we used star count data, this second component cannot be an effect of scattered light and extended PSFs, which might lead to similar detection in integrated light studies, if not modeled carefully \citep{dejong08,sandin14}.
We fit the whole profile with two sech$^2$ components. In both galaxies the second components are very faint, their central surface \new{star count densities $n_0$} are only 0.6\% (IC\,5052) and 0.08\% (NGC\,4244) of the main disk central surface brightness. As a result of their larger scale heights ($z_{0,ext}=2.30\pm0.69$\,kpc for IC\,5052 and $z_{0,ext}=3.04\pm2.29$\,kpc for NGC\,4244), the second components dominate at $|z|>2.2$\,kpc and $|z|>2.8$\,kpc, respectively. \new{If we assume that radial structure of the extended component is similar to the main disk, we can calculate the star count ratio (and thus the stellar mass ratio) of the two components. It is $M\propto z_0 n_0$, which leads to a mass ratio of $M_{ext}/M_{thin}$=0.017 for IC\,5052 and $M_{ext}/M_{thin}$=0.003 for NGC\,4244.}

\new{In NGC\,5023 we did not detect an extended component, but
neither did we reach as faint as in the other galaxies. The profile of NGC\,5023 is based on a single SNAP observation, and its CMD reaches only 1.5\,mag below the TRGB. This leads to fewer detections of stars and thus a higher noise level. Furthermore, the vertical extent of our observations is limited to $z\approx\pm3$\,kpc (since there are no additional fields on the minor axis), which prevents detecting an extended component. Thus, we cannot rule out the presence of a faint extended component in NGC\,5023.}

We note that the choice of sech$^2$ for fitting the second component is a mere choice of convenience. It is not meant to imply that these components are really disk components. It is possible to fit these components equally well with other functions (e.g., exponential, Sersic, power law), but the sech$^2$ allows for the most direct comparison between the main disk and the second component. The possible nature of the second components is discussed in Sect.~\ref{sec:discussion}.

\subsubsection{Comparison with Spitzer data}
\label{sec:spitzer-obs}

We here used Spitzer images from channel 1 (at 3.6$\mu$m) for comparison with the RGB count profiles. For the transformation from RGB counts to surface brightness we used the the same artificial CMDs as in Sect.~\ref{sec:ageselection} with a constant SFR and a flat MDF. By equating the number of RGB stars in the artificial CMDs with the expected flux of the underlying population, we calculated the counts-to-flux conversion factor. This factor can of course only be a rough estimate of the true conversion because the exact transformation depends on the star formation history, which is unknown for the real galaxies. But as Fig.~\ref{fig:deepvprofiles}
shows, where the Spitzer profiles are shown as light blue and the RGB counts as red lines, the transformation matches the profiles quite well. At larger heights RGB stars are apparently the dominant contributor to the channel 1 light.

While the Spitzer and RGB profiles match well, there are two
main differences in a large part of their overlap regions. 

First, the RGB profiles reach much deeper than the Spitzer profiles. The Spitzer profiles are limited by a relatively bright background. To determine the sky brightness, we measured the mean brightness of a large empty region far away from the galactic disk. The corresponding standard deviation in this region was used as an estimate for the noise and is shown as a horizontal gray line in Fig.~\ref{fig:deepvprofiles}.\footnote{The Spitzer images also contain systematic deviations from a constant background level, e.g., large round areas of increased brightness in the image of NGC\,5023. One of these bright spots lies near the center of NGC\,5023 and can be seen in the vertical profile as the bump at $z\approx-2$\,kpc.} 
Star counts are not affected by sky brightness, but they are limited by the number of contaminants, that is, by faint foreground stars and unresolved background galaxies. The density of these contaminants is so low that our star counts can reach an equivalent surface brightness below 30\,[$3.6\mu$m]-mag/arcsec$^2$.

Second, the Spitzer profiles have a sharper peak in the midplane regions than the star count profiles. The reason for this could be that the Spitzer data contain a mixture of all stellar populations, as well as dust and PAH emission \citep{meidt12}. The younger populations, which have a much smaller scale height than the RGB stars (see Sect.~\ref{sec:lowmass-verticalprofs}), will add extra light to the central regions of the profile. It could also be that we miss some of the RGB stars in the central regions due to crowding effects or dust. While we did correct for a decreasing completeness with higher star count densities, we did not correct for dust effects that will influence our optical data (where the star counts are made) more than the infrared Spitzer data.

In Table~\ref{tab:2Dfit_results} the scale heights and -lengths of the five populations and the Spitzer data are listed. The Spitzer scale heights lie within the range of the scale heights of the resolved populations \new{and are close to the value of the AGB scale heights}\footnote{\new{Table~\ref{tab:2Dfit_results} lists the results of a sech$^2$ fit. Since the Spitzer profiles are more peaked than the stellar profiles, they are better fit by a generalized sech$^{2/n}$ profile with $n>1$. This changes the best-fit Spitzer scale heights to be close to the RGB scale
height (see Appendix~\ref{sec:spitzer-scaleheights}).}}, supporting a interpretation as a combination of these different populations.
The Spitzer scale lengths (of NGC\,4244 and NGC\,5023), on the
other hand, are shorter than those of any population, which precludes an interpretation of the Spitzer data as a simple combination of the stellar populations. From the two possible reasons for this mismatch, incompleteness and/or dust, we asses the former as unlikely. Through the extensive artificial star test we have a good estimate of the completeness, and we have corrected for it. Furthermore, all regions with a completeness below 50\% are excluded from the analysis, which leads to small uncertainties in the final star counts. On the other hand, we do see signs of dust. Its effects are discussed in Sect.~\ref{sec:dusteffects}.

\subsection{Summarizing results}
Before we move on to the discussion, we summarize all our results from this section: All three of the analyzed galaxies show the following trends in their structural parameters:
\begin{itemize}
\item The scale height of stellar populations increases with age, but only on timescales longer than about 300\,Myr.
\item The scale height of the young stellar populations is essentially constant along the disk, while AGB and RGB stars exhibit a mild flaring.
\item The scale length of stellar populations is essentially independent of the height above the plane.
\item The break radius is independent of stellar population age.
\end{itemize}
Two out of three galaxies (NGC\,4244 and NGC\,5023) have similar trends in their radial profiles:
\begin{itemize}
\item The scale length of stellar populations decreases with age.
\item The break strength, that is, the ratio of inner to outer scale length, decreases with age.
\end{itemize}
A third galaxy (IC\,5052) has an increasing scale length with age, which leads to a constant break strength. This galaxy also has an additional overdensity of old stars, which lies off the center of the disk. 

\section{Discussion}
\label{sec:discussion}

\subsection{Possible effects of extinction by gas and dust}
\label{sec:dusteffects}

In the previous section we have used a sech$^2$ profile for the vertical and a Bessel K$_1$ function for the radial profile of the star count surface density distribution of our galaxies. By doing this, we have implicitly assumed that the galaxies are transparent, meaning that the surface density is the result of the density integrated along the line of sight through the whole galaxy. The presence of dust will lead to a deviation from this idealized assumption.

Determining the real distribution of dust in external galaxies is very difficult and beyond the scope of this paper; so we cannot correct for extinction effects directly, but we can model the effects of different dust distributions on star count maps of an idealized galaxy model. 

We modeled the density of both the stars and the dust as exponential disks with a sech$^2$ distribution in the vertical direction. We calculated the observed surface brightness by solving the standard differential equation for the 1D radiative transfer along the line of sight,
\begin{align}
\frac{dI}{dx} = -\kappa I + \epsilon,
\end{align}
with $\kappa\propto\rho_{dust}(x,y,z)$ and $\epsilon \propto \rho_{star}(x,y,z)$. 

It might be questioned whether a radiative transfer equation is appropriate for treating star counts. But in fact the effects of absorption are similar: It will attenuate and redden the light of a star until it falls below the detection limit of our observations (or out of our CMD selection boxes); and the higher the absorption, the higher the fraction of stars that are removed from the CMD. Furthermore, kostenlose email teststar counts have the advantage that they are not affected by scattered or reflected light: the light of a star that is absorbed will not be re-emitted to become part of our star counts; thus the emissivity $\epsilon$ is proportional only to the local stellar density and has no scattered light term. The details of the effects of absorption on star counts do, of course, depend on the distribution of stars in the CMD, that
is, on the luminosity function and, thus, on the star formation history. In this sense, our approach is very simplified, but a full extinction modeling is beyond the scope of this paper.

\begin{figure*}[!tb]
  \begin{minipage}[t]{0.33\textwidth}
    \centering
    \includegraphics[width=0.99\columnwidth]{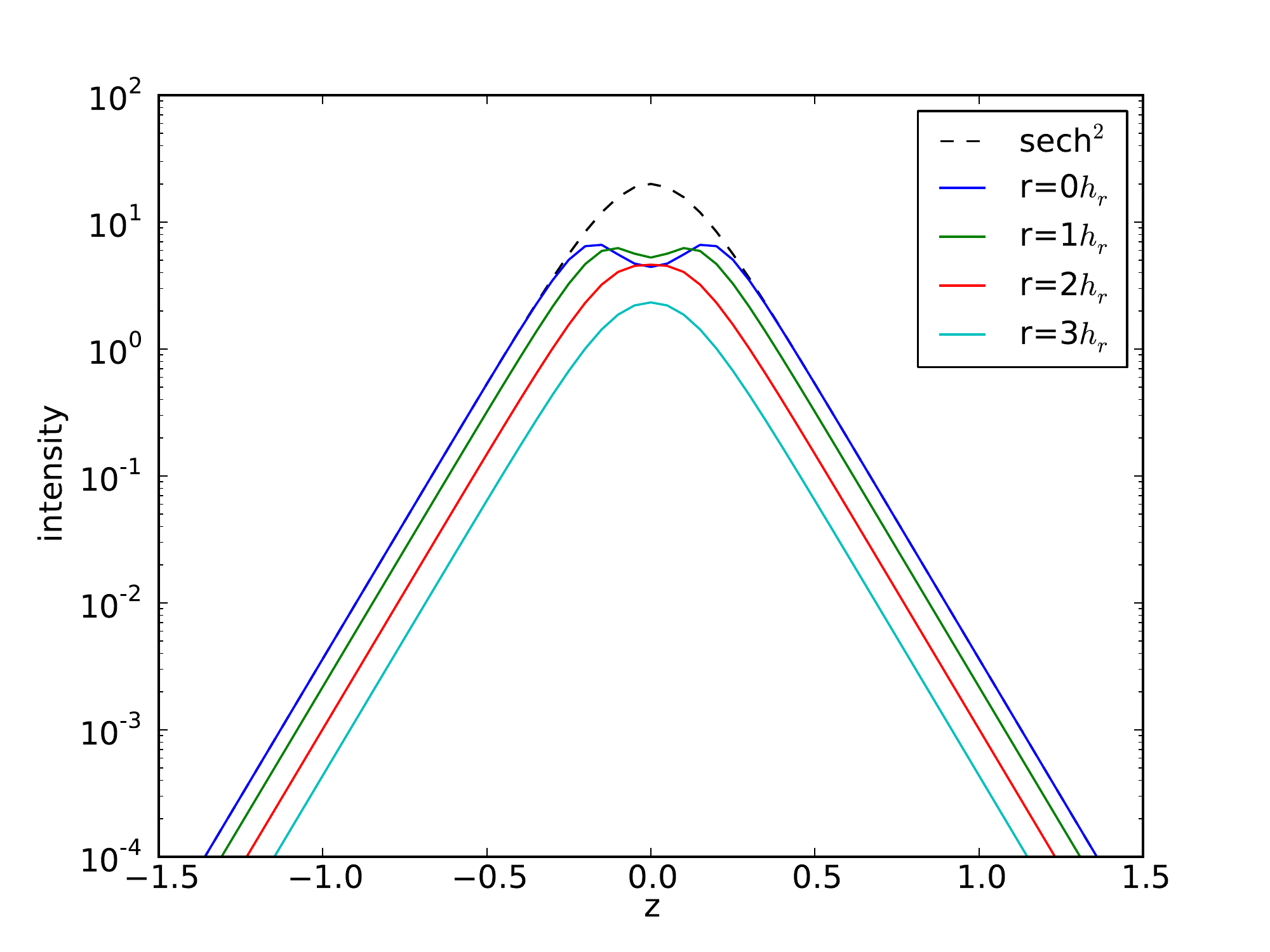}
  \end{minipage}
  \begin{minipage}[t]{0.33\textwidth}
    \centering
    \includegraphics[width=0.99\columnwidth]{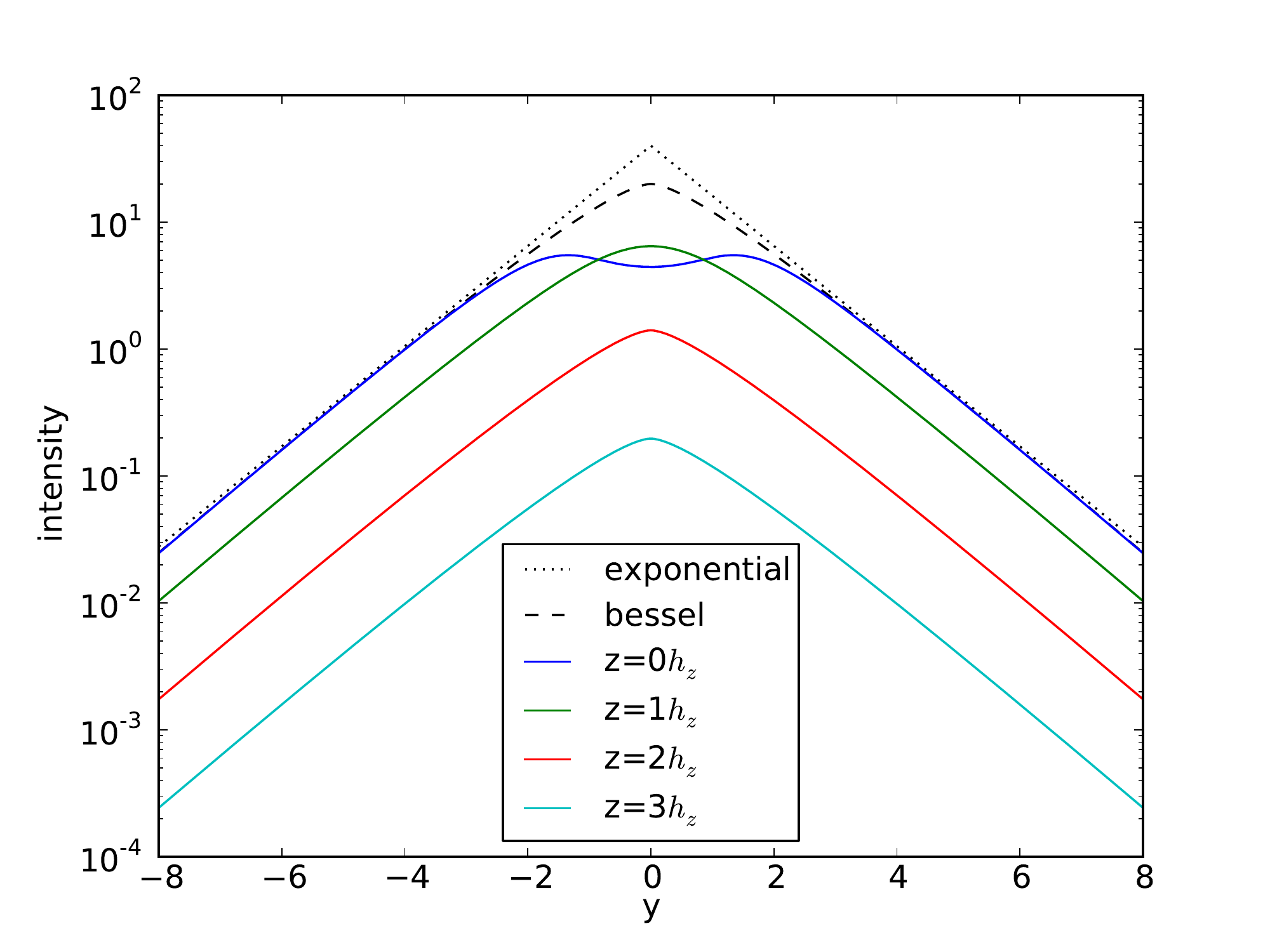}
  \end{minipage}
  \begin{minipage}[t]{0.33\textwidth}
    \centering
    \includegraphics[width=0.99\columnwidth]{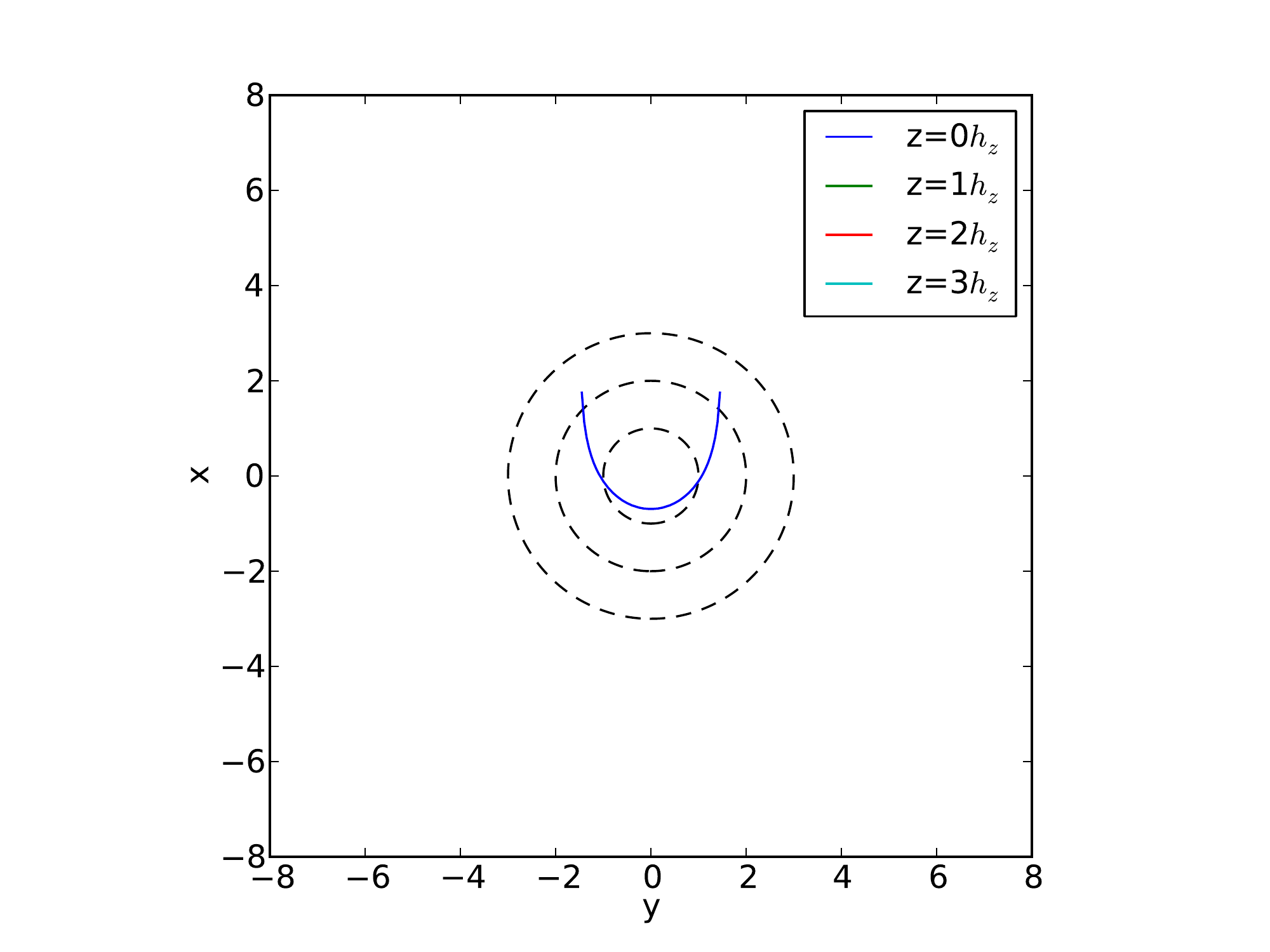}
  \end{minipage}
  \begin{minipage}[t]{0.33\textwidth}
    \centering
    \includegraphics[width=0.99\columnwidth]{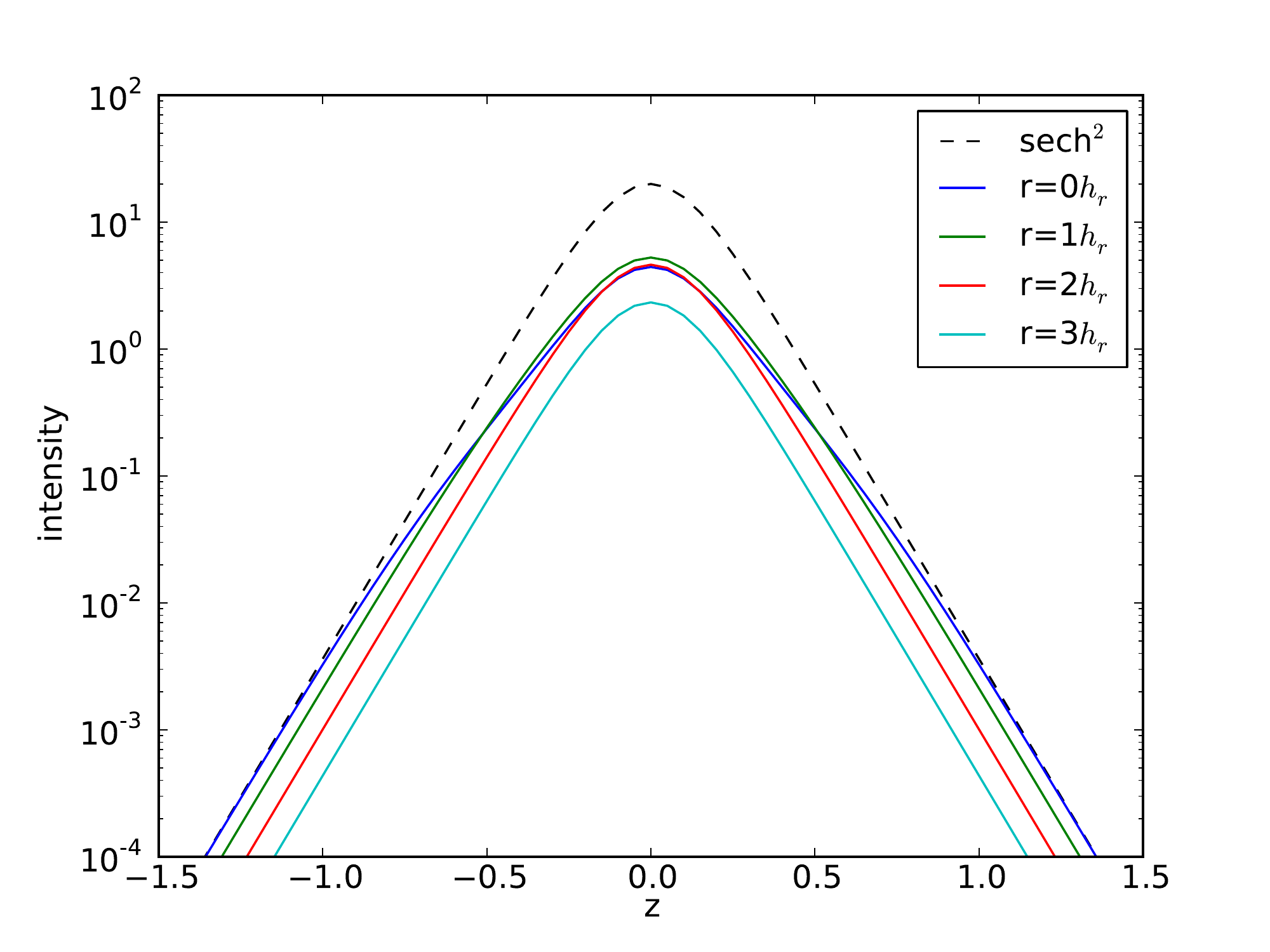}
  \end{minipage}
  \begin{minipage}[t]{0.33\textwidth}
    \centering
    \includegraphics[width=0.99\columnwidth]{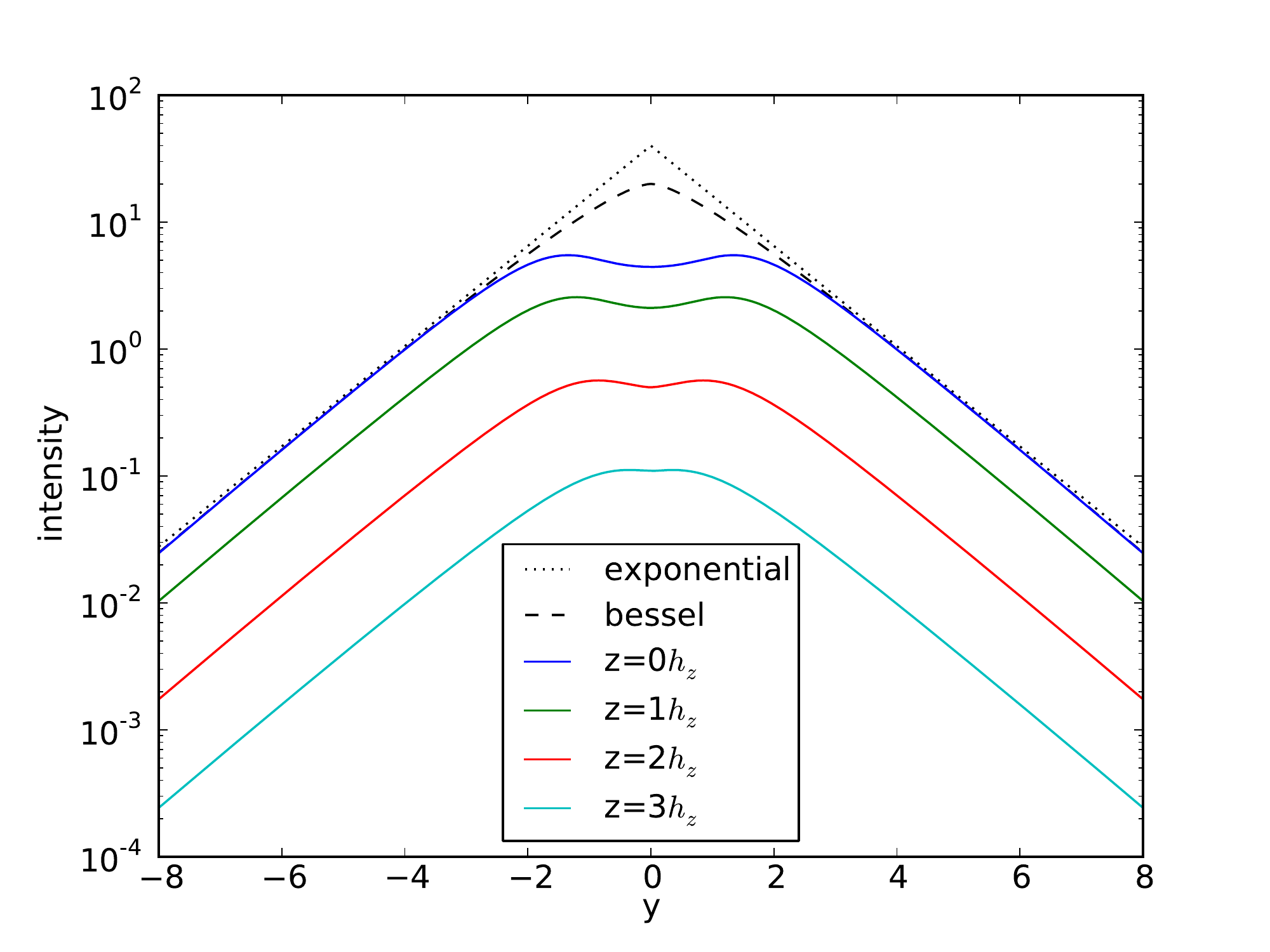}
  \end{minipage}
  \begin{minipage}[t]{0.33\textwidth}
    \centering
    \includegraphics[width=0.99\columnwidth]{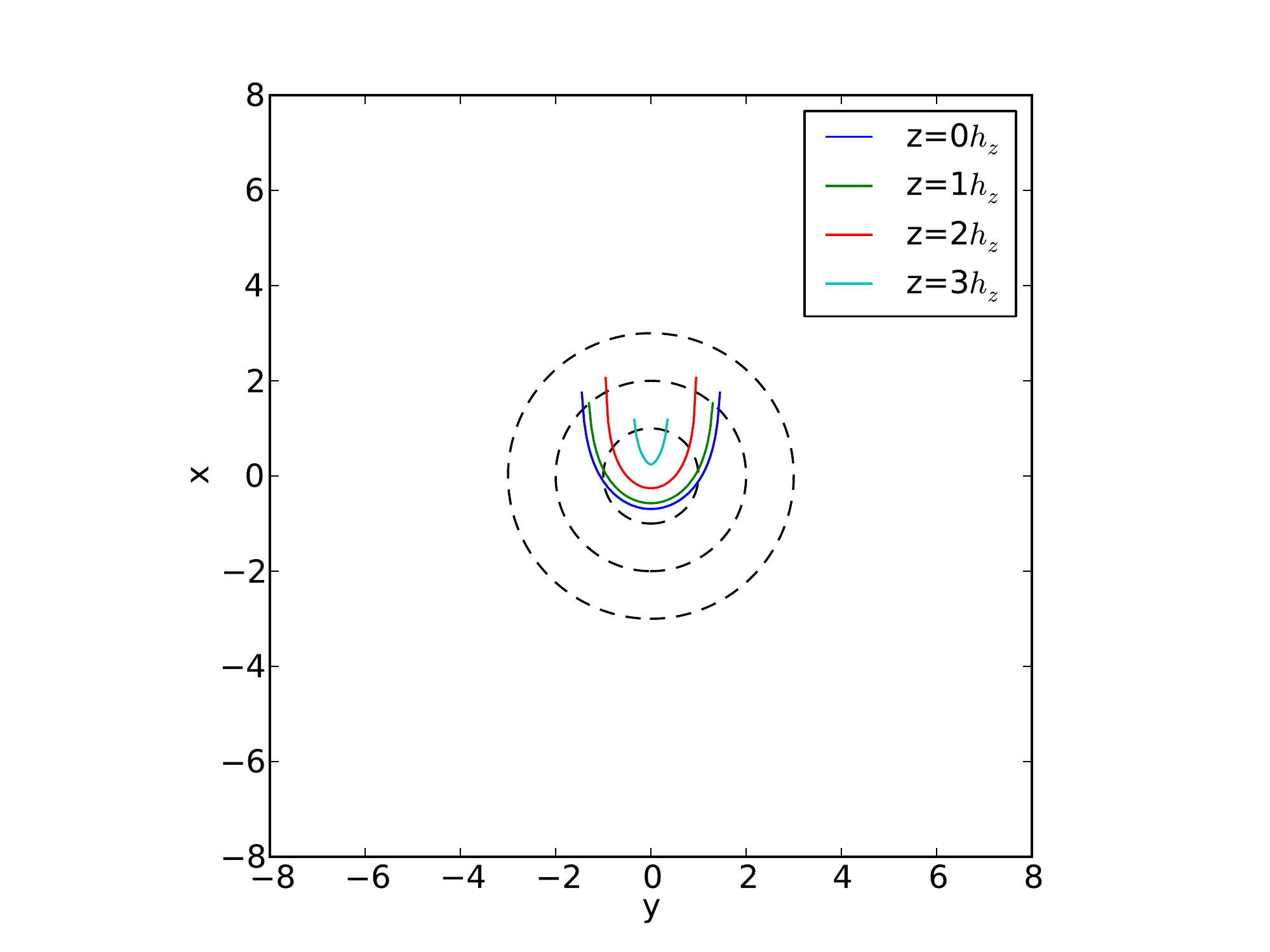}
  \end{minipage}
  \begin{minipage}[t]{0.33\textwidth}
    \centering
    \includegraphics[width=0.99\columnwidth]{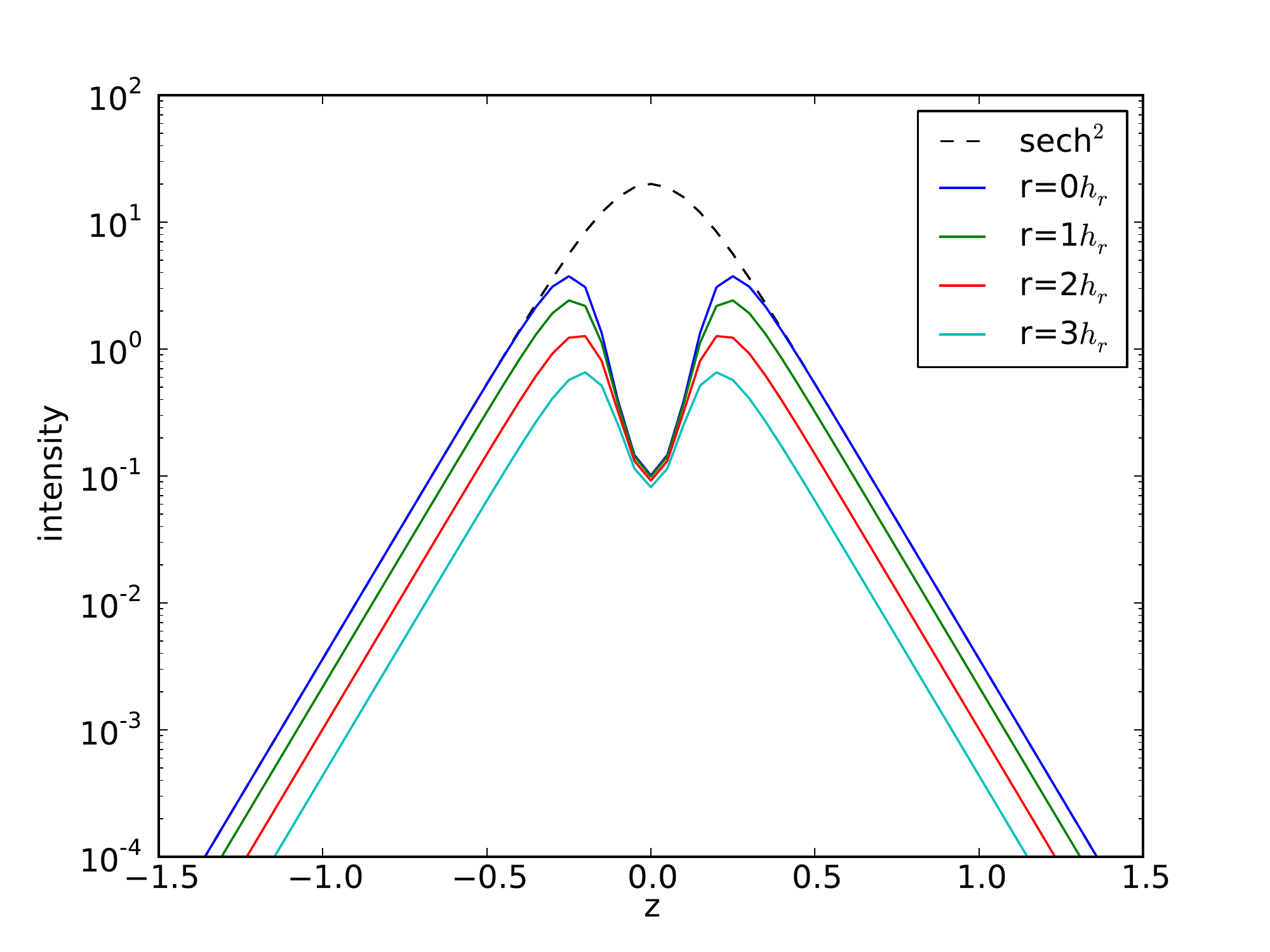}
  \end{minipage}
  \begin{minipage}[t]{0.33\textwidth}
    \centering
    \includegraphics[width=0.99\columnwidth]{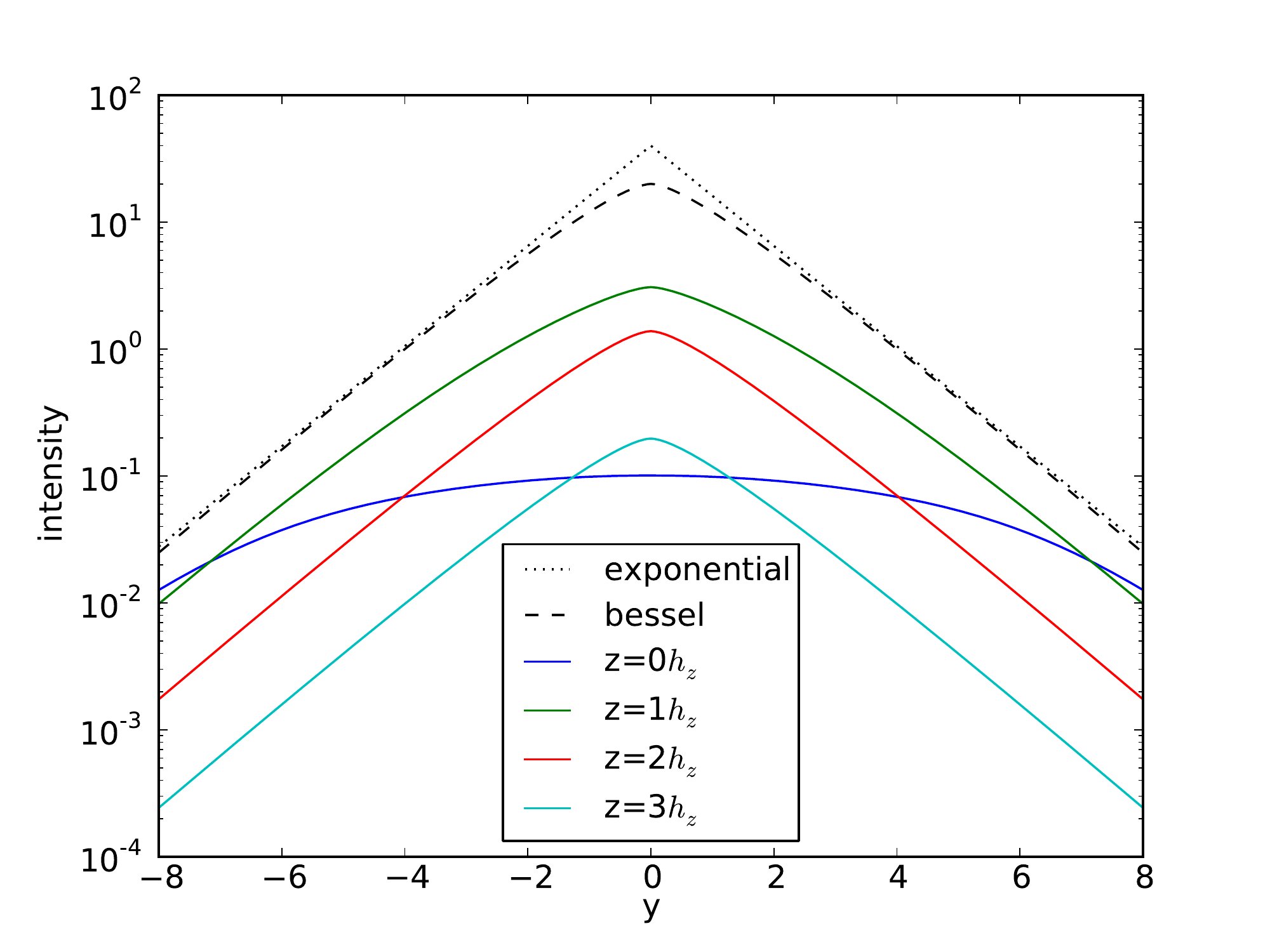}
  \end{minipage}
  \begin{minipage}[t]{0.33\textwidth}
    \centering
    \includegraphics[width=0.99\columnwidth]{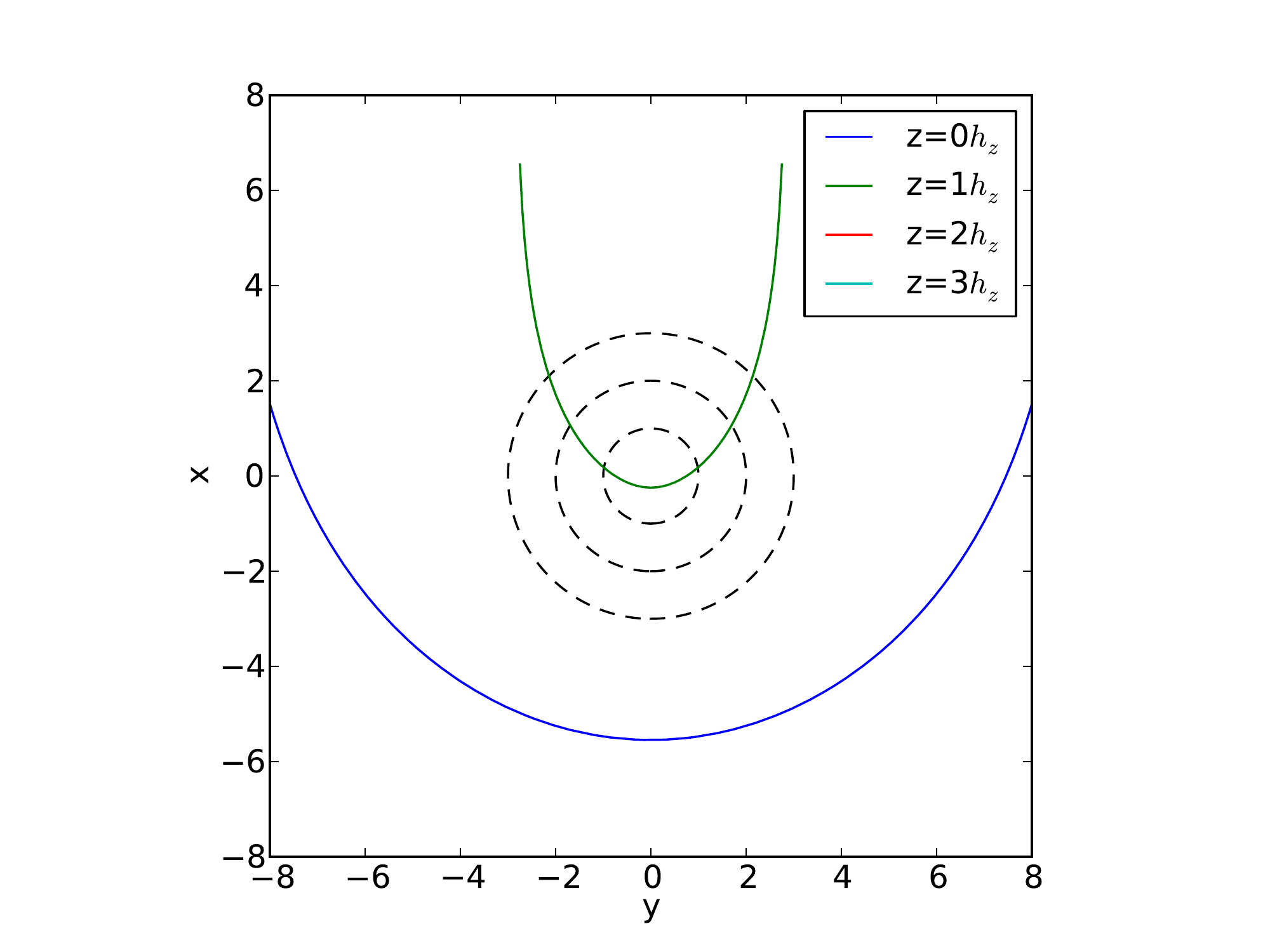}
  \end{minipage}
  \begin{minipage}[t]{0.33\textwidth}
    \centering
    \includegraphics[width=0.99\columnwidth]{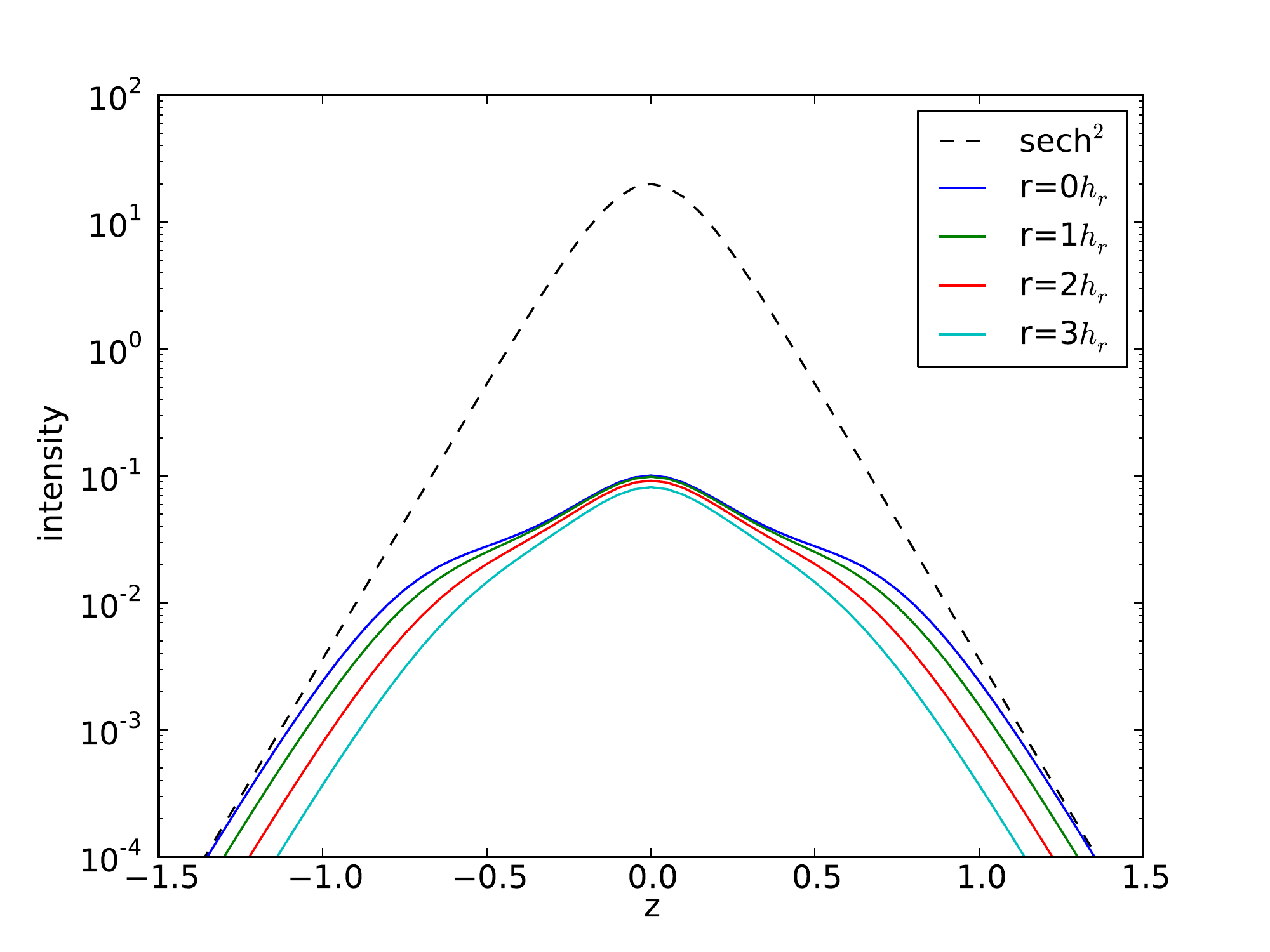}
  \end{minipage}
  \begin{minipage}[t]{0.33\textwidth}
    \centering
    \includegraphics[width=0.99\columnwidth]{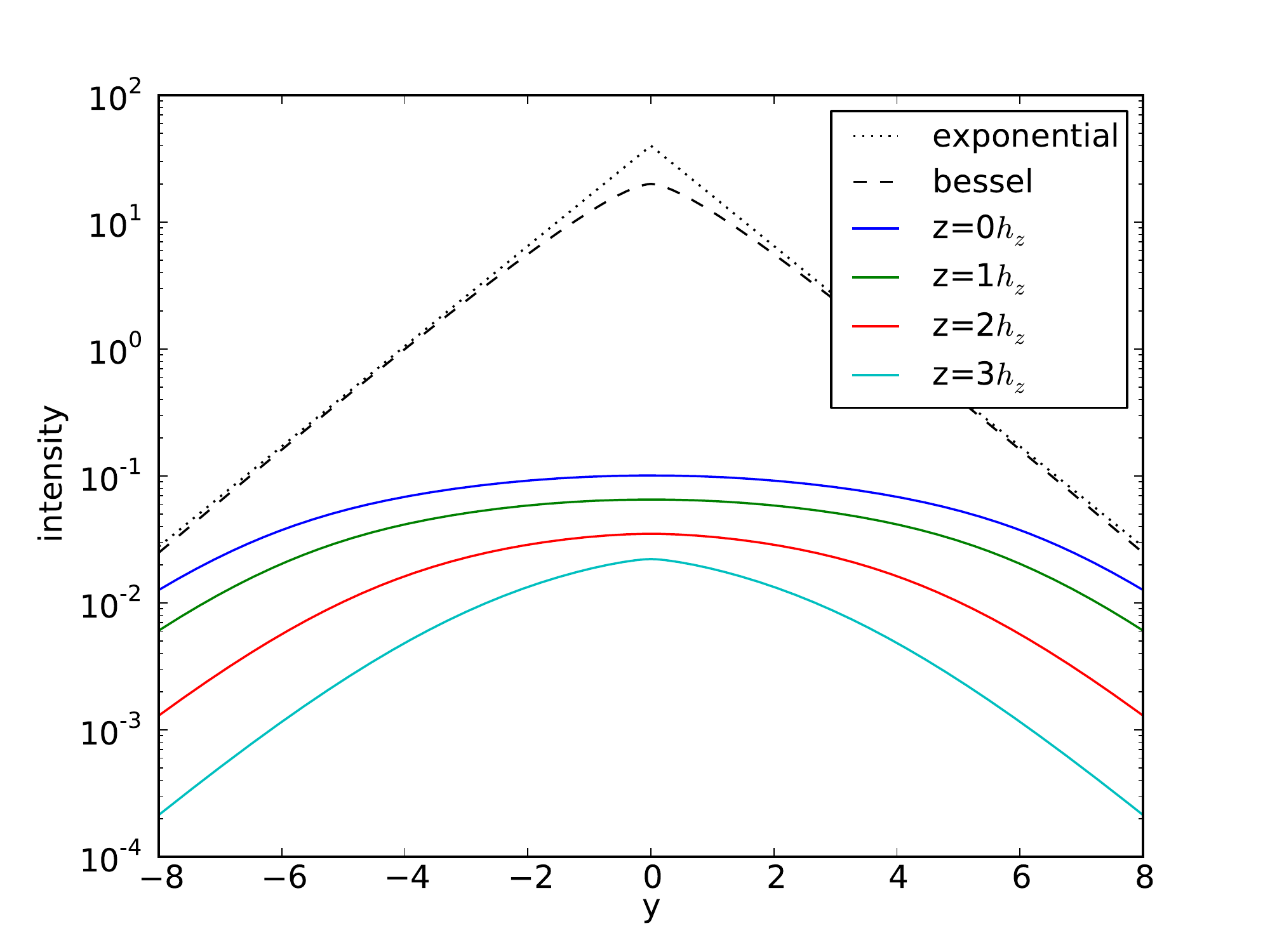}
  \end{minipage}
  \begin{minipage}[t]{0.33\textwidth}
    \centering
    \includegraphics[width=0.99\columnwidth]{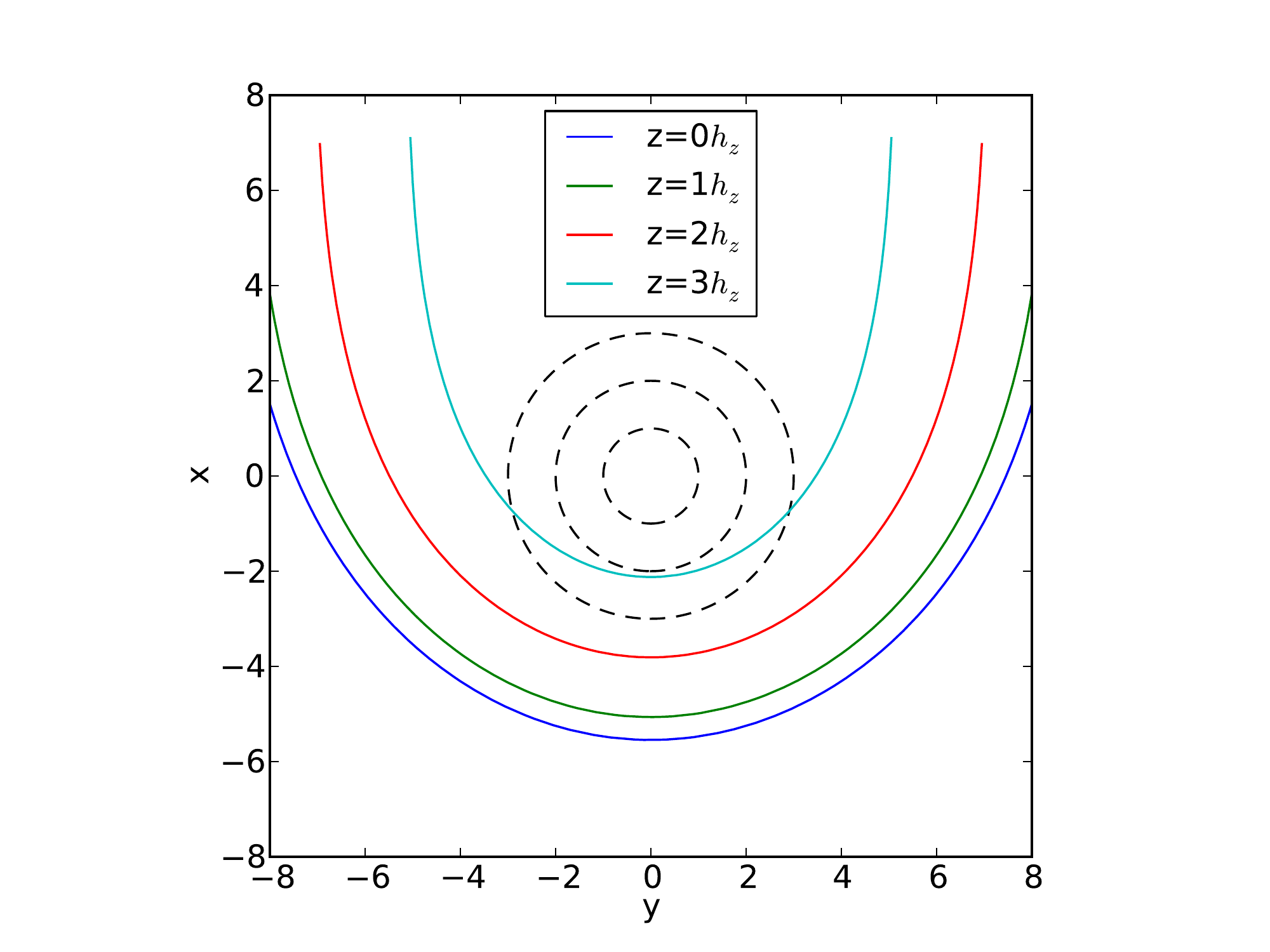}
  \end{minipage}
  \caption[Effects of dust on idealized galaxy models]{Effects of dust on ideal galaxy models; Each row shows a model of different dust properties, from top to bottom they are defined by $(h_{z,dust},h_{r,dust})=$ (0.1,0.5) (0.4,0.5), (0.1,2.), and (0.4,2.). \textit{Left column:} vertical profiles at different radial positions (as given in the legend in terms of stellar scale length); \textit{middle column:} radial profile at different heights above the plane (see legend, given in stellar scale heights); \textit{right column:} face-on view of the position where the optical depth reaches one ($\tau=1$) for different heights above the plane (see legend, given in stellar scale heights); the dashed lines are circles of one, two, and three scale lengths of the stellar component.}
  \label{fig:dustmodels}
\end{figure*}

We used three parameters for the dust and stars to model the 3D distribution: central density $\rho_0$, scale height $z_0$ and scale length $h_r$. The density is then given by
\begin{align}
\rho(x,y,z) = \rho_0\exp(-\sqrt{x^2+y^2}/h_r)\text{sech}^2(z/z_0).
\end{align}
The parameters of the stellar disk were kept fixed, with $h_r=1$ and $z_0=0.2$. The scales of the dust were varied to be either larger or smaller than the stellar ones. We analyzed four models with all possible combinations of $h_r\in \{0.5,2.\}$ and $z_0\in \{0.1,0.4\}$. The resulting vertical and radial profiles are shown in Fig.~\ref{fig:dustmodels}.

The qualitative effect of the dust on the vertical and radial profiles is similar: If the scale height (scale length) of the dust is smaller than that of the stars, there is a dip in the vertical (radial) profile. If the scale height (scale length) of the dust is larger than that of the stars, the vertical (radial) profile is flattened in the center, but no dip can be seen. The depth of the dip or the flattening depends on the dust mass \citep[see also][for a similar analysis]{seth05b}.\footnote{All four dust models have the same central density. Therefore models with larger scale heights or scale lengths have higher total dust masses.}

We compared the results of our simple dust models (Fig.~\ref{fig:dustmodels}) with the observed vertical profiles (Fig.~\ref{fig:vprofs1}, and also Figs.~\ref{fig:ic5052-2Dfits} to~\ref{fig:ngc5023-2Dfits}). The observed profiles have small dips only in the RGB profiles, while the profiles of the AGB and young stars still have peaks in the midplane. This suggests that the dust scale height is smaller then the RGB scale height, but larger than the AGB scale
height. This is consistent with our measurements of the scale heights of the PAH emission, which is near or above the AGB scale
height (see Table~\ref{tab:2Dfit_results}). \new{There is also good agreement between our PAH scale height and the dust scale
height in NGC\,4244 as measured by \citet{holwerda12} using far-IR observations with Herschel.}

Such large dust scale heights agree well with the results of \citet{seth05b}, who came to the same conclusion for a sample of low-mass galaxies (including NGC\,4244), and with the results of \citet{dalcanton04}, who found the ISM of low-mass galaxies ($V_{rot}<120$\,km/s) to be more diffuse and vertically extended than in massive galaxies.

The dust also affects the radial profiles and can lead to a dip or a flattening (see Fig.~\ref{fig:dustmodels}). In the data we do not see any central dip in our radial profiles (Fig.~\ref{fig:vprofs1}), but a flattening cannot be ruled out. Taking into account the smaller scale length of the Spitzer 3.6$\mu m$ data, it appears even very likely that the scale length of the star count data is affected by dust. Thus, the true scale length of the underlying populations might be smaller than given here.

\subsection{Direct comparison to other structural measurements in these galaxies}

\paragraph{Star count measurements}
\citet{seth05b} analyzed resolved stellar populations in the same three galaxies and divided them into three age groups (young, intermediate=AGB, old=RGB). At that time, they did not have enough data sets to study scale lengths or flaring, but they were able
to measure scale heights. Their results are given in Table~\ref{tab:literaturescaleheights}; they agree with our measurements.

In a series of papers, \citeauthor{tikhonov05b} studied the resolved stellar populations of a number of galaxies. \citet{tikhonov05b} analyzed NGC\,4244, among other galaxies. They used one central ACS field (which in our field list is Field01) and two WFPC2 fields. In agreement with our results, they found an increasing thickness of the stellar populations from blue to AGB and RGB stars, but they did not quantify this. In one of the WFPC2 fields, they claimed to detect the transition from a thick disk to a halo in form of a flattening of the vertical gradient of RGB star density at a height of z$=2.7$\,kpc. This is a similar height as we found for the transition to a weak halo in Sect.~\ref{sec:thickdisk?}, but a detailed comparison is not possible because their paper lacks a description of the selection of the stars, completeness correction, and treatment of unresolved background galaxies.

\citet{tikhonov06a} have analyzed the stellar populations of IC\,5052 and NGC\,5023. They also found an increase of thickness with increasing stellar age and a break in the vertical density gradient of RGB stars at z$=1.9$\,kpc and z$=1.6$\,kpc for IC\,5052 and NGC\,5023, respectively. While for IC\,5052 this could be the same transition we found in Sect.~\ref{sec:thickdisk?}, we do not see such a break in NGC\,5023. A more detailed comparison is not possible due to the only rough description of their data and missing figures in their manuscript.

\begin{table}[!ht]
\centering
\caption{Scale heights from the literature.}
\label{tab:literaturescaleheights}
 \begin{tabular}{lcccr}
\hline \hline
           & IC\,5052 & NGC\,4244  & NGC\,5023 &  Ref.   \\
\hline
counts $z_0$&         &          &          &   \\
- young      &   0.26   &   0.33   &   0.20   & 1 \\
- AGB        &   0.47   &   0.44   &   0.29   & 1 \\
- RGB        &   0.66   &   0.55   &   0.39   & 1 \\
\hline
NIR $z_0$  & 0.27/0.47 & 0.60      & 0.38      &  2  \\
           & 0.25/0.73 &           &           &  3  \\
           & 0.19/0.65 &           &           &  3  \\
           &           & 0.60/0.74 &           &  4    \\
\hline
optical $z_0$&         & 0.58      & 0.46      &  5  \\
\hline
dust $z_0$&         & 0.6     &       &  6  \\

 \end{tabular}
\tablefoot{All values in kpc. All scale heights are scaled to our assumed distances.}
\tablebib{(1)~\citet{seth05b};(2)~\citet{salo13}; (3)~\citet{comeron11b}; (4)~\citet{comeron11a}; (5)~\citet{vanderkruit81a,vanderkruit82a}; (6)~\citet{holwerda12}} 
\end{table}

\paragraph{Infrared measurements}
The galaxies are also listed in the $S^4G$ survey \citep{sheth10,regan13}, which also includes structure decompositions \citep{salo13}. \citet{salo13} found somewhat larger scale heights (IC\,5052: 270/740\,pc, NGC\,4244: 600\,pc, and NGC\,5023: 380\,pc) than we did for the 3.6$\mu$m images that range near our RGB scale heights. 

IC\,5052 is included in the sample of \citet{comeron11b}, who also used $S^4G$ data. They fit a thin (indexed with t) and a thick (indexed with T) disk with $z_{0,t}=250$\,pc and $z_{0,T}=730$\,pc in the outer part and $z_{0,t}=190$\,pc and $z_{0,T}=650$\,pc in the inner part of the galaxy\footnote{The original values in their paper are different (180\,pc, 530\,pc, 140\,pc, and
470\,pc); we have converted them to our scale: While they used exponential scale heights, we used sech$^2$, which differs by a factor of 2. Furthermore, \citeauthor{comeron11b} assumed a larger distance (8.1\,Mpc). We have corrected for this.}. They obtained a ratio of column mass densities $\Sigma_T/\Sigma_t=2.1$ in the outer part and $\Sigma_T/\Sigma_t=2.5$ in the inner part. They stated that their data are also compatible with a single-disk structure.

NGC\,4244 was separately analyzed in \citet{comeron11a}. They claimed to have detected a very subtle thick disk in NGC\,4244, but only on one side of the galaxy, while on the other side a single exponential is sufficient to describe the vertical profile. They derived $h_z=350$\,pc on the northern side and $h_{z,t}=300$\,pc and $h_{z,T}=370$\,pc on the southern side, which corresponds to $z_{0}=700$\,pc $z_{0,t}=600$\,pc and $z_{0,t}=740$\,pc for a sech$^2$ fit. This is significantly larger than our results. A possible reason for this discrepancy is the use of different fit functions. While an exponential and a sech$^2$ do coincide for $z\gtrsim 4h_z=2z_0$, they differ for lower values. Because both \citeauthor{comeron11a} and we also fit regions nearer to the plane, the scale heights for the exponential fit are larger
than those for the sech$^2$ fit.

\paragraph{Visual measurements}
\citet{vanderkruit81a,vanderkruit82a} have measured g-band\footnote{The authors denote their band ''J'', but with an effective bandpass from 400\,nm to 540\,nm \citep{vanderkruit79} it lies between the B and V band and covers a wavelength range similar to the SDSS g-band.} scale heights as a function of position in NGC\,4244 and NGC\,5023. They found constant scale heights along the plane with values of $z_0=0.58$\,kpc for NGC\,4244 and $z_0=0.46$\,kpc for NGC\,5023. This agrees with our RGB scale heights.

\subsection{Disk heating timescales and mechanisms}
\label{sec:heating}

The heating of galaxy disks has mostly been studied in the Milky Way, mostly due to the lack of good age indicators in external galaxies. In the Milky Way the heating is usually examined through the age--velocity dispersion relation (AVR), which is equivalent to an age--scale height relation, because scale height $h_z$ and vertical velocity dispersion $\sigma_z$ are connected by 
\begin{align}
\label{equ:velocitydispersion}
\sigma_z^2 = c\pi G\Sigma h_z, 
\end{align}
where G is the gravitational constant, $\Sigma$ the surface mass density, and c a constant that depends on the form of the vertical mass profile\footnote{e.g., for a sech$^2$ it is c=2 and for an exponential c=3/2.}.

Even though the AVR has been studied for decades, there is still no consensus on the form of the AVR. Some authors found a smooth power-law increase of the velocity dispersion with time \citep{wielen77,nordstroem04,koval09}, while others found a flattening for stars older than a few Gyr \citep{carlberg85,soubiran08}, with a possible increase for very old stars ($>9$\,Gyr) in the thick disk \citep{edvardsson93,quillen01,anguiano12a,anguiano12b}. Possible reasons for the discrepancies might be difficulties in age dating stars\footnote{For example, an age uncertainty of only 30\% for the oldest stars, which is much lower than the uncertainties in current data, can erase a sharp increase in velocity dispersion at 9\,Gyr \citep{martig14a}.} or selection effects (dwarf stars vs. giants; including all stars vs. excluding moving groups and thick-disk stars). But the authors' choice of model can likewise influence their conclusions. For example, \citet{seabroke07} showed that although the data of \citet{nordstroem04} are indeed consistent with a power-law increase of $\sigma_z$, neither can they reject a saturation of $\sigma_z$ at ages older than 4.5\,Gyr.

As possible mechanisms for heating the disk, scattering by giant molecular clouds (GMC) were the first to be proposed by \citet{spitzer51}, followed by spiral structure \citep{barbanis67} and massive compact halo objects \citep[MACHOS, e.g., black holes;][]{lacey85}. None of these alone were found to be supported by the observations \citep[see, e.g.,][and references therein]{binneytremaine}. Other proposed mechanism include minor mergers \citep{toth92}, which can lead to a sudden increase of velocity dispersion, when the satellite is relatively massive \citep{quinn93}, or to a smooth heating, when many small galaxies are accreted over a long time span \citep{velazquez99}, resonances from the bar \citep{kalnajs91} and dissipating star clusters \citep{kroupa02}. We note that the two most accepted models, namely scattering by GMC and spiral arms, would lead to a saturation. This is easily understood for the vertical motion $\sigma_z$, since GMC and spiral arms act on stars in the very midplane of the disk, and any star with a high vertical velocity will spend most of its orbital time far away from the plane. Thus, it would no longer be affected by GMCs or spiral arms.

Our data do not have the necessary age resolution to answer the question of a possible saturation of the heating, but we can constrain heating efficiencies. We fit a power law $z_0\propto t^\beta$ to our data (see Fig.~\ref{fig:scaleheightevolution}, excluding the data point of the youngest stars) and found indices $\beta$ in the range 0.09-0.32. According to Eq.~\ref{equ:velocitydispersion}, this corresponds to a power law for the velocity dispersion $\sigma_z\propto t^\alpha$ with $\alpha=\beta/2$. This is significantly lower than the Milky Way value $\alpha_{\rm{MW}}>0.3$ \citep[e.g.,][]{haenninen02,nordstroem04}\footnote{There is no clear consensus on the value of $\alpha$ for the MW, but almost all authors found values in the range $0.3<\alpha<0.6$, with the exception of \citet{anguiano12b}, who found $\alpha=0.12$ for the MW thin disk.}, but consistent with the values for low-mass galaxies found by \citet{seth05b}. A probable explanation for the lower disk heating in low-mass galaxy is that the two main heating mechanisms, scattering by GMCs and by spiral structure, are less efficient because low-mass galaxy are gravitationally more stable, resulting in a less strong spiral structure, and a more extended, but less dense interstellar medium, resulting in fewer GMCs \citep{dalcanton04}.

\subsection{Structural evolution of stellar populations}
\label{sec:evolutionmodels}

\paragraph{}
\citet{martig14a,martig14b} have analyzed seven simulations of disk galaxies representing a variety of merger histories. They have separated each galaxy into so-called mono-age populations, each representing an age range of 500\,Myr, and analyzed how their structural parameters (scale length and -height, velocity dispersion) change as a function of age (at z=0).

In galaxies with a quiescent merger history they found a smooth decrease of scale length with age and a smooth increase of scale
height (and vertical velocity dispersion). Only for very old stars (age$>8.5$\,Gyr), a jump in vertical velocity dispersion can be seen. These very old stars are centrally more concentrated and show strong flaring; they are more likely associated with a bulge or inner halo component. The physical reason for the smooth evolution of scale lengths is seen in radial migration. Stars are born in an irregular, but relatively flat profile with a break. With time, stars migrate and populate the outer disk, thus erasing the break. At the same time, the central parts contract (due to conservation of angular momentum), which leads to a steepening of the radial profile. The vertical heating in their simulation is caused by spiral arms, overdensities in the disk, the bar and bending waves, but not by radial migration.

As expected, mergers have a significant effect on the evolution of scale height: A merger leads to a sharp increase in the AVR by increasing the velocity dispersion or scale heights of all populations born before the merger. But it also has an interesting effect on the scale length evolution: Near the time of a merger, the normal trend of scale lengths with age is reversed. For a certain age range (usually about $~2$\,Gyr), the scale length increases with age. This leads to V-shaped structures in a scale
 length vs. age plot. During this period, scale height and scale
length are correlated.

The increase (decrease) of scale height (scale length) with age and the corresponding anti-correlation in a $h_z$--$h_r$ plot that \citeauthor{martig14b} see in quiescent galaxies is very similar to what we see in NGC\,4244 and NGC\,5023 (see Fig.~\ref{fig:scales}).

\paragraph{}
The exceptional behavior of IC\,5052 can also be explained in the light of the results of \citeauthor{martig14a}, if we assume that IC\,5052 recently experienced a merger. This assumption is strongly supported by the irregular radial profile of IC\,5052 (see Sect.~\ref{sec:irregularities}); the overdensity of old stars that is found with an offset from the dynamical center of the galaxy could be the remnant of a merging satellite. Furthermore, the residual map from the fit (Fig.~\ref{fig:ic5052-2Dfits}) reveals a stream-like feature that connects the overdensity on the northwestern \new{(in our plots left)} side of the disk with a smaller overdensity at the southeastern \new{(right)} side. 
\citet{martig14a} have also shown that a merger will lead to a strong flaring of the pre-existing disk. That fits the fact well that the old population of IC\,5052 has the strongest hints on flaring (see Table~\ref{tab:flaring}) and the highest heating efficiency (see Fig.~\ref{fig:scaleheightevolution}) of all our galaxies.

\paragraph{}
\citet{bird13} have analyzed a high-resolution cosmological simulation of a Milky Way-mass galaxy. They also separated their simulated galaxy into different age groups and found an increasing scale
height and a decreasing scale length with age. These trends are similar to \citet{martig14a,martig14b} and our observations, but in contrast to \citet{martig14a,martig14b}, these trends are imprinted on the populations during their formation and change only little during the evolution of the galaxy. In their model the age evolution of the structural parameters is a direct consequence of the time evolution of the properties of the gas from which the stars are formed. While the gas at high redshift has a large vertical extent and is mainly dispersion supported, it becomes thinner and radially more extended with time; and the stars follow this behavior. 

Another feature in the galaxy of \citet{bird13} can also be compared
with our observations. Their galaxy has a strong flaring in all age groups and velocity dispersion profiles that are almost constant with radius for each age group. This strongly disagrees with our observations and with integrated light \citep{vanderkruit81a,vanderkruit81b,vanderkruit82a,vanderkruit82b,shaw90,degrijs96}\footnote{Some galaxies in these papers do show a hint of an increase of scale
height at the very edges of the disks. See also \citet{degrijs97a} for a sample of flaring galaxies.} and kinematical observations \citep{vanderkruit84a,vanderkruit86,bottema93,martinsson13}, which all found constant scale heights and exponentially decreasing velocity dispersions. The strong flaring reported by \citet{bird13} is not a consequence of the evolution or interactions, but it is also already imprinted on the gas at all times.

\new{The small amount of flaring that we observe in the three galaxies also contradicts most simulations that model disk formation in a cosmological context. The ubiquity of merger events inevitably produces flaring disks \citep{bournaud09,qu11}. The absence of (strong) flaring could thus indicate a rather quiet merger history of these galaxies. A possible alternative explanation was recently proposed by \citet{minchev15}, who found that the total stellar density can appear unflared even when all the mono-age populations flare. This may occur because the scale lengths change with age and the flare of each population occurs at a different radii. The combined flares of all populations can then be described as a nonflaring thick disk with a radial age gradient. While we do resolve the stars in populations of different ages, the wide age range especially of the RGB component could smear out the flaring effects; but in this case, a separate thick-disk component should be present. The absence of such a thick disk renders this scenario unlikely for our observations.}

\paragraph{}
\citet{stinson13} compared a cosmological simulation of a Milky Way-like galaxy including detailed chemical modeling with the results of \citet{bovy12b}. They also separated their model galaxy into mono-abundance populations and found an anticorrelation between scale height and scale length. They showed that the mono-abundance populations are good approximations of mono-age populations (usually with $1\sigma$ age spreads of less than 1\,Gyr) and that the anticorrelation is indeed caused by an increase of scale height and a decrease of scale length with age. This therefore fully agrees with our observations of NGC\,4244 and NGC\,5023. \citet{stinson13} have neither studied the radial variations of scale heights, nor the physical origins of these relations.

\subsection{Thin- and thick-disk dichotomy?}

Studies of the age evolution of galactic disks beyond the MW (or even the Local Group) similar to the theoretical work in Sect.~\ref{sec:evolutionmodels} were very difficult until recently, and only a few authors have tried it \citep[e.g.,][]{seth05b,tikhonov12b,tikhonov12a}. Thus, most observational data on disk evolution comes form thin- and thick-disk decompositions\footnote{or from color gradients, but degeneracies between age, metallicity, and extinction effects make an interpretation of these gradients ambiguous.}. 

Thick disks are usually found to be thicker by a factor of at least $z_T/z_t>1.6$ \citep{yoachim06}, in most cases it is even $z_T/z_t>2.6$ \citep[][and references therein]{comeron11b,pohlen04}. The scale lengths of the thick components are, with only a few exceptions, also larger than those of the thin disk \citep{pohlen04,yoachim06,comeron11b}. On average, no flaring was found in either the thin or the thick disk \citep{comeron11b}.

The question then is how our disks would fit into these samples
and whether our old disks are comparable to thick disks.

\begin{figure}[!tb]
 \centering
 \includegraphics[width=3.7in]{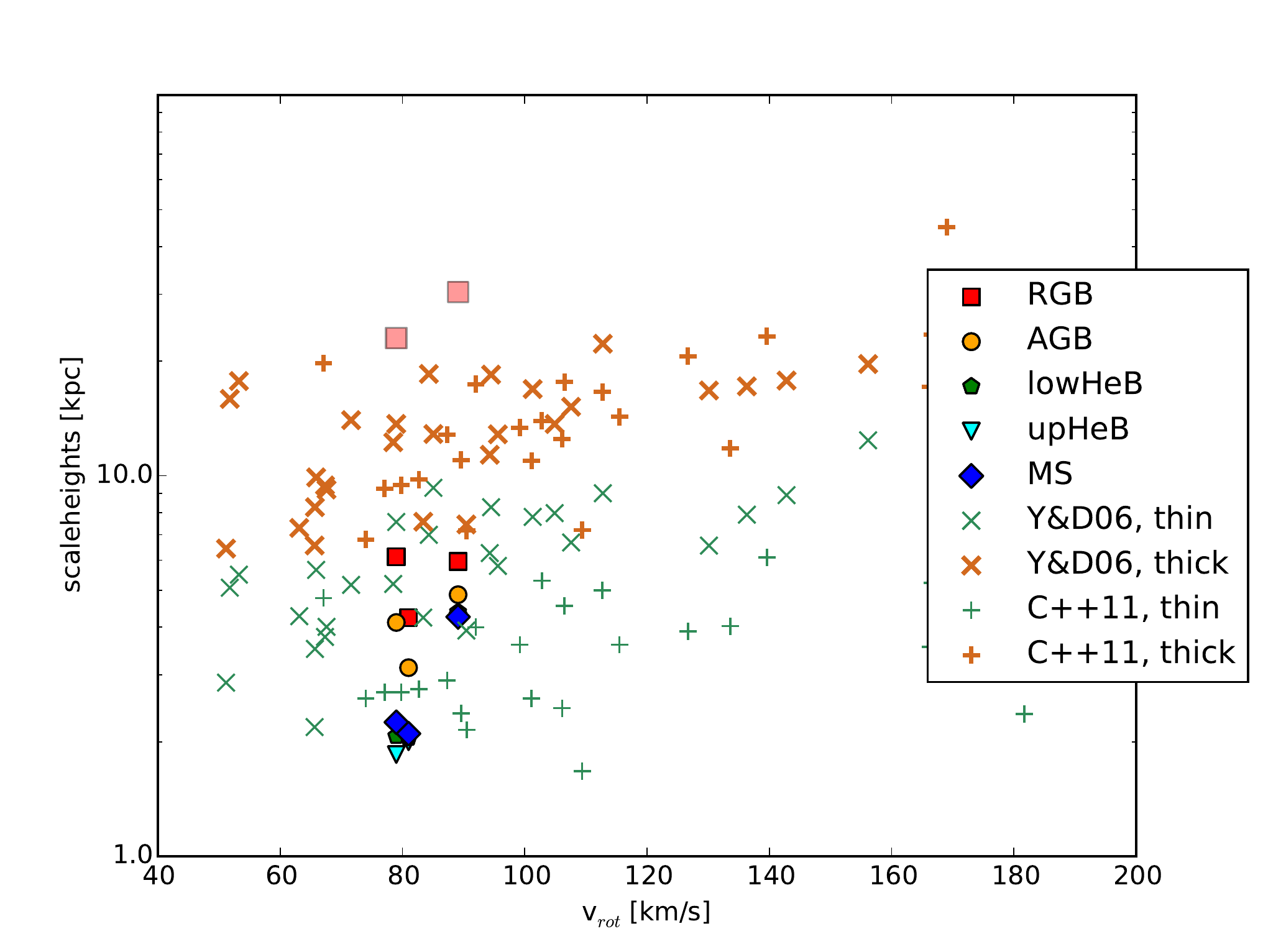}
 \caption[Comparison of our measured scaleheights with literature data]{Comparison of our measured scale heights with data from \citet{yoachim06} and \citet{comeron11b}. The colored circles show our results for the different populations (color code as before); squares show the faint extended component (Sect.~\ref{sec:thickdisk?}).}
 \label{fig:scaleheightcomparison}
\end{figure}

Figure~\ref{fig:scaleheightcomparison} shows the scale heights compared to thin- and thick-disk scale heights reported by \citet{comeron11b} and \citet{yoachim06}. While all our measured disk scale heights are consistent with thin-disk scale heights, the scale heights of the faint extended components in IC\,5053 and NGC\,4244 are larger than typical thick-disk scale heights.

We found old-to-young scale height ratios of about 3, 1.5, and 2 (for IC\,5052, NGC\,4244, and NGC\,5023, respectively). While for IC\,5052 and NGC\,5023 this agrees well with earlier thick-disk studies, it would make the disk of NGC\,4244 the thinnest thick disk (relative to the thin disk) seen so far. These thickness ratios also agree well with the velocity dispersion measurement within the thin disk of the Milky Way (see Sect.~\ref{sec:heating}): The velocity dispersion of stars increases from less than 10\,km/s at an age of 1\,Gyr to more than 20\,km/s at ages of about 6\,Gyr. According to Eq.~\ref{equ:velocitydispersion}, such a doubling of velocity dispersion would result in an increase of scale height by a factor of four, which is already more than we observe. Furthermore, in two of three galaxies (namely NGC\,4244 and NGC\,5023) the old and thicker component has a much shorter scale length than the young populations, in contrast to earlier thick-disk observations in external galaxies. 

Therefore, we conclude that the old disks observed here are really the old thin disks of these galaxies. More extended components are much fainter and cannot contain more than 2\% of the stellar mass (see Sect.~\ref{sec:thickdisk?}).

\subsection{Breaks}

In each galaxy we found the break radius to be the same for all populations. The breaks are sharper, that is, the ratio $h_{r,i}/h_{r,o}$ is higher, for younger populations. This agrees with earlier observations by \citet{dejong07b} and \citet{radburn12}. The conclusion that a constant break radius favors a dynamical origin of the break \citep{dejong07b} was questioned by \citet{roskar08a}. In their simulation, the break was caused by a decrease in the gas density and the break radius moved outward with time, while gas with higher angular momentum was accreted. Despite this change of break radius with time, at the end of the simulation the break radius was the same for populations of all ages. This is due to the radial migration of stars: The oldest stars that have the most time to migrate move farthest outward, shifting their break to larger radii and weakening their strength.

A similar behavior is seen in the test particle simulations of \citet{elmegreen13}. They started with a flat radial density profile of stars with a sharp cutoff at a given radius. This initial disk was perturbed by transient mass clumps in the disk, which led to a double exponential profile with decreasing break strength, until at the end, after several Gyr, the break was erased and a single exponential had formed. During this process, the break radius also moved outward.\footnote{The authors stated that the break radius of the double exponential profile was comparable to the initial cutoff, but a detailed look at their Fig.~2 shows an increase of break radius with time.}

Our findings are in contrast with those of \citet{bakos08}, who argued that the break is only a change of stellar populations, but not a change in the mass density profile. While we also see a change in stellar populations, namely a change in the ratio of young to old stars, which has a maximum near the break, it is also clear from the existence of the break in all populations that the mass profile is also broken.

\section{Conclusions and summary}
\label{sec:summary}

We have mapped the stellar populations of three low mass, edge-on disk galaxies: IC\,5052, NGC\,4244, and NGC\,5023. Through analyzing their CMDs, we separated the stellar content of the galaxies into five populations of distinct ages. We measured the structural parameters (scale heights, inner and outer scale lengths, break radii, and flaring) for each population and analyzed their dependence on stellar age.

Each component of each galaxy was well fit with a single-disk model with constant scale height and a broken exponential radial profile. None of the populations showed a need for second (e.g., thick) disk component. Along the minor axes of IC\,5052 and NGC\,4244 we detected very faint additional components in the old population, which contained at most 1\% of the main disks mass and are thus much fainter than a typical thick disk. These extended components are probably the faint halos of their galaxies.

We estimated the amount of flaring in all the populations. While the younger populations do not flare, the intermediate and old populations show a small increase of scale height by about 10\,pc per kpc in radius. This adds up to a relative flaring of not more than 30\% over the whole observed radial range, which reaches well beyond the break in the radial profiles. While such a low amount of flaring is consistent with most earlier observations, it \new{challenges cosmological simulations of galaxy disks, where interactions and mergers usually lead to a stronger flaring.}

All galaxies have an increase of scale height with age, but the disk heating efficiency, expressed through the index $\alpha$ in the power law of the form $\sigma_z\propto \sqrt{z_0} \propto t^\alpha$ , is lower than in the Milky Way. This can be well understood through the weaker spiral structure (i.e., higher gravitational stability) and more diffuse interstellar medium of low-mass galaxy disks.

The break in the radial profiles appears at the same radius for all ages in each galaxy. In two of the galaxies (NGC\,4244 and NGC\,5023), the break strength decreases with age, which supports break formation models that induce a sharp cutoff of star formation (either through star formation thresholds or limited gas supply), which are then softened by scattering and migration of stars over time.

In IC\,5052 we detected an additional overdensity of old stars, which lies off the center of the disk, and a stream-like feature in the fit residuals. These were interpreted as an indication of a recent merger. The effects of this merger can be seen in the increasing scale length with age, the larger amount of flaring, and the higher heating efficiency in IC\,5052 compared to the other two galaxies.


\begin{acknowledgements}
We thank Peter Erwin for his help with the IMFIT program and for his quick incorporation of our suggestions. We thank Andrew Dolphin for the permission to use the MATCH program and for his help.

D.S. gratefully acknowledges support from the Cusanuswerk through a PhD scholarship and from the Deutsches Zentrum f\"ur Luft- und Raumfahrt (DLR) through grant 50OR1012. 

This work is based on observations made with the NASA/ESA Hubble Space Telescope, obtained at the Space Telescope Science Institute, which is operated by the Association of Universities for Research in Astronomy, Inc., under NASA contract NAS 5-26555. These observations are associated with programs SNAP-9765, SNAP-10523, GO-10889, GO-12196, and GO-13357. Support for these programs was provided by NASA through grants from the Space Telescope Science Institute.

This research has made use of NASA's Astrophysics Data System. We acknowledge the usage of the HyperLeda database (http://leda.univ-lyon1.fr). 
\end{acknowledgements}

\bibliographystyle{aa}
\bibliography{../bibliography.bib}

\onecolumn
\begin{appendix}

\section{Scale heights in the Spitzer [3.6$\mu$] data}
\label{sec:spitzer-scaleheights}

We show here the measurements of scale heights along the major axis of the galaxies for the Spitzer data. We fit both a simple sech$^2$ and generalized sech$^{2/n}$. The Spitzer profiles are usually more peaked than a sech$^2$ (see also Sect.~\ref{sec:spitzer-obs}), and we found $n>1$. The scale heights of the sech$^{2/n}$ fits agree well with the RGB scale heights, and there is no or very little flaring. 

The parameter n of the sech$^{2/n}$ increases toward the center of the galaxies, meaning that the profiles in the inner disks are more peaked than in the outer disks. 
\begin{figure}[!ht]
    \centering
  \begin{minipage}[t]{0.33\textwidth}
    \centering
    \includegraphics[width=0.99\columnwidth]{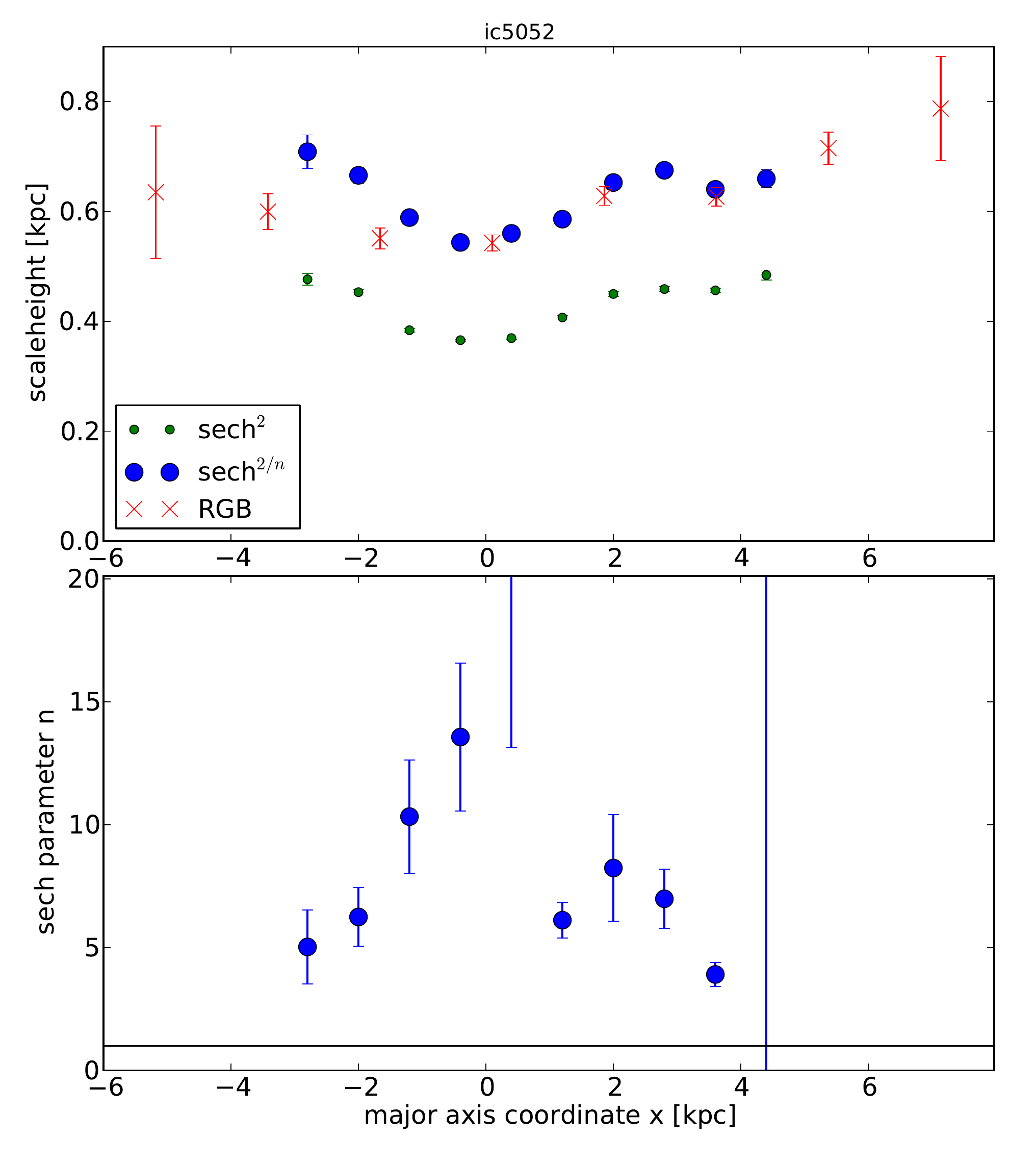}
  \end{minipage}
  \begin{minipage}[t]{0.33\textwidth}
    \centering
    \includegraphics[width=0.99\columnwidth]{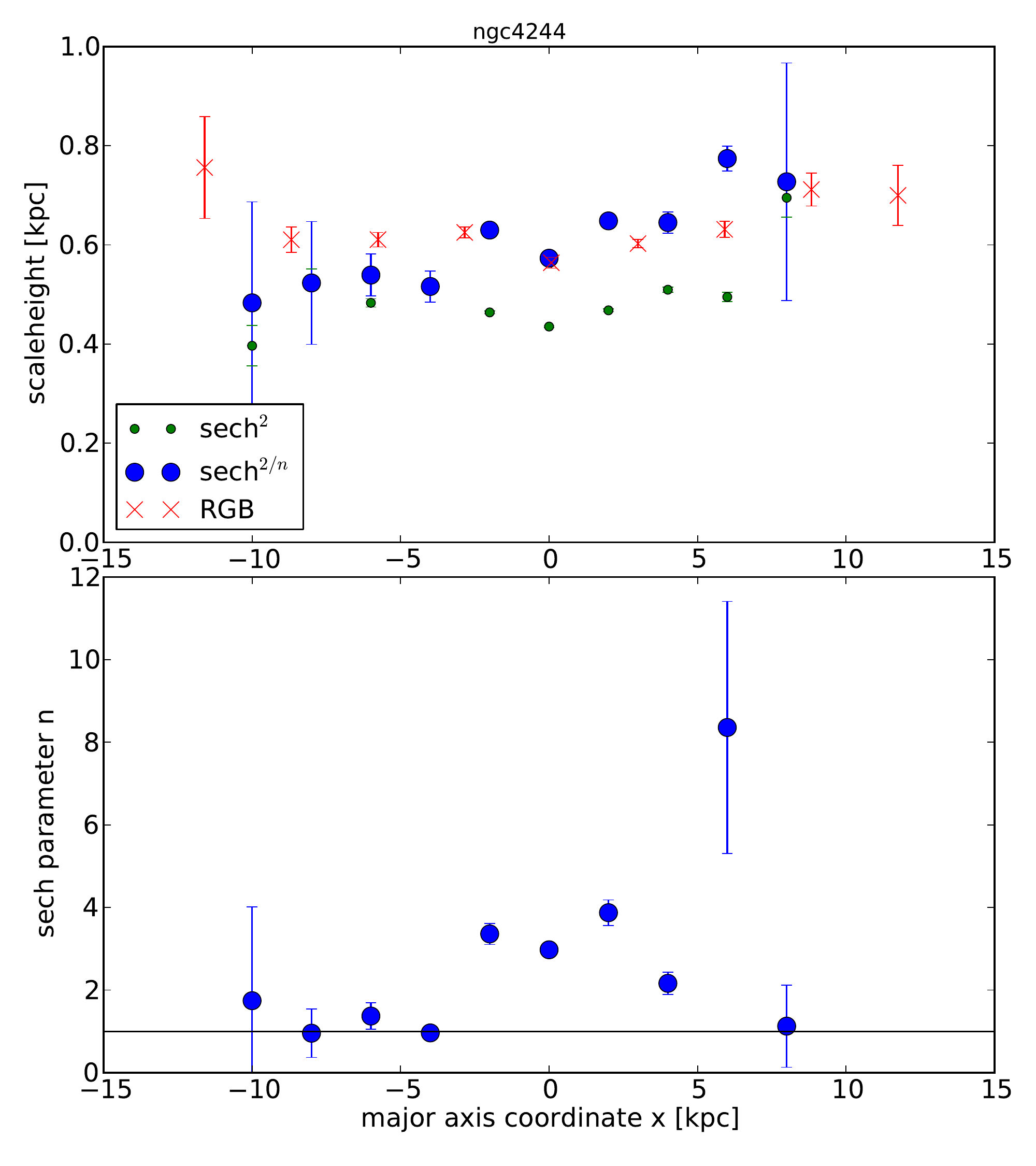}
  \end{minipage}
  \begin{minipage}[t]{0.33\textwidth}
    \centering
    \includegraphics[width=0.99\columnwidth]{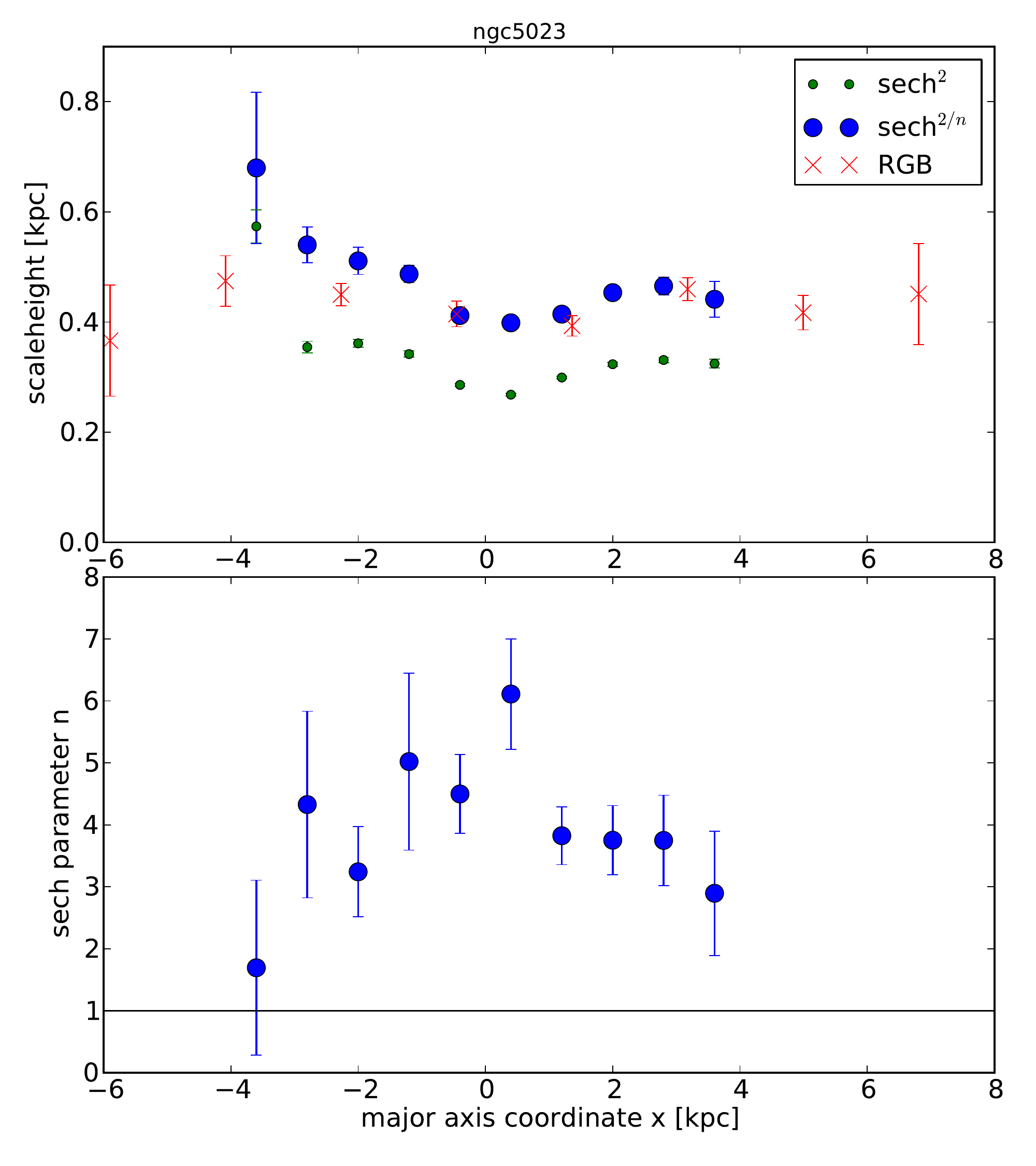}
  \end{minipage}
\caption[Scaleheights in Spitzer channel 1]{\emph{Upper panels}: Scale heights in Spitzer IRAC channel 1 as a function of position along the major axis (from left to right: IC\,5052, NGC\,4244, NGC\,5023). \emph{Lower panels}: parameter n of the generalized sech$^{2/n}$ fits.}
\label{fig:spitzer_scaleheights}
\end{figure}

\section{Color-magnitude diagrams}
\new{We present here a number of color magnitude diagrams to show the population selection boxes and to illustrate the variety of CMD features, stellar densities, and observation depths. For each galaxy we show one CMD from the central disk, the outer disk, and the halo (if available).}

\begin{figure}[!ht]
    \centering
  \begin{minipage}[t]{0.33\textwidth}
    \centering
    \includegraphics[width=0.99\columnwidth]{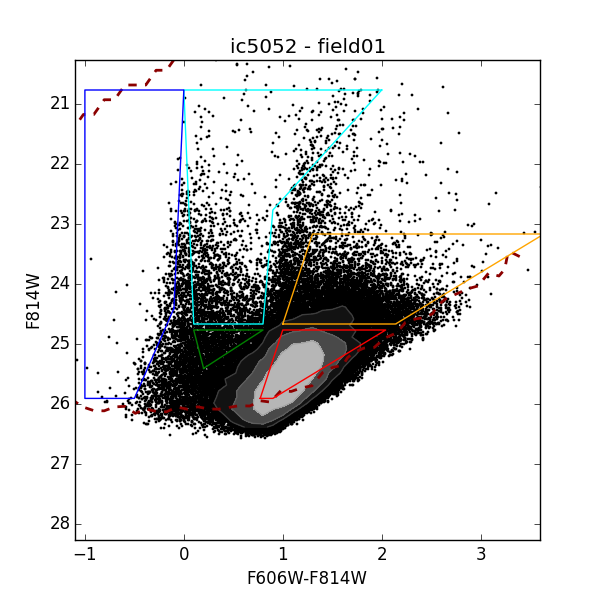}
  \end{minipage}
  \begin{minipage}[t]{0.33\textwidth}
    \centering
    \includegraphics[width=0.99\columnwidth]{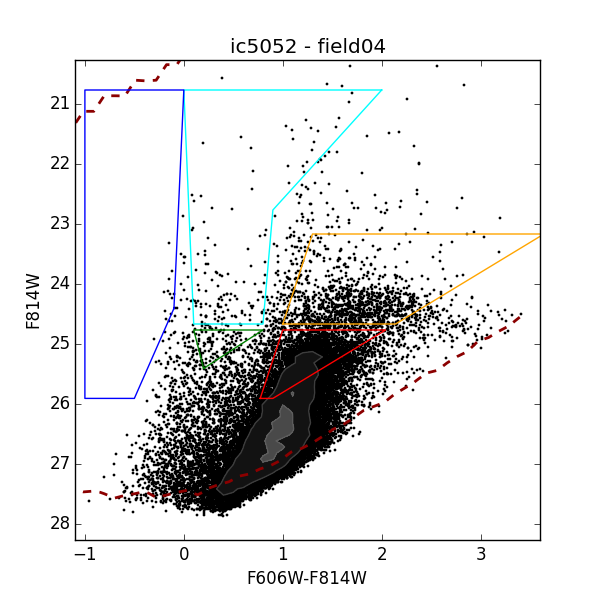}
  \end{minipage}
  \begin{minipage}[t]{0.33\textwidth}
    \centering
    \includegraphics[width=0.99\columnwidth]{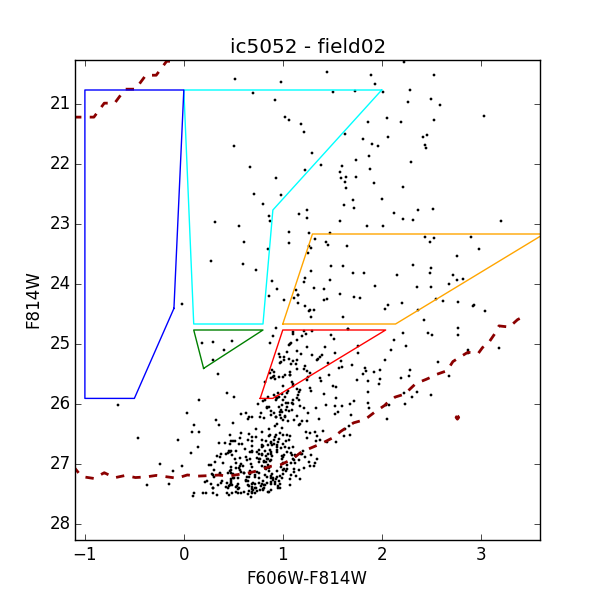}
  \end{minipage}
\caption[CMDs of IC\,5052]{\new{Color-magnitude diagrams from three fields in IC\,5052. We show a field from the central disk (left), the outer disk (middle), and the halo (right). Dense regions in the CMD are plotted as a Hess diagram with contours at 100, 200, 400, and 800 stars per (0.1\,mag)$^2$. Colored boxes are the selection boxes for the populations defined in Sect.~\ref{sec:ageselection},
and the red dashed line is the 50\% completeness level of the observations.}}
\label{fig:CMDs_ic5052}
\end{figure}

\begin{figure}[!ht]
    \centering
  \begin{minipage}[t]{0.33\textwidth}
    \centering
    \includegraphics[width=0.99\columnwidth]{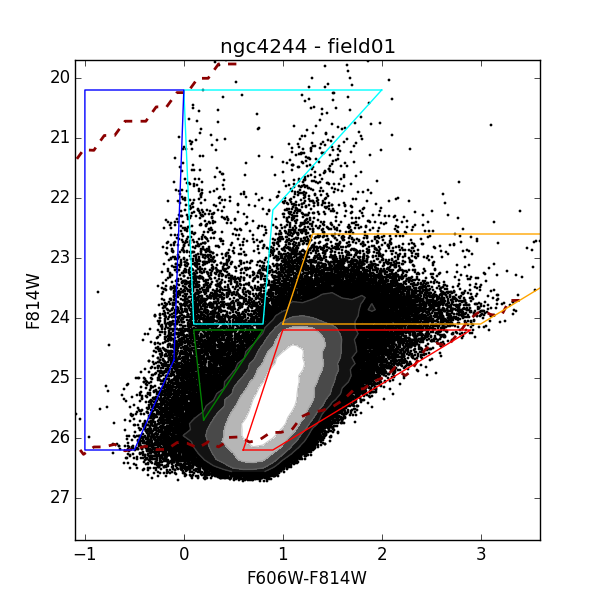}
  \end{minipage}
  \begin{minipage}[t]{0.33\textwidth}
    \centering
    \includegraphics[width=0.99\columnwidth]{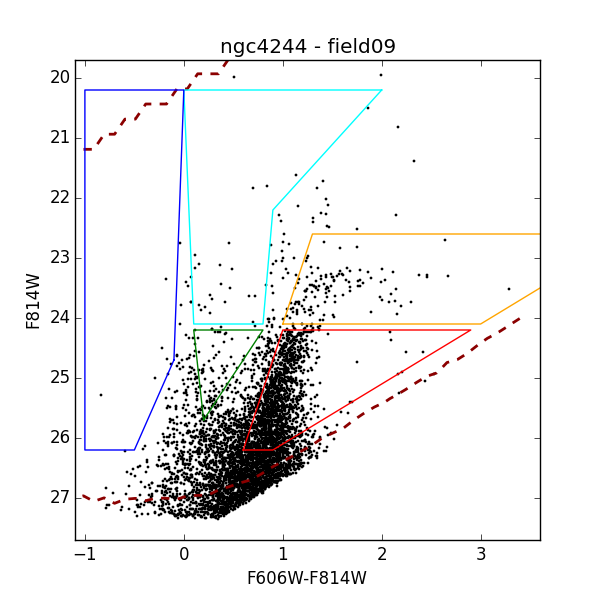}
  \end{minipage}
  \begin{minipage}[t]{0.33\textwidth}
    \centering
    \includegraphics[width=0.99\columnwidth]{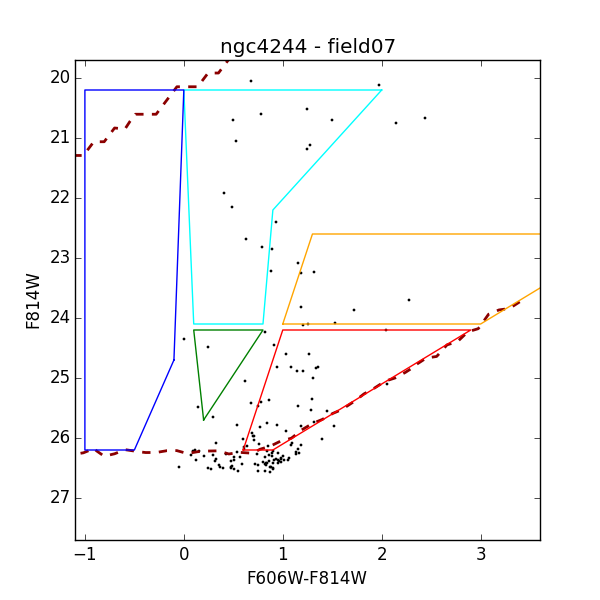}
  \end{minipage}
\caption[CMDs of NGC\,4244]{Color-magnitude diagrams from three fields in NGC\,4244: Details as in Fig~\ref{fig:CMDs_ic5052}.}
\label{fig:CMDs_ngc4244}
\end{figure}

\begin{figure}[!ht]
    \centering
  \begin{minipage}[t]{0.33\textwidth}
    \centering
    \includegraphics[width=0.99\columnwidth]{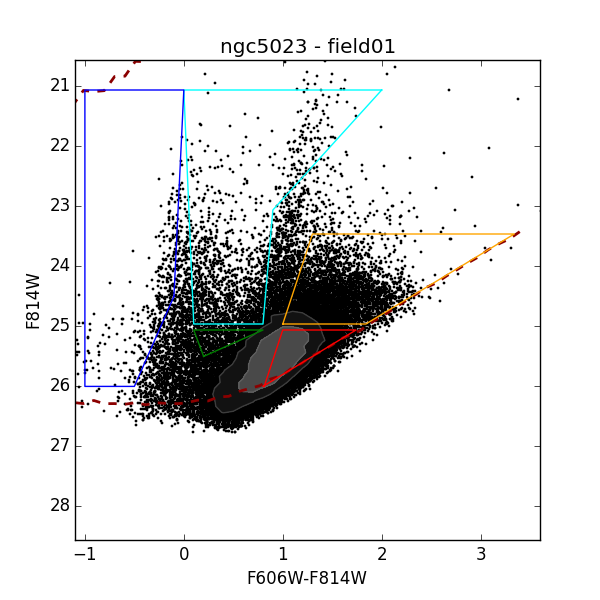}
  \end{minipage}
  \begin{minipage}[t]{0.33\textwidth}
    \centering
    \includegraphics[width=0.99\columnwidth]{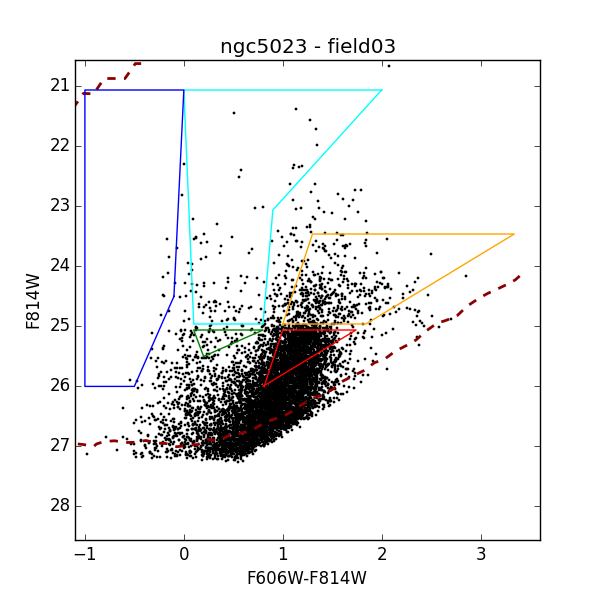}
  \end{minipage}
  \begin{minipage}[t]{0.33\textwidth}
    \rule{0.99\columnwidth}{0mm}
  \end{minipage}
\caption[CMDs of NGC\,5023]{Color-magnitude diagrams from three fields in NGC\,5023: Details as in Fig~\ref{fig:CMDs_ic5052} (for NGC\,5023 no halo field is available).}
\label{fig:CMDs_ngc5023}
\end{figure}

\section{2D Fits}
We present here the details of the two-dimensional fits to the star count maps. For each galaxy we show for each population the data and the model image. We also show the weighted residuals. We defined the residual $wdev_i$ of each pixel in the context of Poisson noise dominated data to be
\begin{equation}
wdev_i = \mathrm{sign}(n_i-m_i)\sqrt{-2(n_i\ln m_i - m_i + n_i(1-\ln n_i))}
\label{equ:residuals}
,\end{equation}
with $n_i$ the pixel value of the data and $m_i$ the pixel value of the model. This is analog to the normal weighted residual $(n_i-m_i)/\sigma_i$ in the sense that it is positive for $n_i>m_i$ and negative for $n_i<m_i$, and it is $PLR = \sum_i(wdev_i^2)$.

\begin{sidewaysfigure*}[!ht]
  \begin{minipage}[t]{0.245\linewidth}
    \centering
    \includegraphics[width=0.99\columnwidth]{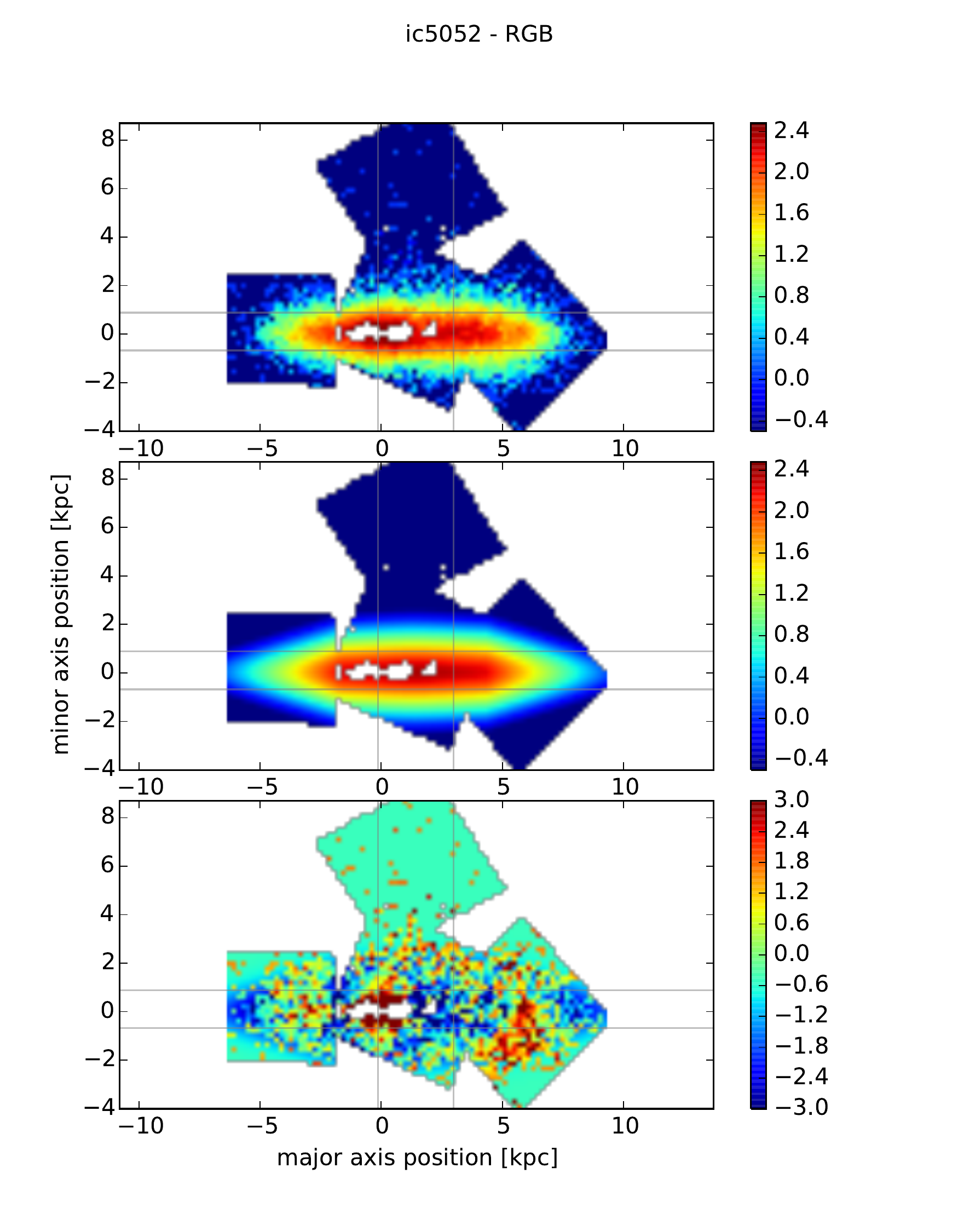}
  \end{minipage}
  \begin{minipage}[t]{0.245\linewidth}
    \centering
    \includegraphics[width=0.99\columnwidth]{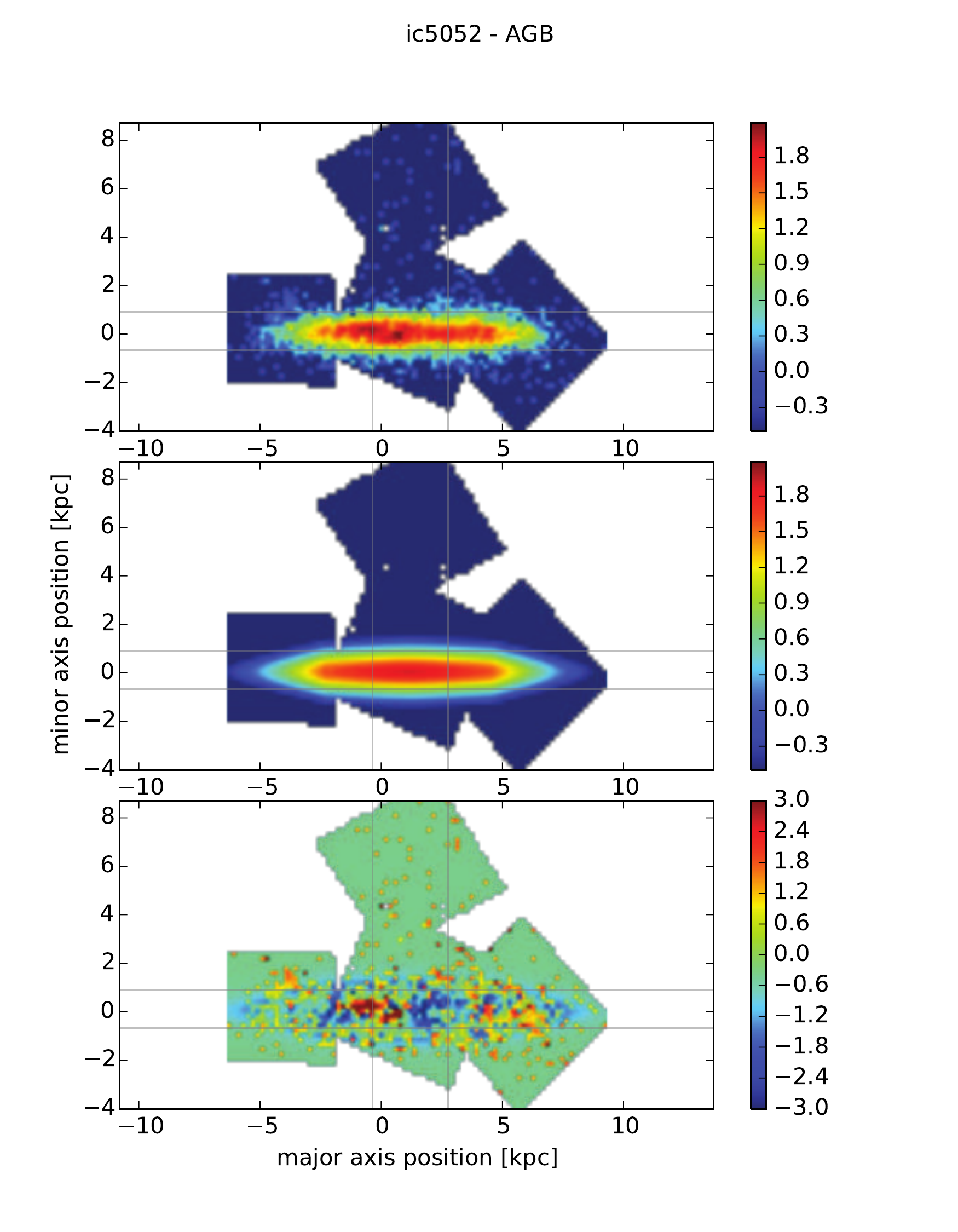}
  \end{minipage}
  \begin{minipage}[t]{0.245\linewidth}
    \centering
    \includegraphics[width=0.99\columnwidth]{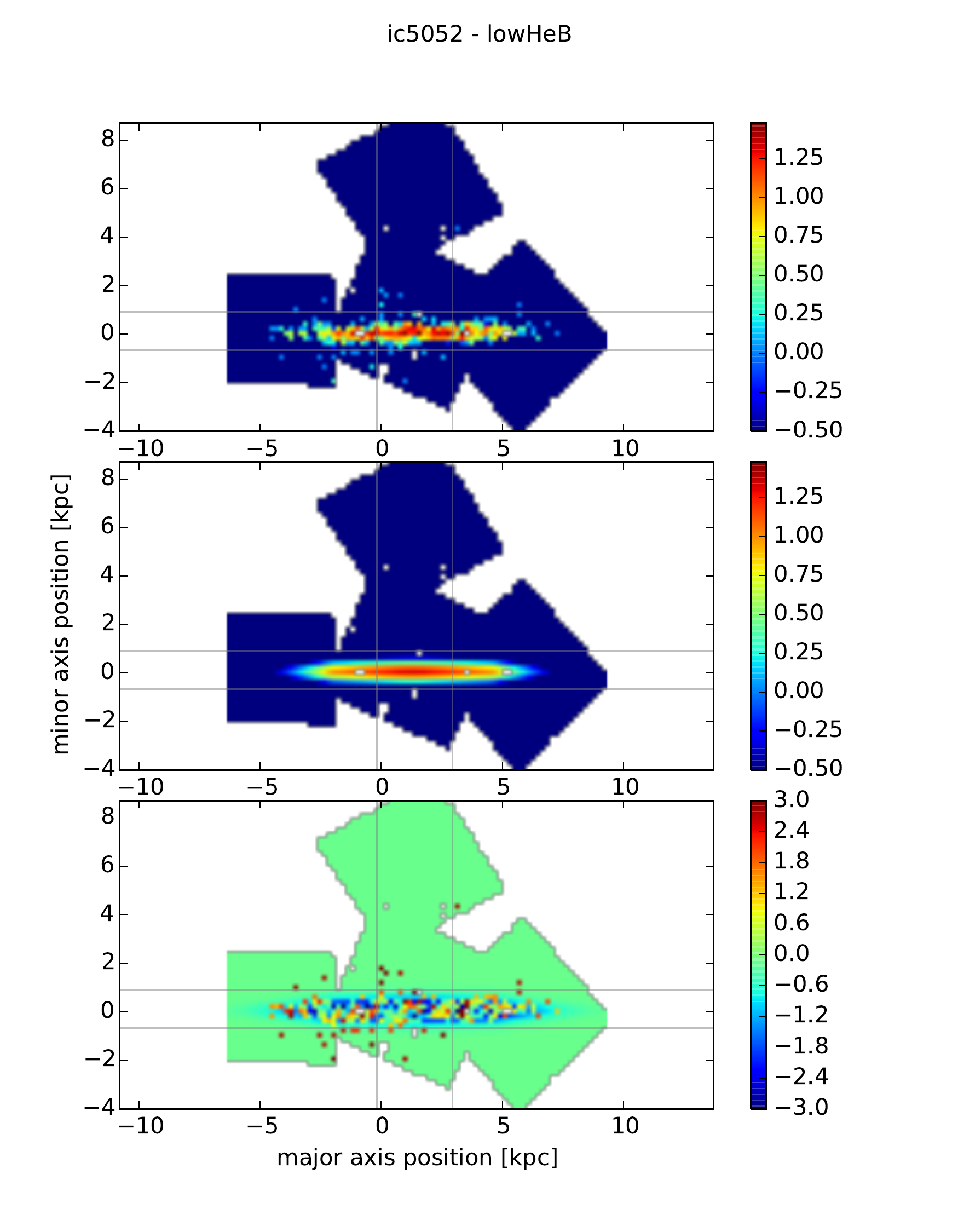}
  \end{minipage}
  \begin{minipage}[t]{0.245\linewidth}
    \centering
    \includegraphics[width=0.99\columnwidth]{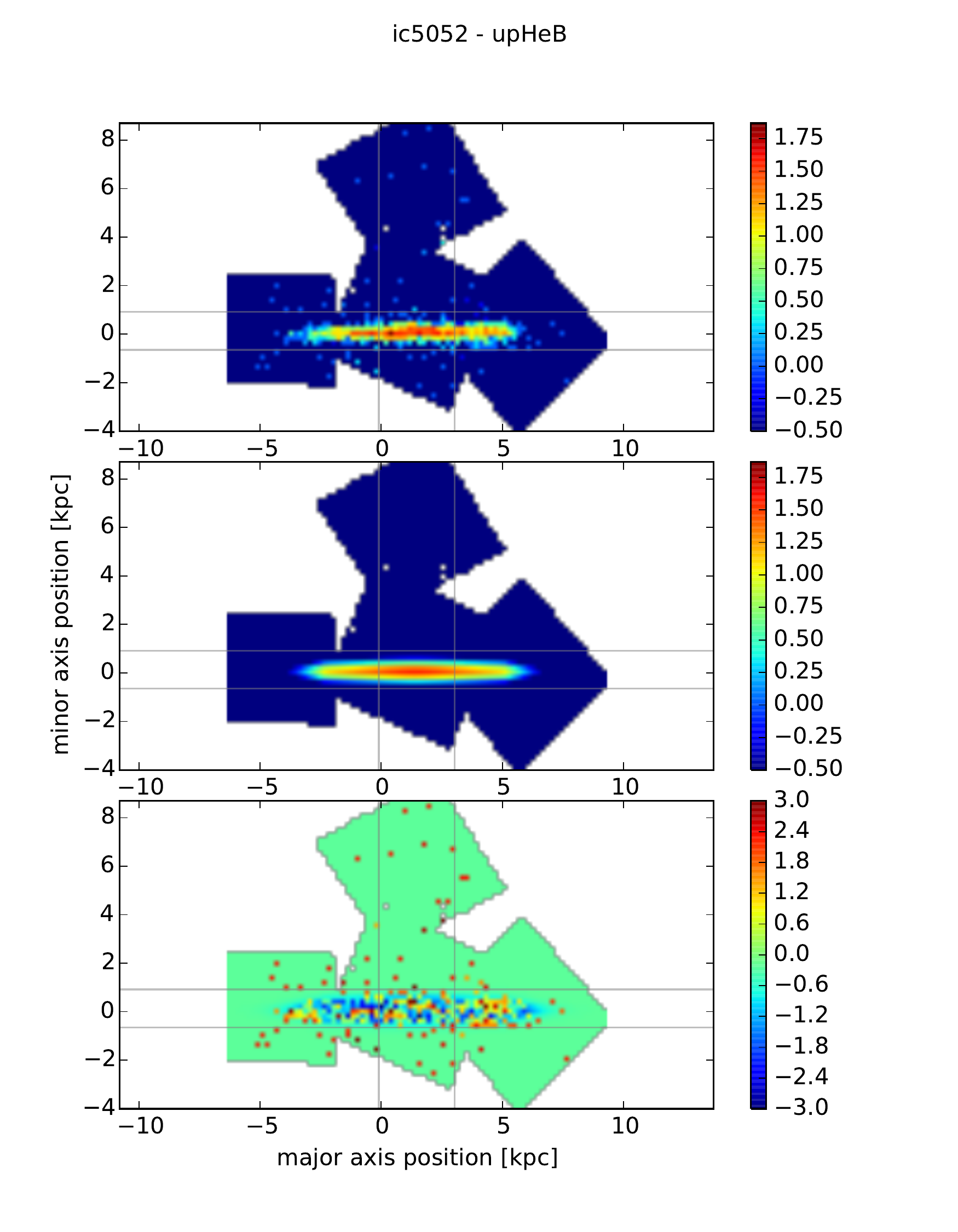}
  \end{minipage}
  \begin{minipage}[t]{0.245\linewidth}
    \centering
    \includegraphics[width=0.99\columnwidth]{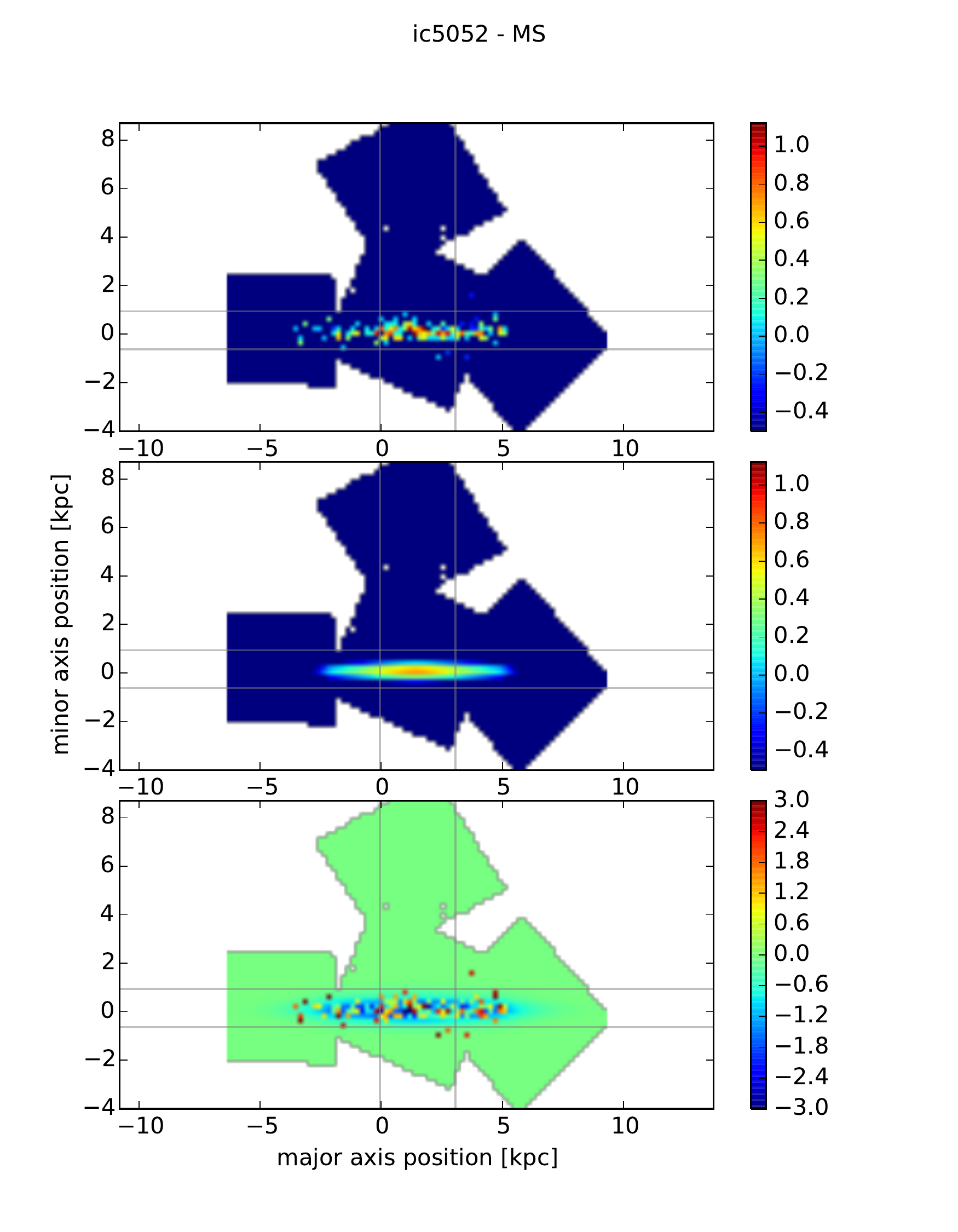}  \end{minipage}
    \centering
  \begin{minipage}[t]{0.37\linewidth}
    \centering
    \includegraphics[width=0.99\columnwidth]{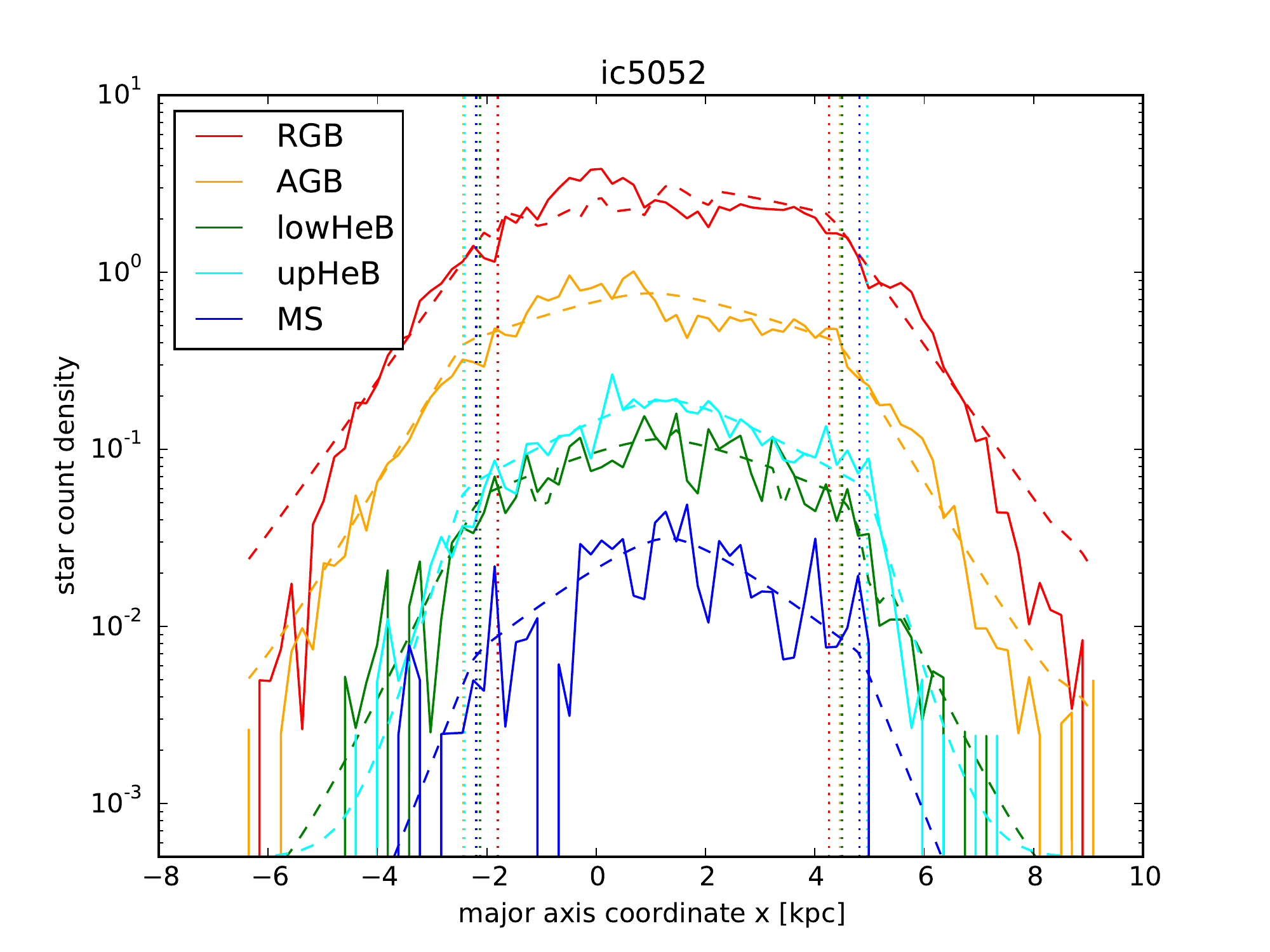}
  \end{minipage}
  \begin{minipage}[t]{0.37\linewidth}
    \centering
    \includegraphics[width=0.99\columnwidth]{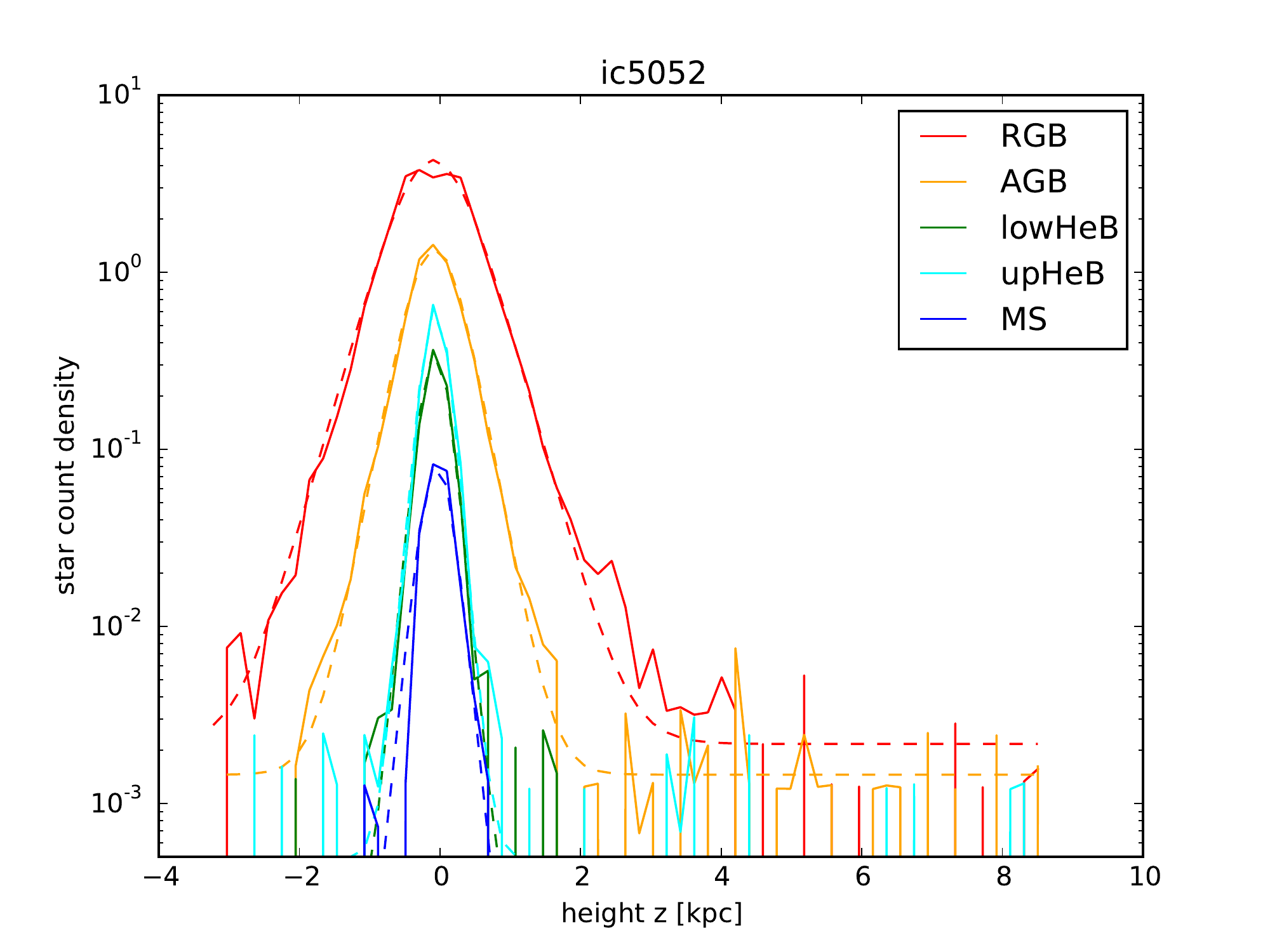}
  \end{minipage}
\caption[2D fits of IC\,5052]{\emph{Top row:} 2D fit maps of the five populations in IC\,5052 (from left to right: RGB, AGB, lowHeB, upHeB, MS). Each panel contains three maps: \emph{top:} stellar density maps, \emph{middle:} best-fit model, \emph{bottom:} weighted residual (data$-$model, see Eq.~\ref{equ:residuals}). The gray lines show the regions from which the minor and major axis profiles are extracted. The two overdensities in the residual map of the fit to the RGB data (left panel); \emph{bottom row (middle and right):} major and minor axis profiles of the data (solid lines) and the 2D models (dashed lines); dotted vertical lines show the break radii. The model lines may show irregularities (as the real data do) due to masked-out regions.}
\label{fig:ic5052-2Dfits}
\end{sidewaysfigure*}

\begin{sidewaysfigure*}[!ht]
  \begin{minipage}[t]{0.245\linewidth}
    \centering
    \includegraphics[width=0.99\columnwidth]{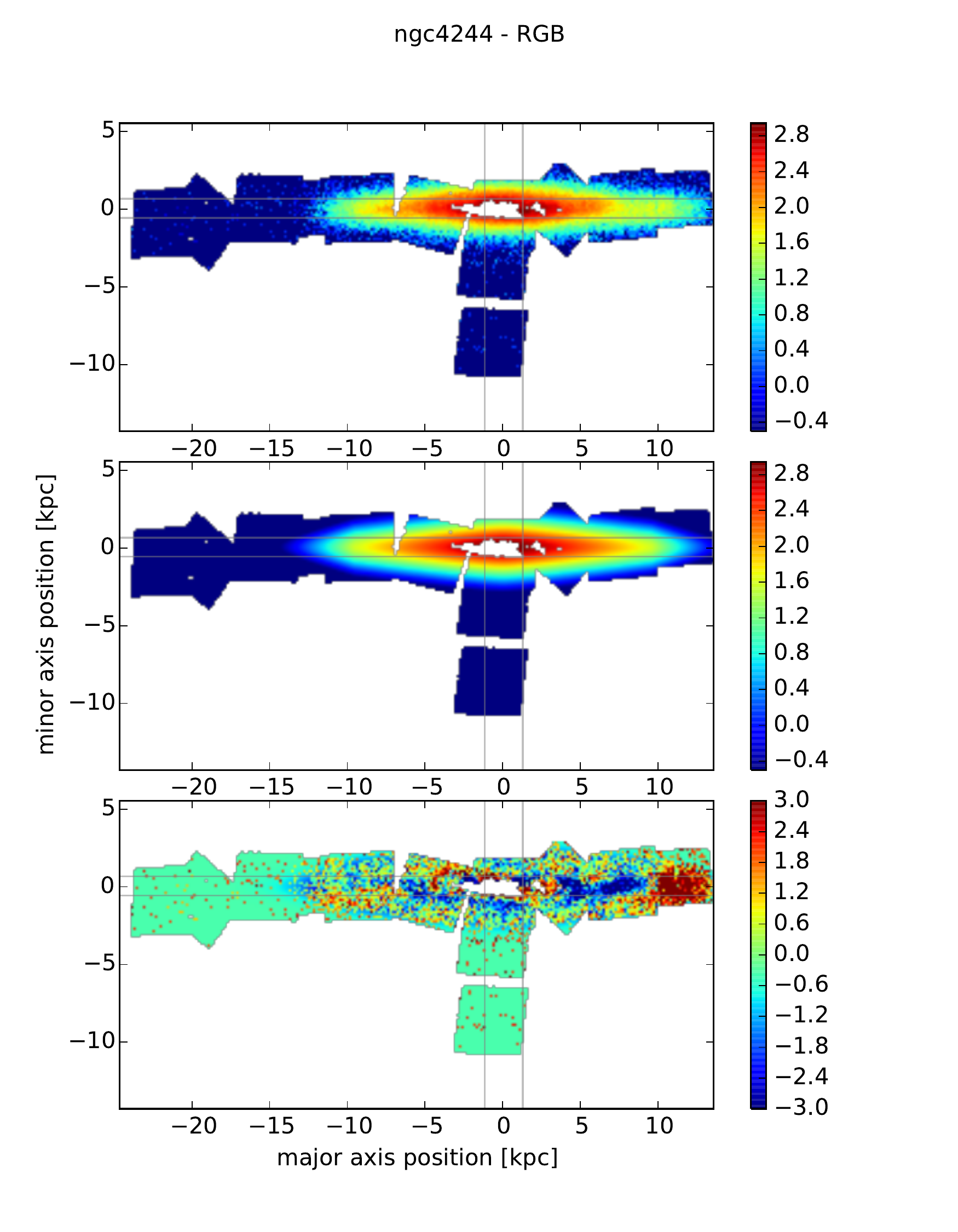}
  \end{minipage}
  \begin{minipage}[t]{0.245\linewidth}
    \centering
    \includegraphics[width=0.99\columnwidth]{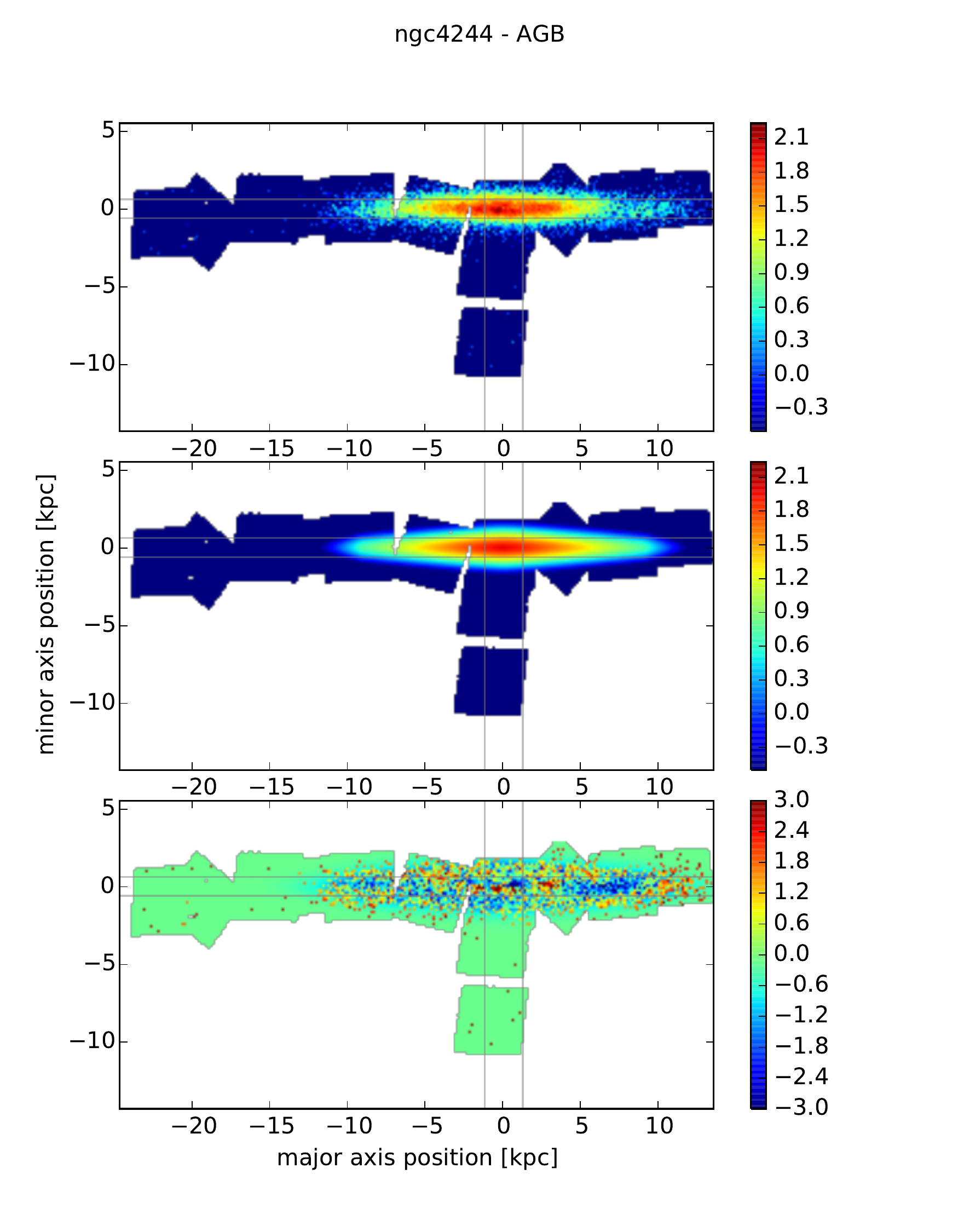}
  \end{minipage}
  \begin{minipage}[t]{0.245\linewidth}
    \centering
    \includegraphics[width=0.99\columnwidth]{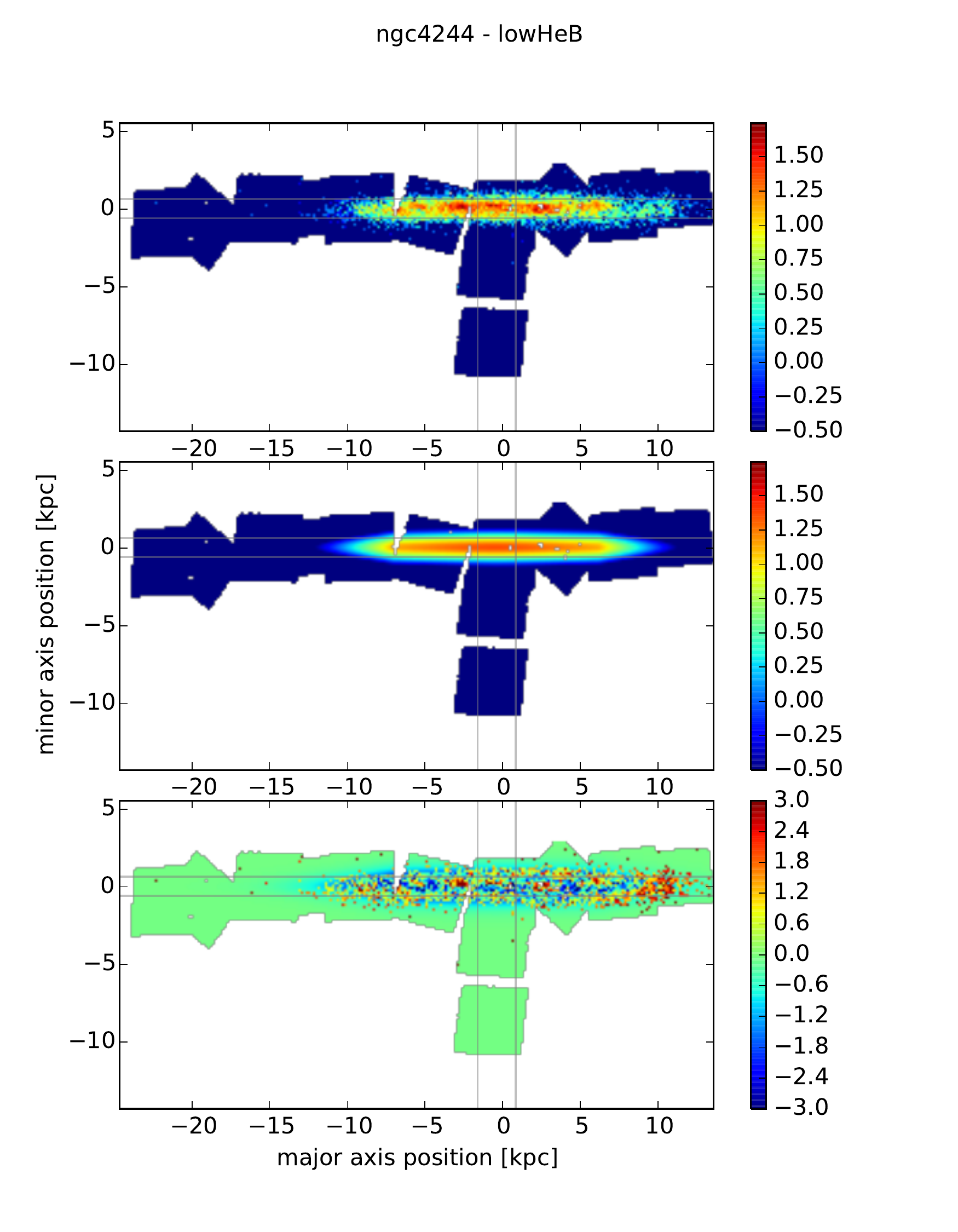}
  \end{minipage}
  \begin{minipage}[t]{0.245\linewidth}
    \centering
    \includegraphics[width=0.99\columnwidth]{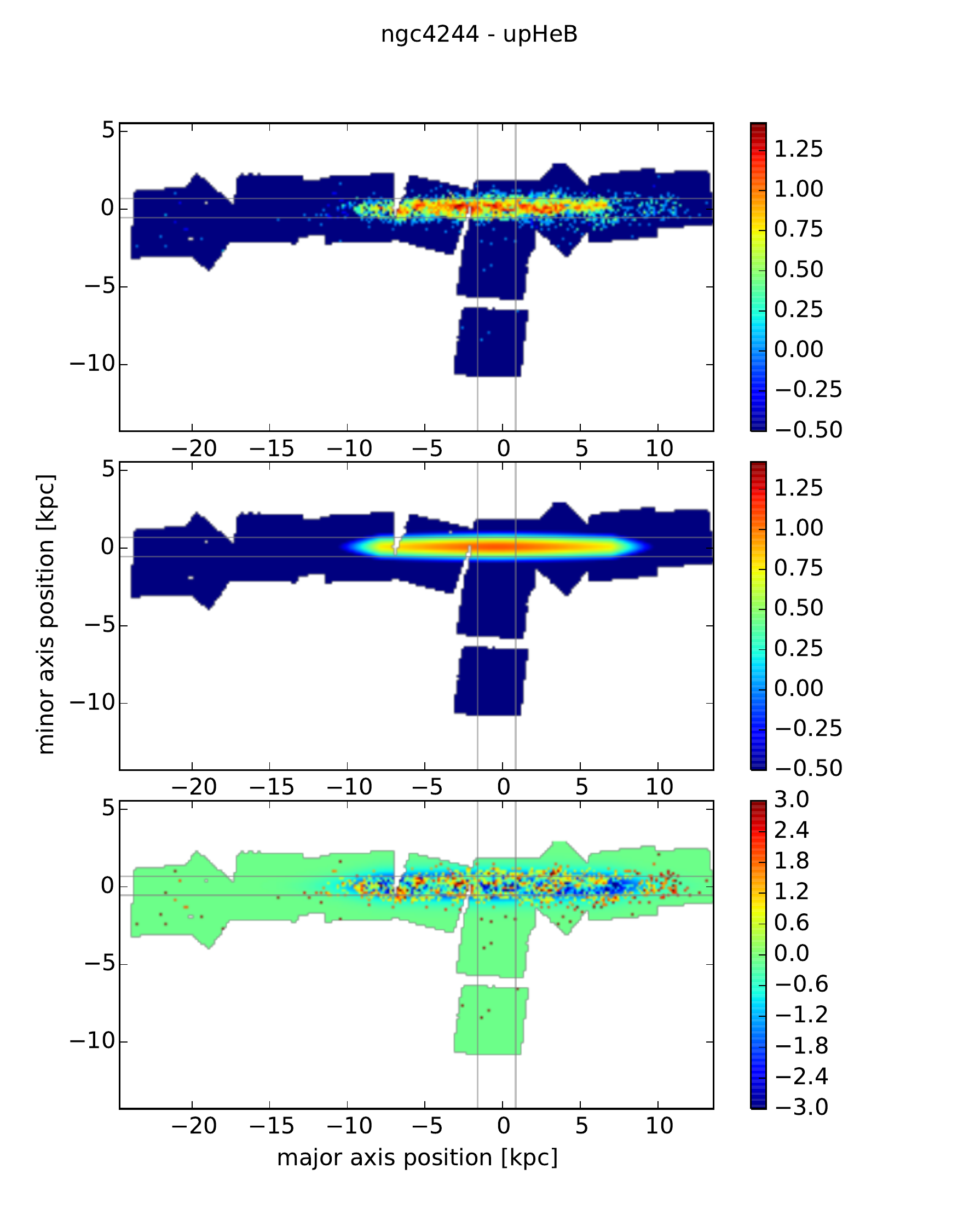}
  \end{minipage}
  \begin{minipage}[t]{0.245\linewidth}
    \centering
    \includegraphics[width=0.99\columnwidth]{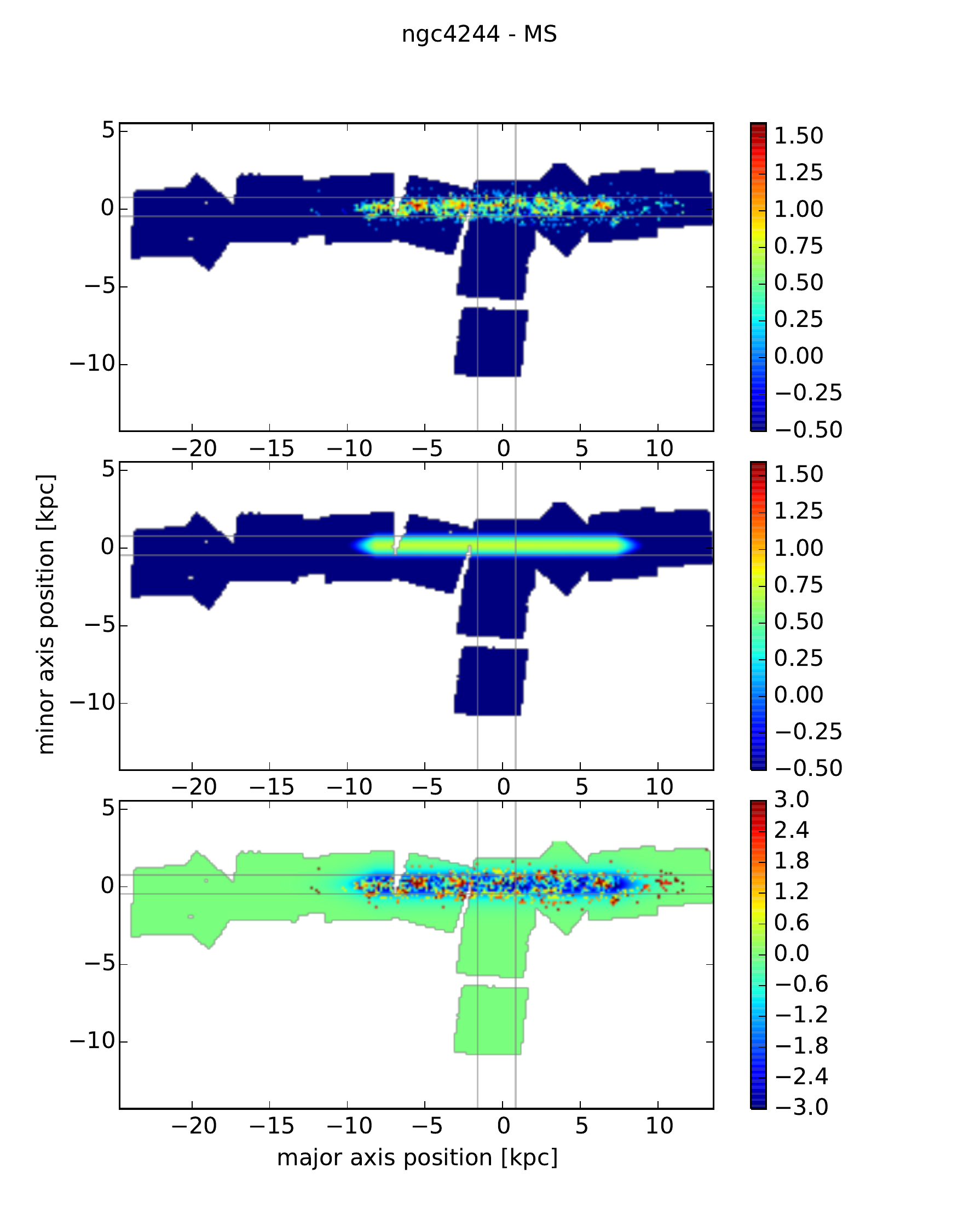}
  \end{minipage}
    \centering
  \begin{minipage}[t]{0.37\linewidth}
    \centering
    \includegraphics[width=0.99\columnwidth]{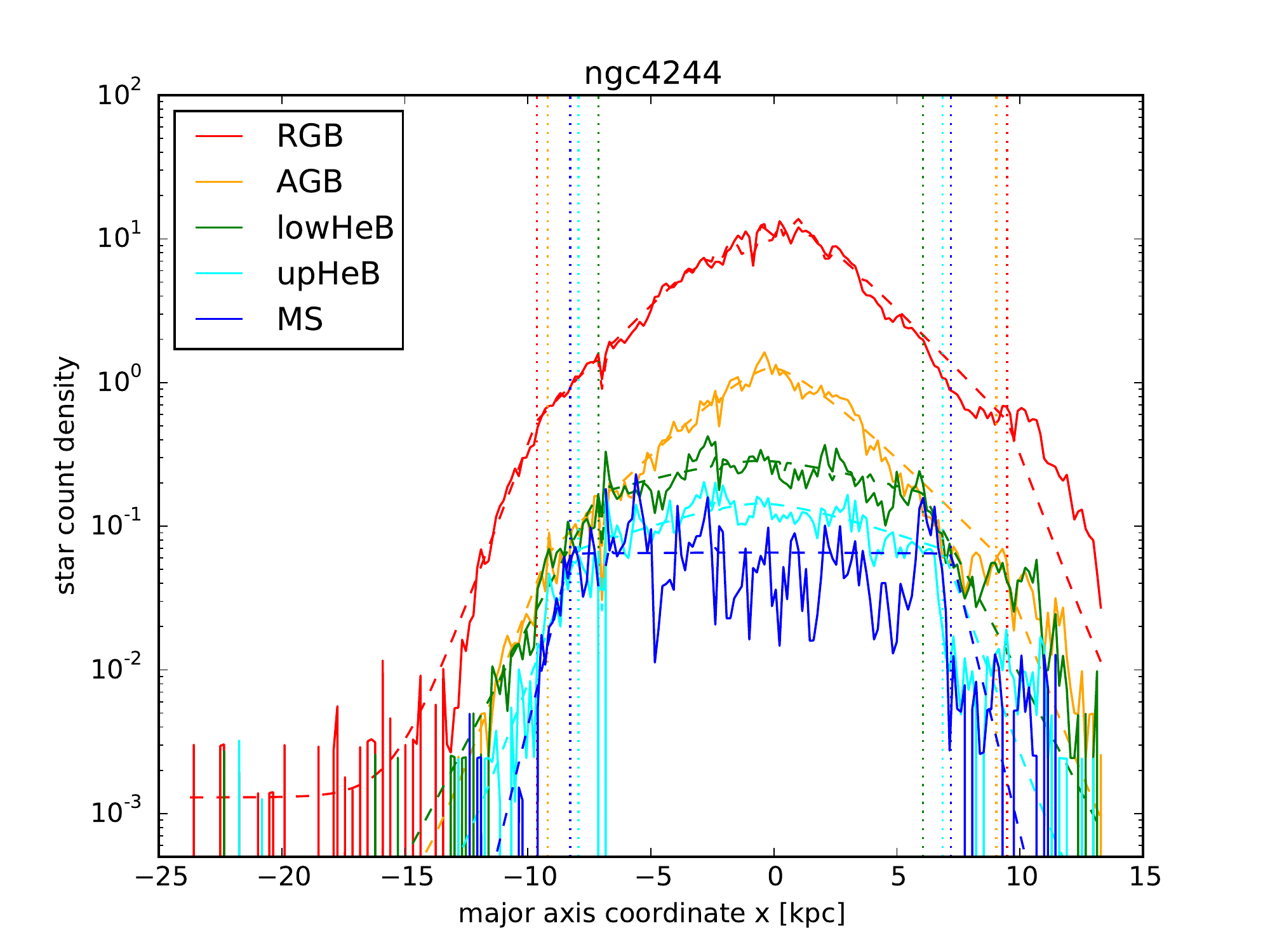}
  \end{minipage}
  \begin{minipage}[t]{0.37\linewidth}
    \centering
    \includegraphics[width=0.99\columnwidth]{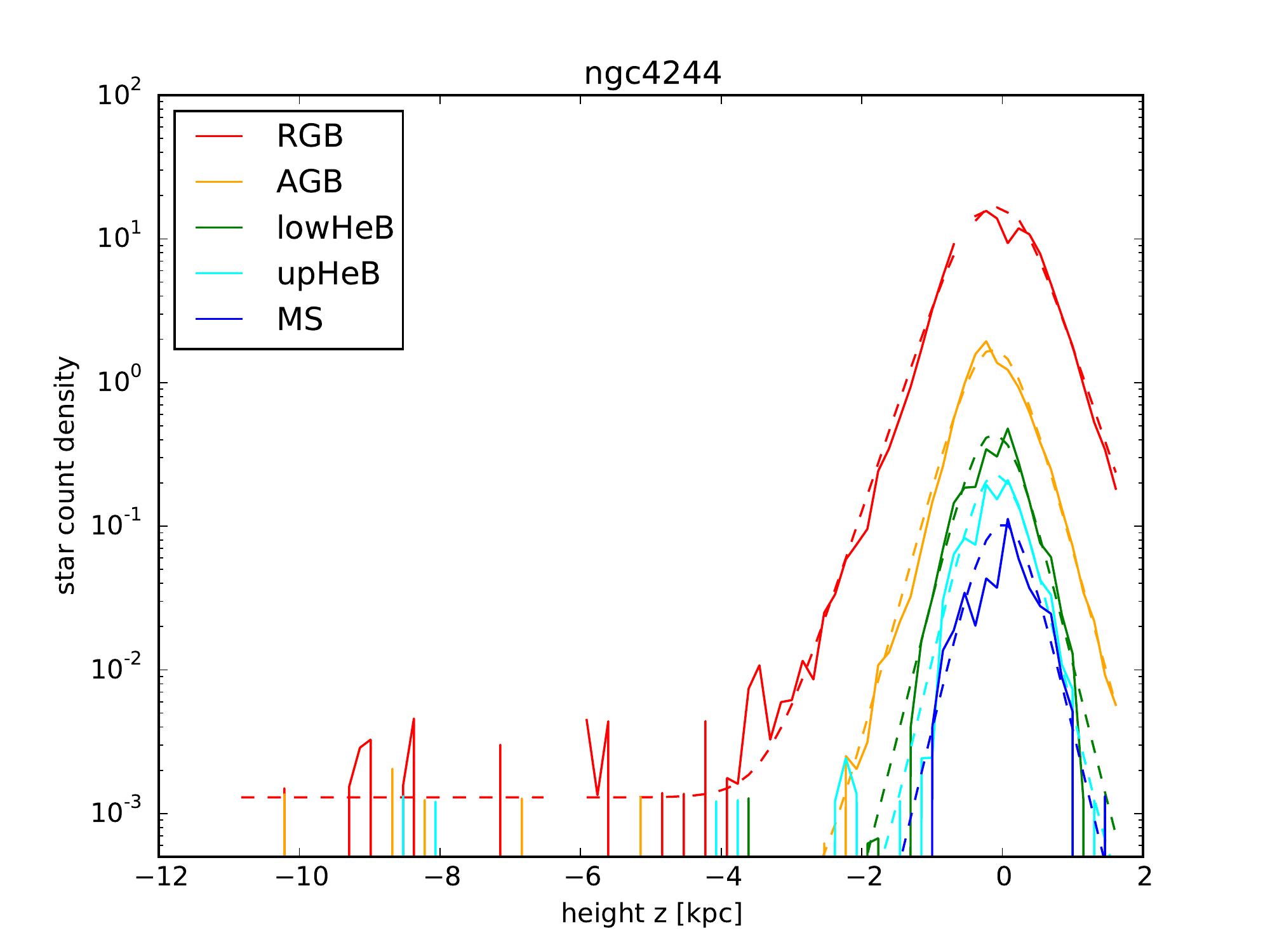}
  \end{minipage}
\caption[2D fits of NGC\,4244]{Same as Fig.~\ref{fig:ic5052-2Dfits}, but for NGC\,4244. We restricted the fits to the left side of the galaxy because of irregularities at the right end.}
\label{fig:ngc4244-2Dfits}
\end{sidewaysfigure*}

\begin{sidewaysfigure*}[!ht]
  \begin{minipage}[t]{0.245\linewidth}
    \centering
    \includegraphics[width=0.99\columnwidth]{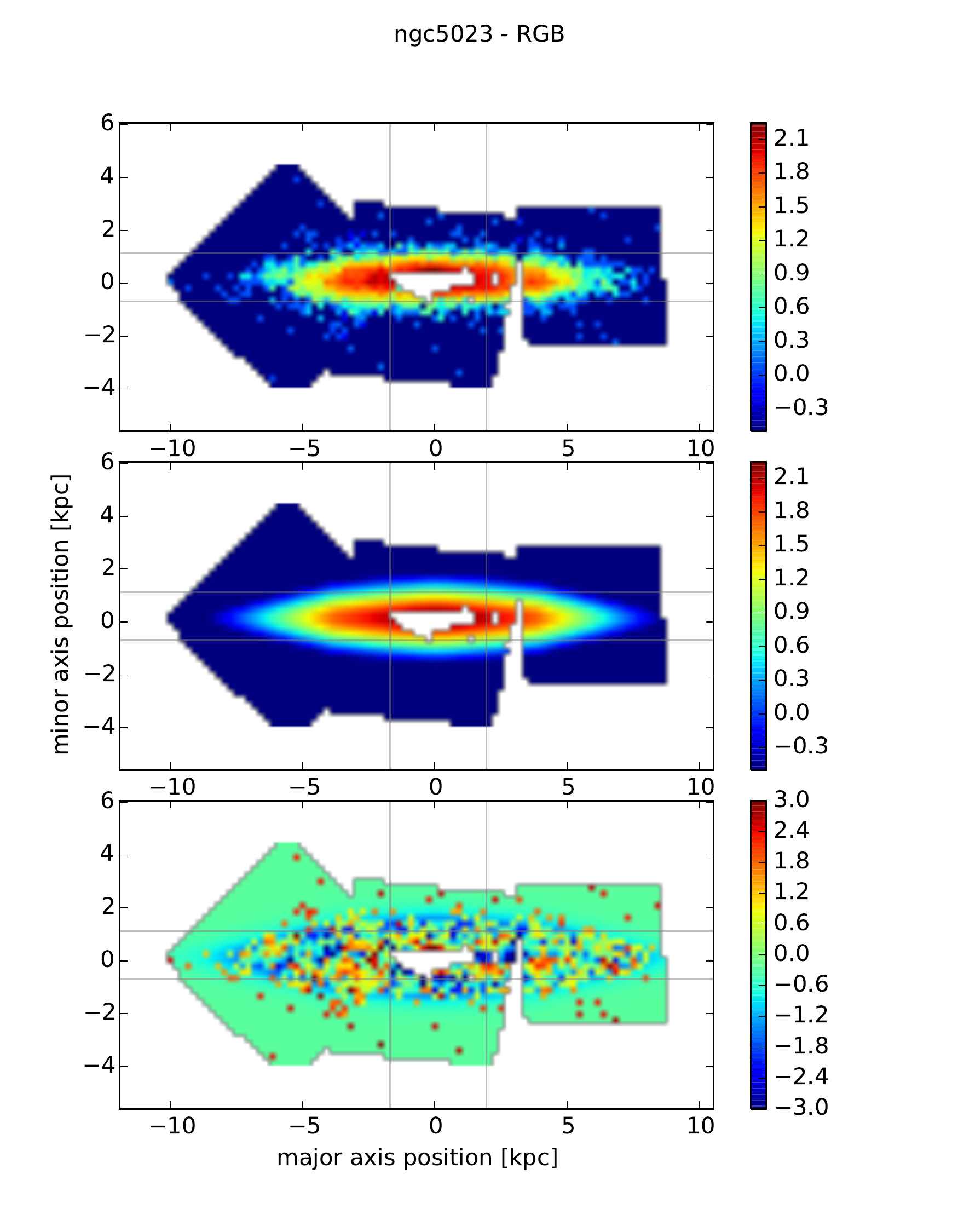}
  \end{minipage}
  \begin{minipage}[t]{0.245\linewidth}
    \centering
    \includegraphics[width=0.99\columnwidth]{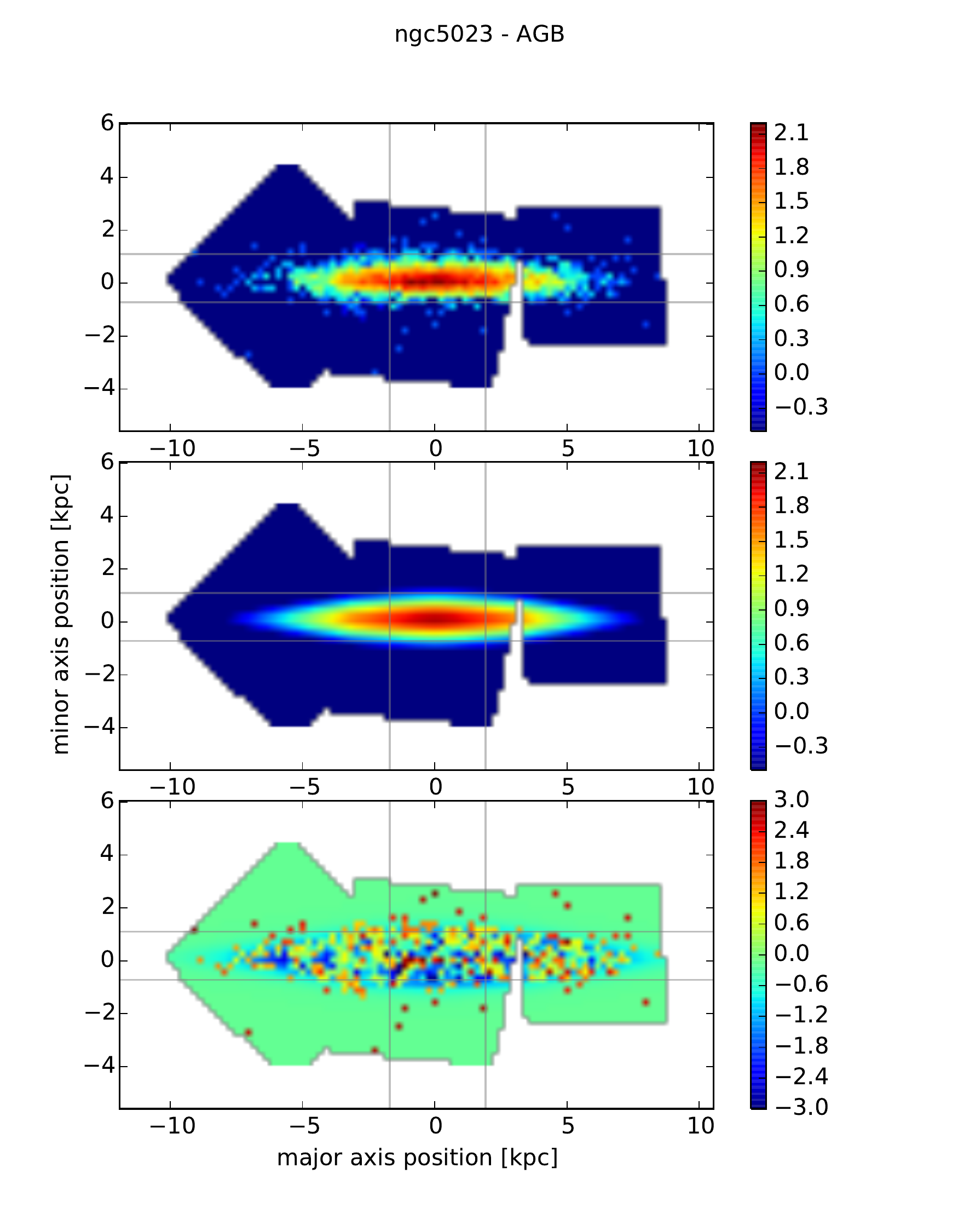}
  \end{minipage}
  \begin{minipage}[t]{0.245\linewidth}
    \centering
    \includegraphics[width=0.99\columnwidth]{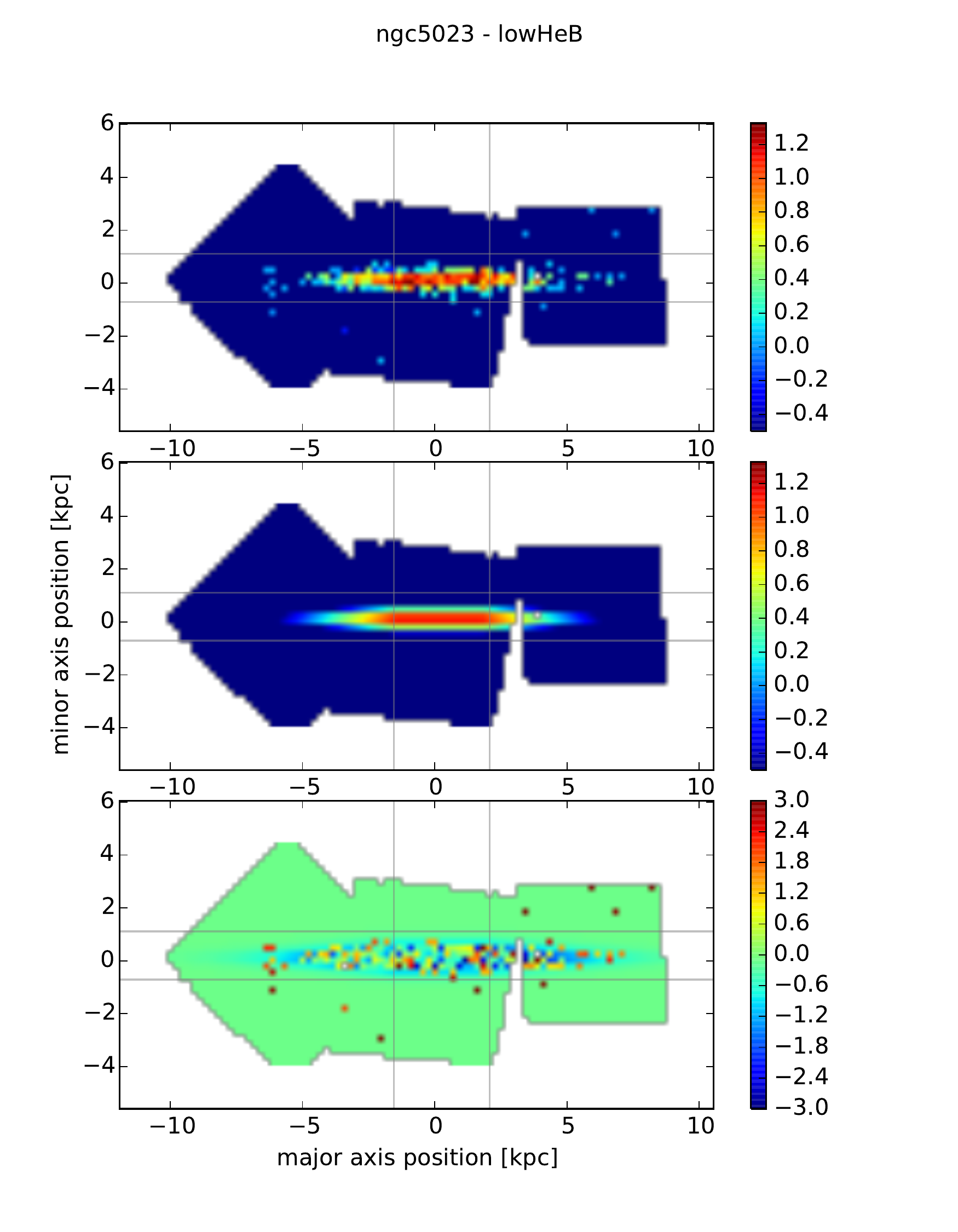}
  \end{minipage}
  \begin{minipage}[t]{0.245\linewidth}
    \centering
    \includegraphics[width=0.99\columnwidth]{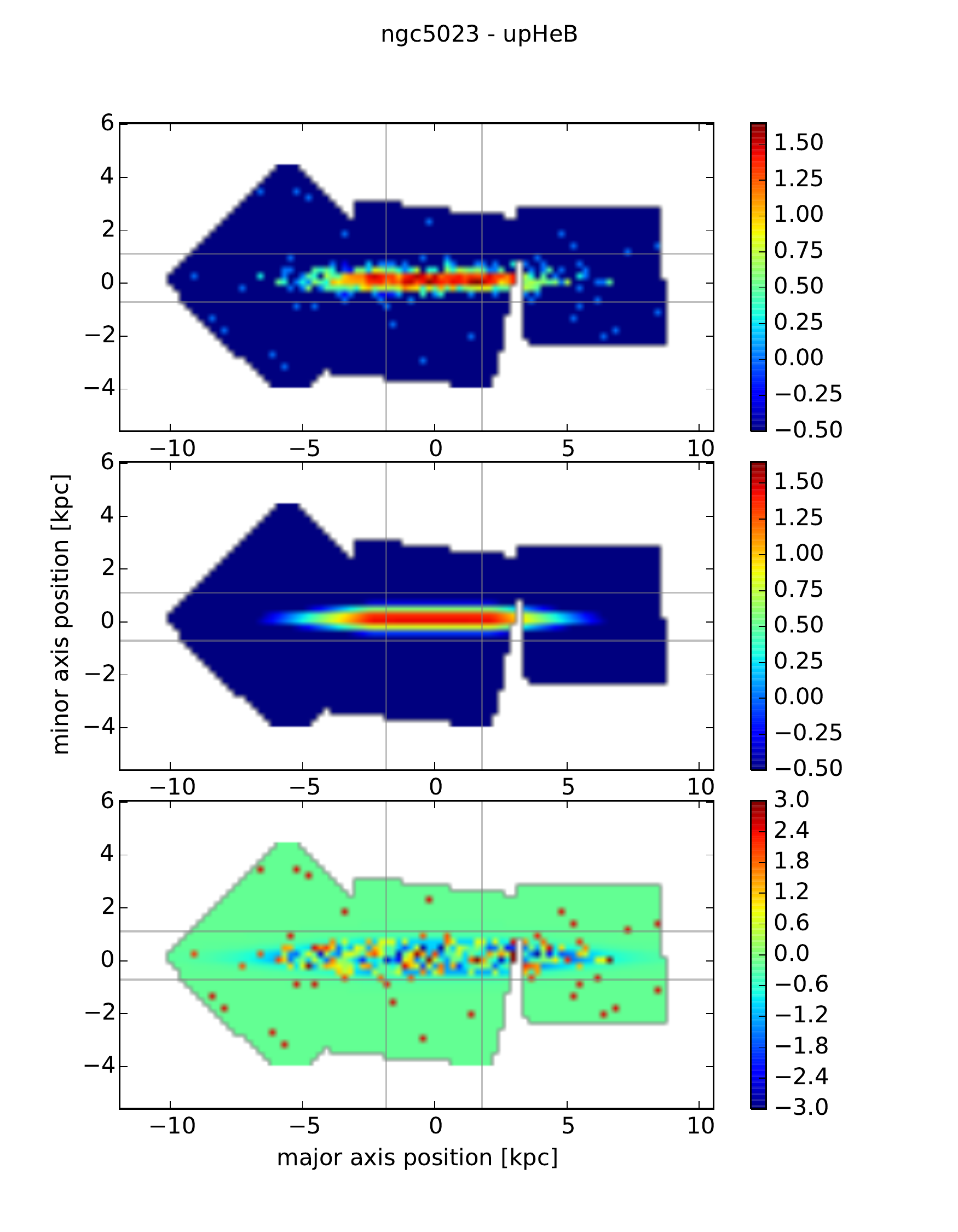}
  \end{minipage}
  \begin{minipage}[t]{0.245\linewidth}
    \centering
    \includegraphics[width=0.99\columnwidth]{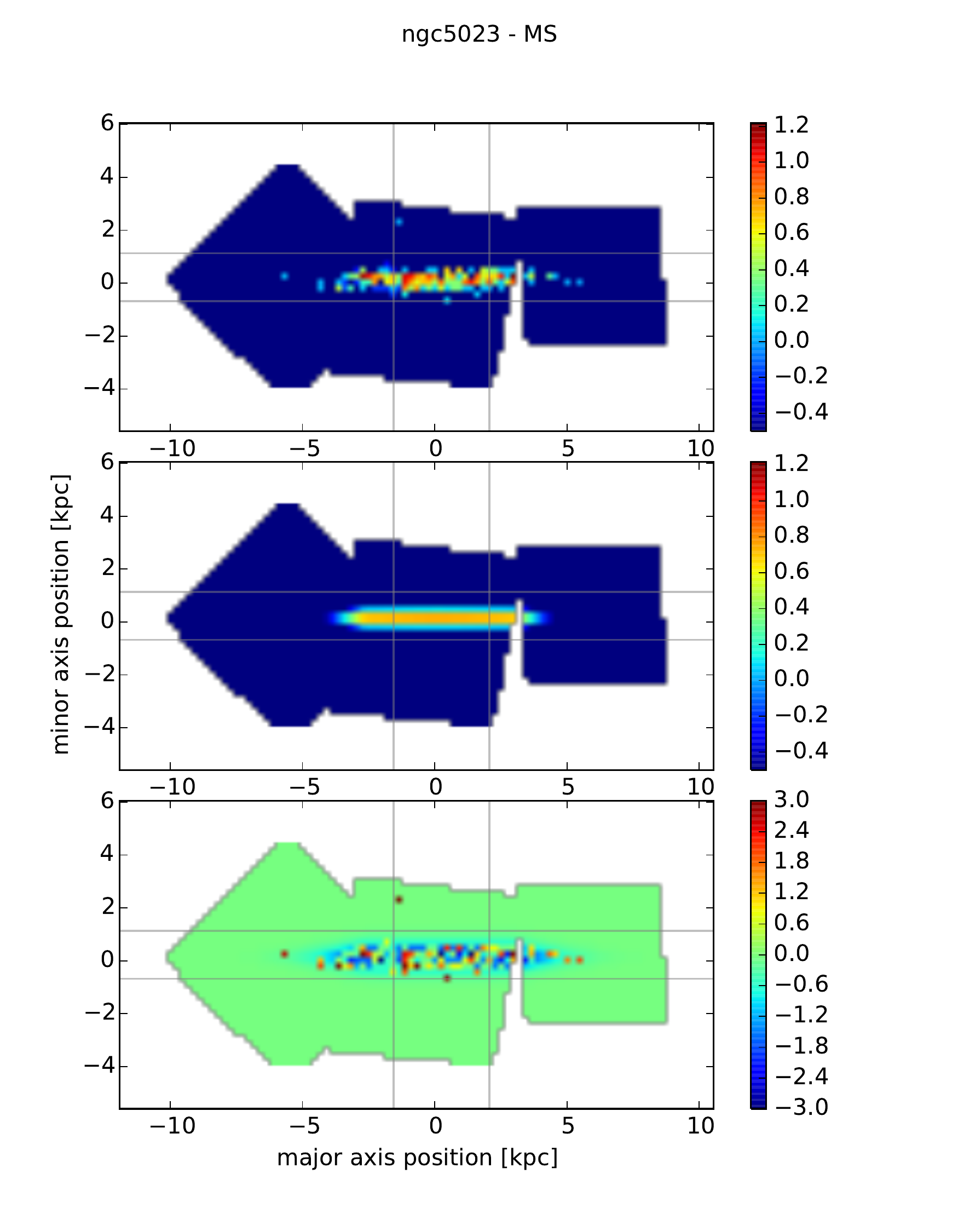}
  \end{minipage}
    \centering
  \begin{minipage}[t]{0.37\linewidth}
    \centering
    \includegraphics[width=0.99\columnwidth]{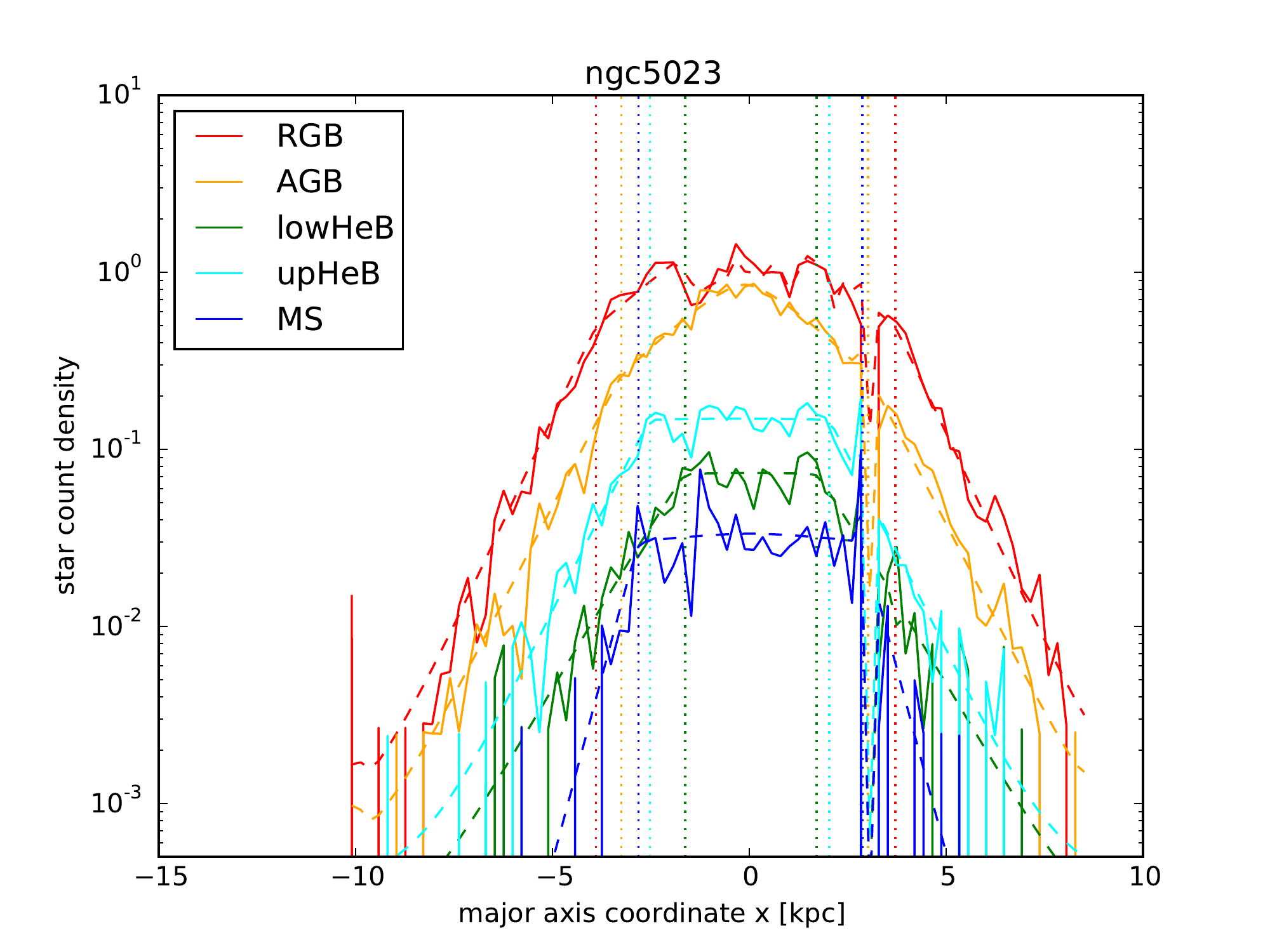}
  \end{minipage}
  \begin{minipage}[t]{0.37\linewidth}
    \centering
    \includegraphics[width=0.99\columnwidth]{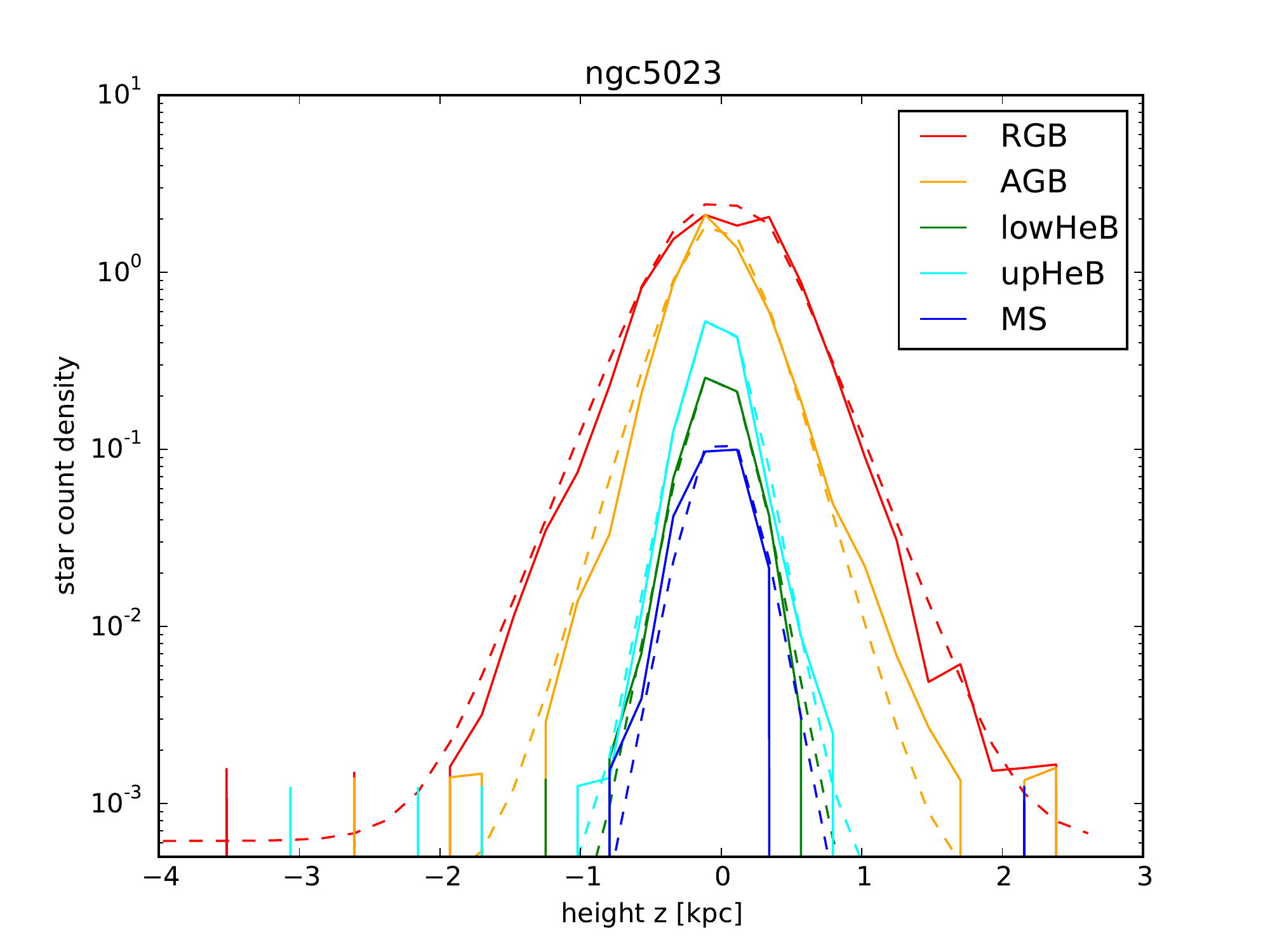}
  \end{minipage}
\caption[2D fits of NGC\,5023]{Same as Fig.~\ref{fig:ic5052-2Dfits}, but for NGC\,5023.}
\label{fig:ngc5023-2Dfits}
\end{sidewaysfigure*}

\end{appendix}

\end{document}